\documentclass{article}
\pdfoutput=1

    \usepackage[final,nonatbib]{neurips_2020}

\usepackage[utf8]{inputenc} 
\usepackage[T1]{fontenc}    
\usepackage{hyperref}       
\usepackage{url}            
\usepackage{booktabs}       
\usepackage{amsfonts}       
\usepackage{nicefrac}       
\usepackage{microtype}      
\usepackage{wrapfig}
\usepackage[ruled,linesnumbered]{algorithm2e}
\usepackage{appendix}
\usepackage{xcolor}
\usepackage{amsmath}
\usepackage{import}
\usepackage{tikz}
\usepackage{caption}
\usepackage{subcaption}
\usepackage{pgf,tikz} 
\usepackage{environ}
\usepackage{graphicx}
\usepackage{multicol}
\usepackage[perpage]{footmisc}
\usepackage{tikz-dependency}
\usepackage{pgfplots}
\usetikzlibrary{automata}
\usetikzlibrary{shapes, decorations, arrows, calc, arrows.meta, fit, positioning}
\usepackage{dsfont}

\usepackage{color, colortbl}
\definecolor{LightGreen}{rgb}{0.6,1,0.66}

\usepackage[utf8]{inputenc}
\usepackage[english]{babel}
\usepackage{kky}
\usepackage{cleveref}
\usepackage{bm}
\usepackage{subcaption}
\usepackage{multirow}
\usepackage{listings}

\definecolor{dkgreen}{rgb}{0,0.6,0}
\definecolor{gray}{rgb}{0.5,0.5,0.5}
\definecolor{mauve}{rgb}{0.58,0,0.82}
\definecolor{snippet_bg}{rgb}{1, 1, 0.9}

\makeatletter
\def\blfootnote{\gdef\@thefnmark{}\@footnotetext}
\makeatother

\usepackage{array}
\newcolumntype{P}[1]{>{\centering\arraybackslash}p{#1}}

\usepackage{algorithm2e}

\crefname{theorem}{theorem}{theorem}
\crefname{corollary}{corollary}{corollary}
\crefname{assumption}{assumption}{assumption}
\crefname{lemma}{lemma}{lemma}
\crefname{remark}{remark}{remark}
\crefname{proposition}{proposition}{proposition}
\crefname{conjecture}{conjecture}{conjecture}
\crefname{definition}{definition}{definition}

\DeclareFontFamily{U}{wncy}{}
\DeclareFontShape{U}{wncy}{m}{n}{<->wncyr10}{}
\DeclareSymbolFont{mcy}{U}{wncy}{m}{n}
\DeclareMathSymbol{\Sh}{\mathord}{mcy}{"58}

\newcommand{\ours}{Pyfectious }

\newcommand{\ar}[1]{{\color{blue} {}}}
\newcommand{\ashkan}[1]{{\color{cyan} {}}}
\newcommand{\amin}[1]{{\color{red} {}}}
\newcommand{\stefan}[1]{{\color{red} {}}}
\newcommand{\bernhard}[1]{{\color{purple} {}}}

\usepackage{pifont}
\newcommand{\cmark}{\ding{51}}%
\newcommand{\xmark}{\ding{55}}%

\DeclareMathSymbol{:}{\mathrel}{operators}{"3A}

\title{Pyfectious: An individual-level simulator to discover optimal containment polices for epidemic diseases}

\author{%
  Arash Mehrjou\textsuperscript{*}\\
  Max Planck Institute for Intelligent Systems\\
  T\"ubingen, Germany \&\\
  ETH Z\"urich, Z\"urich, Switzerland\\
  \texttt{amehrjou@ethz.ch} \\
  \And
  Ashkan Soleymani\textsuperscript{*} \\
  Max Planck Institute for Intelligent Systems \\
  T\"ubingen, Germany \\
  \texttt{ashkan.soleymani@tuebingen.mpg.de} \\
  \And
  Amin Abyaneh \\
  Max Planck Institute for Intelligent Systems \\
  T\"ubingen, Germany \\
  \texttt{amin.abyaneh@tuebingen.mpg.de} \\
  \And
  Samir Bhatt \\
  Faculty of Medicine,  School of Public Health \\
  Imperial College \\
  London, UK  \\
  \texttt{s.bhatt@imperial.ac.uk} \\
  \And
  Bernhard Sch\"olkopf \\
  Max Planck Institute for Intelligent Systems \\
  T\"ubingen, Germany \\
  \texttt{bs@tuebingen.mpg.de} \\
  \And
  Stefan Bauer \\
  Max Planck Institute for Intelligent Systems \& \\ CIFAR Azrieli Global Scholar \\
  T\"ubingen, Germany  \\
  \texttt{stefan.bauer@tuebingen.mpg.de} \\
}

\begin{document}

\maketitle

\begin{abstract}
    Simulating the spread of infectious diseases in human communities is critical for predicting the trajectory of an epidemic and verifying various policies to control the devastating impacts of the outbreak. Many existing simulators are based on compartment models that divide people into a few subsets and simulate the dynamics among those subsets using hypothesized differential equations. However, these models lack the requisite granularity to study the effect of intelligent policies that influence every individual in a particular way. In this work, we introduce a simulator software capable of modeling a population structure and controlling the disease's propagation at an individualistic level. In order to estimate the confidence of the conclusions drawn from the simulator, we employ a comprehensive probabilistic approach where the entire population is constructed as a hierarchical random variable. This approach makes the inferred conclusions more robust against sampling artifacts and gives confidence bounds for decisions based on the simulation results. To showcase potential applications, the simulator parameters are set based on the formal statistics of the COVID-19 pandemic, and the outcome of a wide range of control measures is investigated. Furthermore, the simulator is used as the environment of a reinforcement learning problem to find the optimal policies to control the pandemic. The obtained experimental results indicate the simulator's adaptability and capacity in making sound predictions and a successful policy derivation example based on real-world data. As an exemplary application, our results show that the proposed policy discovery method can lead to control measures that produce significantly fewer infected individuals in the population and protect the health system against saturation.\blfootnote{Correspondence to: Arash Mehrjou (\texttt{amehrjou@ethz.ch})}\blfootnote{$^*$ Equal contributions.}
\end{abstract}

\section{Introduction}
\label{sec:introduction}

The approaches to control an epidemic disease such as COVID-19 are divided into two main categories: {\bf 1)} pharmaceutical and {\bf 2)} non-pharmaceutical. While the first category involves medication and vaccination, the second approach, which is the main interest of the current work, concerns interventions on human communities to slow down the spread of the disease~\cite{specktor2020coronavirus}. The objective of non-pharmaceutical methods is to reduce the growth rate of the infection to prevent collapsing the healthcare systems that are widely known as~\emph{flattening the curve}~\cite{specktor2020coronavirus}. When infectious diseases cause an epidemic, strict control measures such as bans on crowded public events, travel restrictions, limiting public transportation, and minimizing physical contacts are commonly adopted by countries to control the number of infections and prevent their healthcare system overburdening~\cite{ferguson2020report}. These methods are collectively known as~\emph{social distancing}. 

Social distancing policies are often based on expert's common sense and previous experiences in partially similar conditions~\cite{specktor2020coronavirus}. For instance, vaccination~\cite{tseng2012immunization}, travel restrictions~\cite{chinazzi2020effect}, school closure~\cite{viner2020school}, and wearing protective instruments such as masks~\cite{leung2020respiratory} were used during SARS-CoV\footnote{Severe Acute Respiratory Syndrome Corona Virus} in $2003$ whose effect were evaluated through several studies.

A detailed study of the effect of various policies after an epidemic breaks out requires a precise population model. Models are developed at different levels of abstraction. Compartment models divide the population into sub-groups and model the spread of the disease as a system of differential equations whose states are the size of each sub-group~\cite{wang2020four}. By defining more fine-grained sub-groups, the model becomes more accurate and realistic, while at the same time, it becomes computationally more demanding. In a more detailed extreme, the model states are the health conditions of every individual in the population.

There are two primary benefits in modeling the population at the level of individuals: {\bf 1)} They can be utilized to verify more abstract models; that is, the collective behavior observed in compartment models must be aligned with the aggregation of the states of the fine-grained models. {\bf 2)} They allow investigating the policies influencing every individual uniquely. Therefore, more sophisticated policies can be proposed compared to compartment-level policies that equally affect all members of a particular community. For example, superspreaders are known to have a critical role in driving a pandemic. Investigating the effect of controlling them is crucial to direct the limited control resources more effectively. This study and similar ones of fine-grained control measures are not feasible in compartment models.

We have developed \ours, a light-weight python-based individual-level simulator software together with a probabilistic generative model of the structured population of an arbitrary city. This software's components are developed by having the ultimate idea that it is going to be used as a reinforcement learning environment that is fast yet detailed enough to discover non-trivial control policies for real-world pandemics in the future. We hope that the same role that the Chess rules played for AlphaZero to learn a superhuman chess player~\cite{silver2017mastering}, \ours would play for a general-purpose reinforcement learning algorithm to learn the best policy to control epidemic diseases. The software and its accompanying examples are available at \href{https://github.com/amehrjou/Pyfectious}{\color{blue}{https://github.com/amehrjou/Pyfectious}}.

Therefore, our work's main contributions compared to existing simulators in different aspects such as population generation, simulation, policy enforcement, and runtime details are discussed below.

\begin{itemize}
    \item \textbf{Population model:} We enumerate below various aspects of generating the society by creating the individuals. These aspects include attributes of each individual, their roles in the society, their daily schedule, the types of interaction with each other, etc. (See~\Cref{results_table_population_model} for comparison).
    
    \begin{itemize}
        \item \textbf{Generality:} This criterion determines the detail level of the population, that is, the granularity of the simulator in modeling the attributes of individuals and their interactions in the real world. For instance, in a school, the simulator's ability in modeling students, teachers, cleaning crew, and other roles is pivotal to be faithful to the real-world dynamics of schools. Here, we exemplified some details that \ours take into account when simulating schools at this granularity level. The population of a school community is divided into roles such as students and teachers based on the individuals' attributes. For example, students of the same class belong to the same age group. The other important factor is the type of interactions among individuals of different groups. For example, the interactions among the students of the same class are more frequent and effective in transmitting the disease than interactions among the students of different classes. Moreover, unlike the simulators that allow individuals to take simple roles, \ours is capable of modeling individuals with complicated and compound roles. For example, an individual may have a job, be a member of some friend gatherings, use public transport, eat regularly at some restaurants, be a member of a gym, and so on. The design of \ours is entirely suitable to model these details.
        
        \item \textbf{Extendability: } The roles of the individuals, their corresponding daily schedule, the interactions among them, and their personal attributes are all manually configurable in \ours. The format of providing these parameters as the input to \ours is detailed in~\Cref{manual-simulation}. This level of flexibility allows \ours to be configured for any arbitrary city based on the real-world statistics of that city's population structure. Having maximum flexibility has been one of the initial design goals for \ours to make it suitable for a wide range of applications, including discovering the optimal control measures for any target city.
        
        \item \textbf{Probabilisticity: } Another fundamental design principle of \ours is to make the conclusions derived from the simulation's outcomes robust against slight change in the provided settings. Therefore, every parameter of the simulation is assumed to be a realization of a distribution that can be provided as the input configuration. Therefore, the entire structured population of the city is a large hierarchical random variable. Many cities with almost the same population structure can be sampled from this random variable to derive confidence bounds on the simulation's obtained results.
        
    \end{itemize}
    
    \item \textbf{Simulation:} The configuration parameters of the simulator (See ~\Cref{results_table_simulation}) that are explained below determine the accuracy vs. computation trade-off of the simulation.
    
    \begin{itemize}
        \item \textbf{Event/clock-based simulation: } The simulators of temporal events often belong to one of two categories: event-based or clock-based. In an event-based simulation, queues of planned events are executed by their order of occurrences. In a clock-based simulation, a series of periodic changes to simulator modules occur at specified time intervals. To gain more computational efficiency for the purpose of this work, we combine these two methods into a clock/event-based approach. Briefly speaking, it acts as an event-based simulation equipped with a running background clock. Even though the events in the queue are executed by the specified order, the event of interest (transmission of the disease) occurs only at the clock edges. The detailed description of this mixed proposed simulation method and its advantages come in~\Cref{sec:timer}.
        
        \item \textbf{Multiresolution time: } The settable clock/event method that was briefly described above allows \ours to be multiresolution. The resolution is controlled by the running background clock period and is adjustable according to the available computational resources.
        
        \item \textbf{Interaction model: } The transmission of the disease is through an underlying graph that determines the structure of the population. This structure allows a detailed simulation of the individuals' mobility and their interactions. The daily schedule of each individual and her exposure time to other individuals can be thoroughly modeled. The further details, such as the transmission of disease by touching a surface that an infected individual already touches, can also be modeled thanks to the flexibility of the underlying connectivity graph.
    \end{itemize}
    
    \item \textbf{Policy enforcement: } To mitigate the spread of the virus and mortality, governments enforce policies to limit the potential ways of disease transmission. A special feature of \ours is the possibility to construct smart policies that can act on each individual differently according to her condition. A mixture of policies with different levels of granularity can be enforced during the simulation; at the same time, the statistics of the disease are fed back to the controller and update the policy concurrently. A diverse set of built-in policies is provided with \ours, and they can be a good starting point for a user to modify and investigate the result of an arbitrary policy. This feature makes \ours a fast and full-fledged environment for a reinforcement learning algorithm to infer optimal policies given a specified cost function such as mortality, active infected cases, etc. To showcase this possibility, an experiment for policy discovery is provided in~\Cref{section:finding_an_optimal_policy}. The features mentioned above are summarized in~\Cref{results_table_policy}.
    
    \begin{itemize}
        \item \textbf{Flexibility: } The design architecture of \ours gives maximum flexibility for the policy design in terms of what features can be altered by a policy. Any dynamic change in the connectivity graph, testing resources, disease, and individual attributes are possible. Hence, in addition to common real-world policies such as contact tracing, social distancing, testing, vaccination, closures, quarantines, full or partial lockdowns, more complicated and individual-specific policies can be easily studied.
        
        \item \textbf{Probabilistic control: } Probabilistic control measures are allowed in \ours. For example, a policy may enforce quarantining randomly chosen $50\%$ of school's students each day.
        
        \item \textbf{Conditional policy: } Conditional policies are critical for smart and efficient control of an epidemic disease. In the terminology of control theory, this resembles feedback controllers where the action applied to the system depends on the observations from the systems' states. This feature allows modeling real-world policies such as the closure of public places when the number of active cases increases and re-opening them when it decreases. Examples of these feedback policies are discussed in~\Cref{section:automated_restrictions}.
        
        \item \textbf{Extendability: } A simple user-friendly language is developed to write policies of interest or extend the wide set of built-in policies. Each policy consists of two components: The condition that triggers the policy and the control measure, which is the action taken by the policy when the triggering condition is satisfied.
        
        \item \textbf{Policy discovery: } By defining a cost function, e,g, the peak of the confirmed infected cases, a wide range of policies can be tested in parallel at each round and use their outcome as a learning signal for the RL agent to move in the space of feasible policies towards those with more desirable results. As an example, in~\Cref{section:finding_an_optimal_policy}, the optimal policy is inferred using Bayesian optimization when the peak of the curve of confirmed cases is taken as the cost function.
        
    \end{itemize}
    
    \item \textbf{Implementation overview: } Thanks to the aforementioned novelties in the implementation at multiple levels from algorithms to software architecture, \ours achieves superior scalability in the size of the simulated population and also simulation's duration compared with other simulators (see~\Cref{results_table_technical}). Comprehensive documentation and code-snippets of the use cases are provided as Jupyter notebooks to facilitate a quick start in running experiments with an arbitrary setting for epidemic researchers or policymakers.

\end{itemize}

This paper aims to describe the novelties of the proposed light-weight and scalable simulator software called \ours. The software is designed with the ultimate goal to become a fast environment for a reinforcement learning agent to discover detailed and effective individual-level policies to control the spread of the disease in a structured population. An extensive set of experiments shows the application of \ours in simulating the disease's dynamics in the population, testing the effectiveness of expert-designed policies, and automatic discovery of effective policies.

\setlength{\tabcolsep}{4pt}

\begin{table}[t]
	\renewcommand{\arraystretch}{1.3}
	\caption{Comparing the population model in various epidemic simulation softwares.}
	\label{results_table_population_model}
	\centering
	\scriptsize
	\begin{tabular}{|l | c c c c c | } 
		
		\hline
		 & \multicolumn{5}{c|}{Population Model} \\
		\hline
		 \multirow{2}{*}{Method} & \multirow{2}{*}{Type} & \multirow{2}{*}{Detail Level} & \multirow{2}{*}{Probabilistic} & \multirow{2}{*}{Generative} & Real-world Data\\
		  & & & & & Awareness \\
		\hline
		\hline
		SNDS ~\cite{block2020social} & Social Network & Moderate & \cmark & \cmark & \xmark \\
		\hline
		
		Age-structured SEIR ~\cite{prem2020effect} & Compartment Level & Low & \xmark & \xmark & \cmark\\
		\hline
		
		EpiFlex ~\cite{hanley2006object} & Individual Level & High & \cmark & \cmark & \cmark\\
		\hline
		
		EpiSimS ~\cite{mniszewski2008episims, stroud2007spatial} & Individual Level & High & \cmark & \cmark & \cmark\\
		\hline
		
		EpiModel ~\cite{jenness2018epimodel} & Network Model & Moderate & \cmark & \cmark & \cmark\\
		\hline
		
		SAMSCE ~\cite{hoertel2020stochastic} & Individual Level & Moderate & \cmark & \cmark & \cmark \\
		\hline
		
		CHIME ~\cite{weissman2020locally} & Compartment Level & Low & \xmark & \xmark & \cmark \\
		\hline
		
		TDCO ~\cite{weissman2020locally} & Compartment Level & Low & \xmark & \xmark & \cmark \\
		\hline
		
		SCEDS ~\cite{carcione2020simulation} & Compartment Level & Low & \xmark & \xmark & \cmark \\
		\hline
		
		Diamond Princess Analysis ~\cite{fang2020many} & Individual Level & Moderate & \xmark & \xmark & \cmark \\
		\hline
		
		Epidemiology Workbench ~\cite{nunez2020epidemiology} & Individual Level & Moderate & \cmark & \cmark & \cmark \\
		\hline
		
		How to Restart? ~\cite{d2020restart} & Individual Level & Moderate & \cmark & \xmark & \cmark \\
		\hline
		
		SoEcNetwork Heterogenity ~\cite{akbarpour2020socioeconomic} & Individual Level & High & \cmark & \cmark & \cmark \\
		\hline
		
		QECTTC ~\cite{lorch2020quantifying} & Individual Level & High & \cmark & \cmark & \cmark \\
		\hline
		\hline
		
		\textbf{Pyfectious (ours)} & \textbf{Individual Level} & \textbf{High} & \textbf{\cmark} & \textbf{\cmark} & \textbf{\cmark} \\
		\hline
	\end{tabular}
\end{table}

\begin{table}[!t]
	\renewcommand{\arraystretch}{1.3}
	\caption{Comparing the employed methods to simulate the dynamics of epidemic disease in various epidemic simulation softwares.}
	\label{results_table_simulation}
	\centering
	\scriptsize
	\begin{tabular}{|l | P{3.5cm} P{3.5cm} c c | } 
		
		\hline
		 & \multicolumn{4}{c|}{Simulation Model} \\
		\hline
		 Method & Type & Interactions types for disease Transmission & Detail level & Multi-resolution\\
		\hline
		\hline
		SNDS ~\cite{block2020social} & Actor stepwise & Interaction in network & Moderate & \xmark\\
	    \hline
	    
	    Age-structured SEIR ~\cite{prem2020effect} & Continuous-time & Location-based contacts & Low & \xmark\\
	    \hline
	    
	    EpiFlex ~\cite{hanley2006object} & Event-based model &  Location-based contacts & Moderate & \xmark\\
	    \hline
	    
	    EpiSimS ~\cite{mniszewski2008episims, stroud2007spatial} & Event-based model & Location-based contacts & High & \xmark\\
	    \hline
	    
	    EpiModel ~\cite{jenness2018epimodel} & Clock-based model & Interaction in network & Moderate & \xmark\\
	    \hline
	    
	    SAMSCE ~\cite{hoertel2020stochastic} & Clock-based model & Network/Location contacts & Moderate & \xmark\\
	    \hline
	    
	    CHIME ~\cite{weissman2020locally} & Clock-based model & Contacts & Low & \xmark\\
	    \hline
	    
	    TDCO ~\cite{weissman2020locally} & Clock-based model & Contacts & Low & \xmark\\
	    \hline
	    
	    SCEDS ~\cite{carcione2020simulation} & Clock-based model & Contacts & Low & \xmark\\
	    \hline
	    
	    Diamond Princess Analysis ~\cite{fang2020many} & Event-based model & Location-based contacts & Moderate & \xmark\\
	    \hline
	    
	    Epidemiology Workbench ~\cite{nunez2020epidemiology} & Clock-based model & Location-based (Lattice) contacts & High & \xmark\\
	    \hline
	    
	    How to Restart? ~\cite{d2020restart} &  Clock-based model & Proximity-based and Exposure-time-based contacts & Moderate & \xmark\\
	    \hline
	    
	    SoEcNetwork Heterogenity ~\cite{akbarpour2020socioeconomic} &  Clock-based model & Contact matrices derived from contact network & High & \xmark\\
	    \hline
	    
	    QECTTC ~\cite{lorch2020quantifying} & Event-based model & Mobility-based contacts & High & \xmark \\
	    \hline
	    \hline
	    
	    \textbf{Pyfectious (ours)} &  \textbf{Clock/Event-based model} & \textbf{Graph/Location-based contacts} & \textbf{High} & \textbf{\cmark}\\
	    \hline
	    
	\end{tabular}
\end{table}

\begin{table}[!t]
	\renewcommand{\arraystretch}{1.3}
	\caption{Comparing simulation softwares in terms of the policy enforcement methods. Entries indicated by "-" for EpiModel~\cite{jenness2018epimodel} show that polices are not implemented yet.  
	}
	\label{results_table_policy}
	\centering
	\scriptsize
	\begin{tabular}{|l | p{3.5cm} c c c c c| } 
		
		\hline
		 & \multicolumn{6}{c|}{Policy Enforcement} \\
		\hline
		 \multirow{2}{*}{Method} & \multirow{2}{*}{Types}  & \multirow{2}{*}{Flexibility} & Probabilistic & Conditional & \multirow{2}{*}{Extendability} & Policy\\
		  & & & Policies & Policies & & Discovery\\
		\hline
		\hline
		\multirow{3}{*}{SNDS~\cite{block2020social}} & -seek similarity & \multirow{3}{*}{Moderate} & \multirow{3}{*}{\cmark} & \multirow{3}{*}{\xmark} & \multirow{3}{*}{\xmark} & \multirow{3}{*}{\xmark}\\
		 & -strengthen community & & & & &\\
		 & -repeat-contact bubble & & & & &\\
		\hline
		
		\multirow{2}{*}{Age-structured SEIR ~\cite{prem2020effect}} & -school break and holidays & \multirow{2}{*}{Low} & \multirow{2}{*}{\xmark} & \multirow{2}{*}{\xmark} & \multirow{2}{*}{\xmark} & \multirow{2}{*}{\xmark}\\
		 & -school closure, stop $90\%$ workforce  & & & & &\\
		\hline
		
		\multirow{1}{*}{EpiFlex ~\cite{hanley2006object}} & -decrease infection prob. & \multirow{1}{*}{Very Low} & \multirow{1}{*}{\xmark} & \multirow{1}{*}{\xmark} & \multirow{1}{*}{\xmark} & \multirow{1}{*}{\xmark}\\
		\hline
		
		\multirow{6}{*}{EpiSimS ~\cite{mniszewski2008episims, stroud2007spatial}} & -household quarantine & \multirow{6}{*}{Moderate} & \multirow{6}{*}{\xmark} & \multirow{6}{*}{\xmark} & \multirow{6}{*}{\xmark} & \multirow{6}{*}{\xmark}\\
		& -therapeutic treatment & & & & &\\
		& -school closures & & & & &\\
		& -social distancing & & & & &\\
		& -vaccination  & & & & &\\
		& -contact tracing  & & & & &\\
		\hline
		
		\multirow{1}{*}{EpiModel ~\cite{jenness2018epimodel}} & Not Implemented Yet & \multirow{1}{*}{Moderate} & \multirow{1}{*}{-} & \multirow{1}{*}{-} & \multirow{1}{*}{\cmark} & \multirow{1}{*}{-}\\
		\hline
		
		\multirow{4}{*}{SAMSCE ~\cite{hoertel2020stochastic}} & -lockdown & \multirow{4}{*}{Moderate} & \multirow{4}{*}{\xmark} & \multirow{4}{*}{\xmark} & \multirow{4}{*}{\xmark} & \multirow{4}{*}{\xmark}\\
		& -physical distancing & & & & &\\
		& -mask-wearing  & & & & &\\
		& -shielding of the population at risk  & & & & &\\
		\hline
		
		\multirow{1}{*}{CHIME ~\cite{weissman2020locally}} & -social distancing & \multirow{1}{*}{Low} & \multirow{1}{*}{\xmark} & \multirow{1}{*}{\xmark} & \multirow{1}{*}{\xmark} & \multirow{1}{*}{\xmark}\\
		\hline
		
		\multirow{2}{*}{TDCO ~\cite{weissman2020locally}} & -quarantine individual & \multirow{2}{*}{Very Low} & \multirow{2}{*}{\xmark} & \multirow{2}{*}{\xmark} & \multirow{2}{*}{\xmark} & \multirow{2}{*}{\xmark}\\
		& -government control  & & & & &\\
		\hline
		
		\multirow{2}{*}{SCEDS ~\cite{carcione2020simulation}} & -isolation measures & \multirow{2}{*}{Very Low} & \multirow{2}{*}{\xmark} & \multirow{2}{*}{\xmark} & \multirow{2}{*}{\xmark} & \multirow{2}{*}{\xmark}\\
		& -social distancing  & & & & &\\
		\hline
		
		\multirow{2}{*}{Diamond Princess Analysis ~\cite{fang2020many}} & -self-protection scenarios & \multirow{2}{*}{Moderate} & \multirow{2}{*}{\xmark} & \multirow{2}{*}{\xmark} & \multirow{2}{*}{\xmark} & \multirow{2}{*}{\xmark}\\
		& -control scenarios  & & & & &\\
		\hline
		
		\multirow{4}{*}{Epidemiology Workbench ~\cite{nunez2020epidemiology}} & -self-isolation & \multirow{4}{*}{Moderate} & \multirow{4}{*}{\xmark} & \multirow{4}{*}{\xmark} & \multirow{4}{*}{\xmark} & \multirow{4}{*}{\xmark}\\
		& -social distancing & & & & &\\
		& -testing  & & & & &\\
		& -contact tracing  & & & & &\\
		\hline

		\multirow{3}{*}{How to Restart? ~\cite{d2020restart}} & -social distancing solutions & \multirow{3}{*}{Moderate} & \multirow{3}{*}{\xmark} & \multirow{3}{*}{\xmark} & \multirow{3}{*}{\xmark} & \multirow{3}{*}{\xmark}\\
		& -use of respiratory protective devices & & & & &\\
		& -control of COVID-19 infectors & & & & &\\
		\hline

		\multirow{1}{*}{SoEcNetwork Heterogenity ~\cite{akbarpour2020socioeconomic}} & social distance policies (change the structure of network) & \multirow{1}{*}{High} & \multirow{1}{*}{\xmark} & \multirow{1}{*}{\xmark} & \multirow{1}{*}{\xmark} & \multirow{1}{*}{\xmark}\\
		\hline
		
		\multirow{3}{*}{QECTTC ~\cite{lorch2020quantifying}} & -lockdown & \multirow{3}{*}{Moderate} & \multirow{3}{*}{\xmark} & \multirow{3}{*}{\xmark} & \multirow{3}{*}{\xmark} & \multirow{3}{*}{\xmark}\\
		& -contact tracing & & & & &\\
		& -localized interventions & & & & &\\
		\hline
		\hline

		\textbf{Pyfectious (ours)} & \textbf{Almost Every Possible Policy} & \textbf{Very High} & \textbf{\cmark} & \textbf{\cmark} & \textbf{\cmark} & \textbf{\cmark} \\
		\hline
	\end{tabular}
\end{table}

\begin{table}[!t]
	\renewcommand{\arraystretch}{1.3}
	\caption{A comparison of the simulators considering different technical aspects.}
	\label{results_table_technical}
	\centering
	\scriptsize
	\begin{tabular}{|l | c c c| } 
		
		\hline
		 & \multicolumn{3}{c|}{Technical Details} \\
		\hline
		 Method & Sizewise Scalable & Timewise Scalable & Language\\
		\hline
		\hline
		SNDS ~\cite{block2020social} & \xmark (500-4000) & \xmark  & R\\
		\hline
		
		Age-structured SEIR ~\cite{prem2020effect} & \cmark & \cmark & R\\
		\hline
		
		EpiFlex ~\cite{hanley2006object} & \cmark & \cmark & Windows Software written in C++\\
		\hline
		
		EpiSimS ~\cite{mniszewski2008episims, stroud2007spatial} & \cmark & \cmark & C++\\
		\hline
		
		EpiModel ~\cite{jenness2018epimodel} & - & - & R\\
		\hline
		
	    SAMSCE ~\cite{hoertel2020stochastic} & \cmark & \cmark & C++\\
		\hline
		
		CHIME ~\cite{weissman2020locally} & \cmark & \xmark & ?\\
		\hline
		
		TDCO ~\cite{weissman2020locally} &  \cmark & \cmark  & Python \\
		\hline
		
		SCEDS ~\cite{carcione2020simulation} &  \cmark & \cmark  & Fortran \\
		\hline
		
		Diamond Princess Analysis ~\cite{fang2020many} & \xmark & \xmark & ?\\
		\hline
		
		Epidemiology Workbench ~\cite{nunez2020epidemiology} & \cmark & \cmark  & Python \\
		\hline
		
		How to Restart? ~\cite{d2020restart} & \xmark & \xmark & R \\
		\hline
		
		SoEcNetwork Heterogenity ~\cite{akbarpour2020socioeconomic} & \cmark & \cmark & R \\
		\hline
		
		QECTTC ~\cite{lorch2020quantifying} & \cmark & \cmark & Python \\
		\hline
		\hline
		
		\textbf{Pyfectious (ours)} & \textbf{\cmark} & \textbf{\cmark} & \textbf{Python} \\
		\hline
	\end{tabular}
\end{table}

We have separated two processes in the developed software package. The first process, called~\emph{Population Model}, concerns the constant part of the simulation process. It creates individuals with specified features and divides them into subsets to model the population structure of the city of interest. The other process, called~\emph{Propagation Model}, takes the properties of the disease and the dynamic interactions among the generated individuals to evolve the states of the population model in time.

\section{Population and the propagation models}

\label{sec:population_generative_model}
Many existing simulators have been developed for a particular population that is determined by their hyper-parameters. Even though it might be possible to change the hyper-parameters of a simulated city manually, it may not be straightforward to transfer the simulator to populations from which we only have partial knowledge or uncertain about some of their features. We develop a probabilistic model for every feature of the population, rendering it a fully probabilistic generative model.

To explain the logic behind the developed generative model for the population, we first introduce some terms and their role in the software. The full description of each term is discussed in \Cref{sec:population_generation}.

\begin{enumerate}
    \item \texttt{Population Generator}: This object is instantiated from a class called \texttt{Population Generator} and will be a primary container that stores the information required to generate a population, e.g., people, families, and communities. The~\texttt{Population Generator} is a wrapper around the items inscribed below.

    \item \texttt{Person}: The most fundamental object acting as the building block of the population by representing an individual is an object instantiated from the class called \texttt{Person}. The object also contains the attributes related to an individual, e.g., age, gender, and health condition.
    
    \item \texttt{Family}: Every family is an instance of the class called \texttt{Family} and is a group of multiple individuals (modeled by \texttt{Person} objects) that live together in the same house. Each family's general composition is described by a family pattern that is itself an instance of the class \texttt{Family Pattern}. A family pattern object comprises necessary attributes to generate a \texttt{Family}, such as the number of family members and their gender, age, and health condition. Similar to any other attribute in \ours, these attributes are also provided as probability distributions rather than single values. Therefore, every family pattern can be sampled multiple times to generate a set of families with an almost similar pattern but distinct values for their members' attributes. 
    
    \item \texttt{Community}: An instance of the \texttt{Community} class that describes a social unit consisting of individuals with a commonality, particularly in time and location. It is defined by a community type object and inductively by its smaller social units called subcommunities. A community type object is an instance of a class called \texttt{Community Type} that describes the attributes and subcommunities and the community's connectivity graph. A subcommunity is an instance from the \texttt{SubCommunity} class and comprises people with the same role in the community (for example, the subcommunity of teachers in a school community). A connectivity graph is an instance from the class named \texttt{Connectivity Graph} that represents the possible interactions among the individuals in the community.
\end{enumerate}

    Once the population is generated, the individuals' dynamic interactions and the features of the disease yield a model of the propagation of infection in the population. The essential factors in the propagation model are listed below and will be explained in more detail in the subsequent sections.

    \begin{enumerate}

    \item \texttt{Disease Properties}: Maintains the information related to the key characteristics of the disease that affects its propagation in a structured population. These properties include infection rate, immunity rate, mortality rate, incubation period, and disease period. These quantities are typically sampled from predetermined probability distributions leading to a stochastic representation of the disease.
    
    \item \texttt{Simulator}: The simulator employs the propagation features of the disease and the population model to evolve the simulation in time. Moreover, similar to every control task, the containment of an infectious disease demands both measurement and control. To emulate real-world processes, we develop two classes named~\texttt{Command} and~\texttt{Observer}. The former class instances are objects that mimic a single control decision (for example, shutting down schools if the number of infected students surpasses a threshold). The latter class instances mimic the measurement and monitoring processes such as testing to find dormant infected cases. Both command and observer objects need a starting time. As a general solution, we built a class named \texttt{Condition}. Each instance of this class gets activated when a defined condition in the population is met. The binary output of this object can be fed into any control or monitoring option that is supposed to be triggered when this condition is satisfied. 
    
    \item \texttt{Time}: The chronological flow of the simulation is mainly based on a queue of events, and the propagation of the disease occurs at the edges of a background timer whose frequency trades off accuracy versus computational demand.
    
    \begin{enumerate}
         \item \texttt{Event}: The time evolution of the system is implemented in an event-driven paradigm. A sequence of events determines the daily interactions among individuals. The connectivity of the population is updated when an event occurs. Three types of events are defined in \ours:
        \begin{enumerate}
            \item \texttt{Plan-Day Event} occurs at the beginning of each day and sets the schedule of that day for every individual belonging to the population.
            \item \texttt{Transition Event} occurs at times indicated by a plan-day event, and it changes the location of an individual.
            \item \texttt{Virus Spread Event} occurs when the virus propagates to an individual.
            \item \texttt{Infection Event} occurs when the infection of an individual ends.
            \item \texttt{Incubation Event} occurs when the incubation period is over and indicates a transition from the incubation period to the illness period during which the patient becomes infectious.
        \end{enumerate}
        
        \item \texttt{Infection}: This is an object associated with every infected individual and keeps track of the disease-related information. The infection object must not be confused with the infection event. The former is a container that is created for every individual when she gets infected and contains all infection-related information during the course of the disease, including the outcome that can be a recovery or death. On the other hand, an \emph{infection event} is an event object that occurs when an individual's infection ends, and she can no longer infect other individuals.
    \end{enumerate}
\end{enumerate}

In the subsequent sections, the details of the novelties in the architecture of \ours are presented. The detailed description of the probabilistic algorithm that generates the population and the event-based algorithm that evolves the simulation are discussed. To follow the details, knowing the definitions mentioned above of the objects are assumed.

\section{Software architecture of \ours}
\label{sec:simulator}
\begin{figure}[!t]
    \includegraphics[width=0.99\columnwidth]{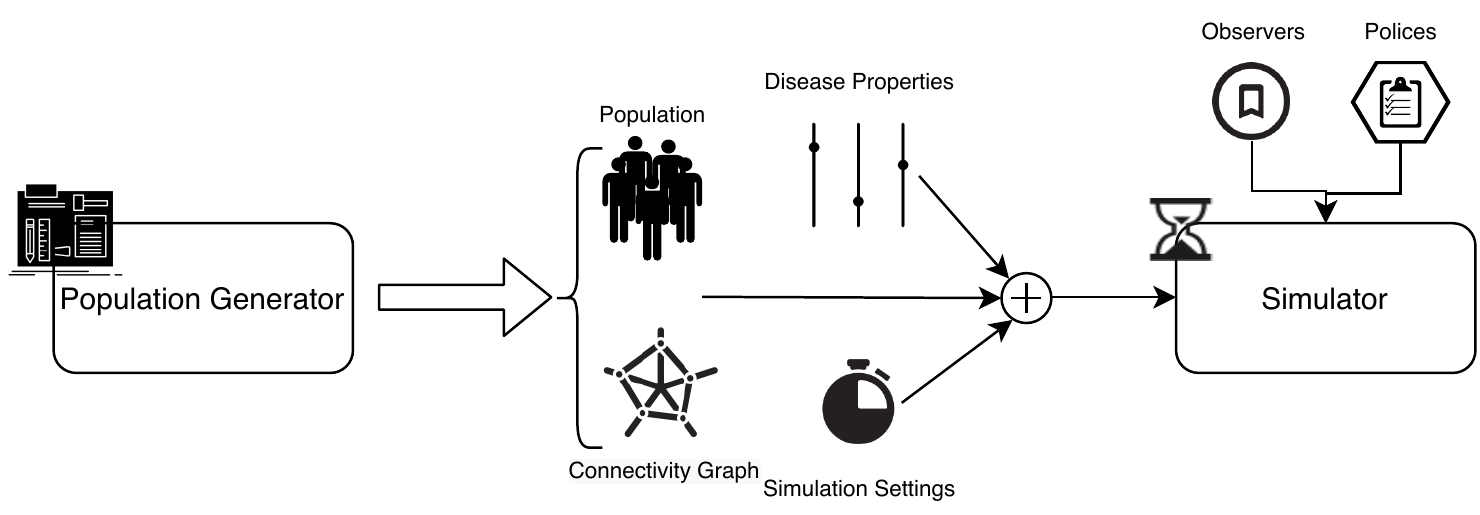}
    \caption{The pipeline of \ours. The population generator creates the individuals and assigns them their roles to form the connectivity graph. The connectivity graph, the disease properties, the clock, and the simulation settings are then fed to the simulator to create the time evolution of the disease. Furthermore, the observers and the policies are provided to the simulator in order to log data and make specific alternations to the simulation to emulate real-world epidemic control measures.}
    \label{fig:overall_pipeline_schematic}
\end{figure}

In this section, the architecture and technical details of the simulator software \ours are presented. We divide the software system into three separate components that can be considered independently: {\bf 1)} population generation, {\bf 2)} disease propagation, and {\bf 3)} time management. The first process generates the structure of a city containing a certain number of individuals that form different communities such as households, schools, shops, etc. The second process determines the properties of the disease and the way it propagates through the population. The third process ties the previous two processes together to produce the evolution of a specified disease in a population with a specified structure. The overall pipeline of the software is illustrated in~\Cref{fig:overall_pipeline_schematic}. Each process is explained in detail in the following subsections.

\subsection{Population generation}
\label{sec:population_generation}
Generating the population consists of creating a structured set of individuals together with a connectivity graph that models the interactions among them. The connectivity graph essentially shows the possible paths where the virus can transmit among individuals. Whether the graph is directed or undirected depends on the properties of the virus for which the simulation is performed. 

\begin{figure}[!t]
	\begin{subfigure}{0.49\columnwidth}
	\centering
		\begin{tikzpicture}[
		bidirected/.style={Latex-Latex},
		inner/.style={inner sep=0.2pt},
        outer/.style={inner sep=4pt}
		]
		\node[{shape=circle, text=black, minimum size=0.1em}] (v) at  (0,-1) {(1)};
		\node (img)  {\includegraphics[width=\columnwidth]{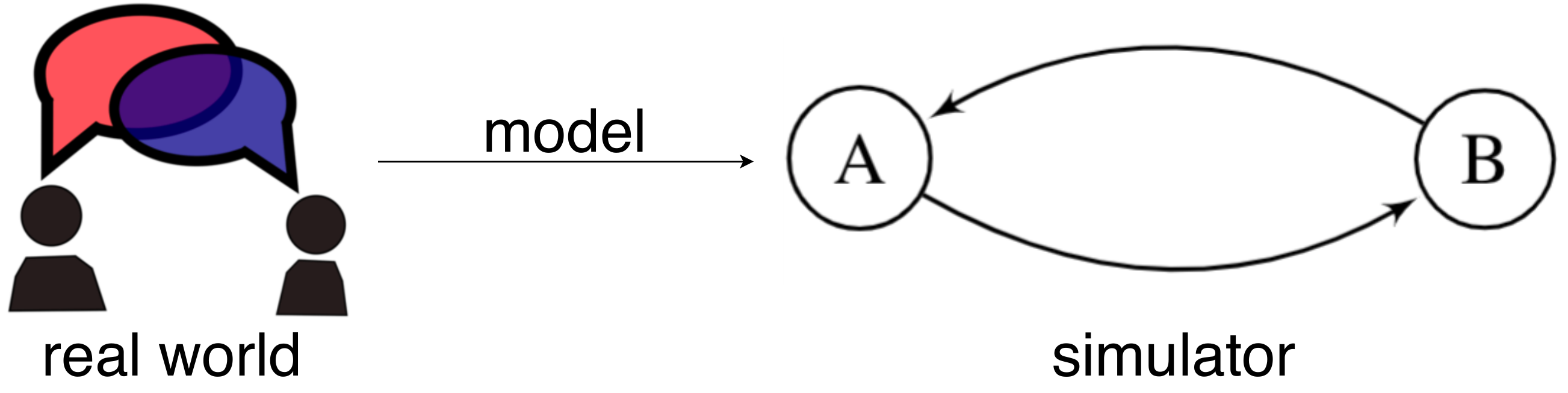}};
		\end{tikzpicture}
	\end{subfigure}
	\hfill
	\begin{subfigure}{0.49\columnwidth}
	\centering
		\begin{tikzpicture}[
		bidirected/.style={Latex-Latex},
		inner/.style={inner sep=0.2pt},
        outer/.style={inner sep=4pt}
		]
		\node[{shape=circle, text=black, minimum size=0.1em}] (v) at  (0,-1) {(2)};
		\node (img)  {\includegraphics[width=\columnwidth]{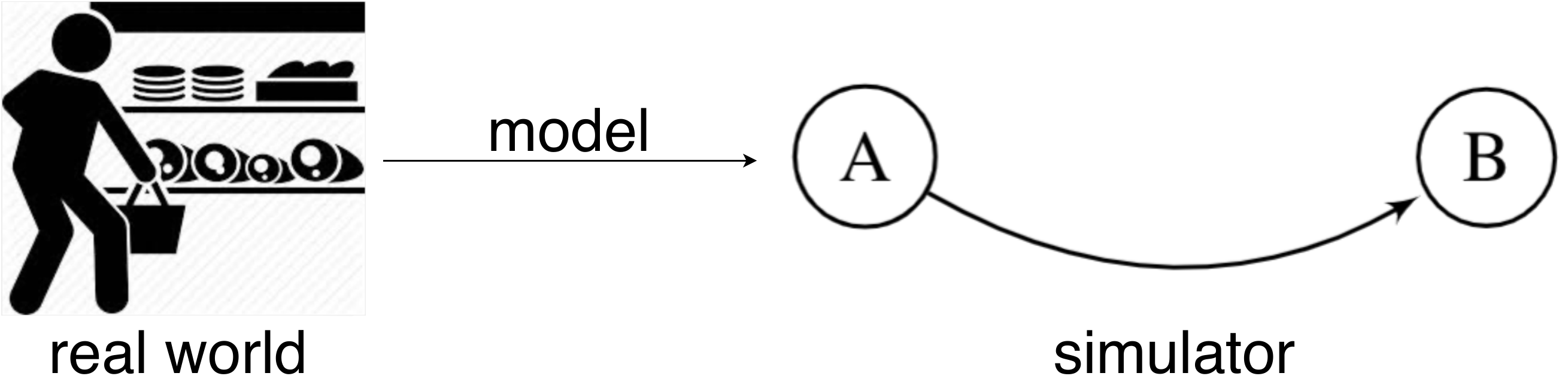}};
		\end{tikzpicture}
	\end{subfigure}
	\caption{1) Two people talking to each other (bi-directional connectivity). Each person can infect the other one 2) A person touching an item that was touched before by another person is modeled as single directional connectivity. Only the first person can infect the second person.} 
	\label{fig:real_world}
\end{figure}

For the type of viruses that are still infectious after being on a surface for a while, the connectivity graph must be directed. The infected person who touches the surface at time $t_0$ can infect the person who touches the surface at a time $t>t_0$, but the other direction of virus transmission is clearly blocked due to the time direction. The more common way of virus transmission, especially for respiratory diseases, is via close interactions while being at the same location. In this case, the connecting edges are symmetric (bi-directional). These two ways of virus transmission are illustrated in~\Cref{fig:real_world}. Regardless of the edge direction, the connectivity of the population graph determines the structure of the city. The creation of such structure is presented in~\Cref{sec:pg:pc,sec:family_pattern,sec:population_generation_procedure,sec:community,sec:community_assignment}. The connectivity details, such as the direction and the strength of edges is then presented in~\Cref{sec:connectivity_graph}.

\subsubsection{Person attributes} \label{sec:pg:pc}
A human individual is realized as an instance of a class named~\texttt{Person}. A person object represents the most fundamental unit of the simulation and has the succeeding attributes: age, gender, and health condition; These attributes are used to determine the role of the individual in the family and the society, her hobbies, and her daily plan. Here, we list the description for these attributes of an individual. Notice that the fundamental logic behind \ours is to create a significant hierarchical probability distribution for the city so that the whole city can be seen as a random variable. Hence every property is indeed a realization of a probability distribution.

\begin{itemize}
    \item age ($a \sim A, a \geq 0.$): A real positive number realized from a probability distribution $A$.
    \item gender ($g \sim G$): A binary value realized from a binomial distribution $G$.
    \item health condition ($ h \sim H, h \in [0, 1]$): A real number realized from a specified distribution $H$ supported on $[0, 1]$ that represents the health condition of the person as an aggregated function of medical variables such as body mass index (BMI), diabetes, background heart disease, blood pressure, and etc. Greater $h$ signals a better aggregated health condition.

\end{itemize}

\subsubsection{Family pattern}
\label{sec:family_pattern}
A class named \texttt{Family Pattern} represents the pattern of the families who live in the society. The pattern of a family consists of three types of information. Firstly, the structural information, i.e., the existence of each of the parents and the number of children. Secondly, the distributions from which the attributes (see \Cref{sec:pg:pc}) of the family members are sampled. Lastly, the distribution from which the living location of the household within the city is sampled. Let $(S, L, \{\varphi_1, \varphi_2, \dots, \varphi_n\})$ represent a family pattern where $S$ is the structural information, $L$ is the distribution of the family location, and $\{\varphi_1, \varphi_2, \dots, \varphi_n\}$ is the set distributions over the attributes of $n$ members of the family. Each $\varphi_i$ is itself a set of distributions of each attribute of a family member, i.e., $\varphi_i = [A_i, G_i, H_i]$. Let $p_i$ refers to the $i$-th member of the family. Its attributes are then are sampled from $\phi_i$ independent of the other members.

\begin{equation}
    (p_i \sim \varphi_i) \equiv (p_i := [a_i\sim A_i, g_i \sim G_i, h_i\sim H_i]).
\end{equation}

\subsubsection{Generating the population}
\label{sec:population_generation_procedure}

To generate the population of the city, two pieces of information are requested from the user: {\bf 1)} total size of the population {\bf 2)} a set of family patterns with the probability of their occurrence. Recall that every level of the city hierarchy is probabilistic in \ours. Therefore, each family pattern can also be regarded as a random variable from which the family instances are realized. The instantiation process continues until the total number of people in the society exceeds the population size provided by the user.

Let $\Pcal=\{\Phi_1, \Phi_2, \dots, \Phi_k\}$ and $\{\pi_1, \pi_2, \ldots, \pi_k\}$ be the set of family patterns and their probabilities respectively. Each family in the society is a realization from one of these patterns. To instantiate a family, first, a family pattern is chosen with its corresponding probability, then a family instance is realized from the chosen pattern. Hence, an instantiated family pattern $\phi$ in the society follows a mixture of distributions of each family pattern in $\Pcal$, that is $\phi\sim \sum_{i=1}^k\pi_i \Phi_i$ where $\{\pi_i\}_{i=1}^{k}$ is a simplex and $\sum_{i=1}^k \pi_i=1$. This indicates that each family pattern $\phi$ follows $\Phi_1$ with probability $\pi_1$, $\Phi_2$ with probability $\pi_2$, ..., and $\Phi_k$ with probability $\pi_k$. Until the sum of the people in the families exceed the provided population of the city, new families are kept adding to the city. Each time one of the $\{\Phi_1, \Phi_2, \dots, \Phi_k\}$ patterns with their corresponding probabilities $\{\pi_1, \pi_2, \dots, \pi_k\}$ is chosen and its members are generated from the corresponding pattern. Unlike~\cite{lorch2020quantifying}, this strategy prioritizes creating families over individuals. The advantages of this approach compared to those that create individuals first and then assign them to families are discussed below.

\paragraph{Family first vs person first.} The structured population of existing real-world society is the end product of passing many generations over the years. Hence, the most accurate approach to model the current state of the society is to emulate the entire time evolution from the very beginning of the formation of the city until the current time. The emulation is possible only if the emulator is given an accurate account of all major events that has occurred over hundreds or thousands of years with a significant impact on the population structure. This information is obviously unavailable at the present time. Hence, to emulate the current structured population of a society, a membership problem needs to be solved at multiple levels. Let's focus on the structure of the population at the level of families and ignore other structures such as workplaces, schools, and etc. Recall that the only information we get from the user is the population size and the family patterns. One approach would be generating as many individuals as the requested population size with attributes defined by the set of family patterns. Once this pool of individuals is created, an immensely heavy importance sampling process needs to be solved to bind individuals that are likely to form a family under the mixture distribution of family patterns. To tackle the computational intractability, we propose an alternative method that puts families first and create individuals that already match a family. This method releases us from the computationally heavy importance sampling process at the cost of having less strict control on the population size. However, the resultant population does not exceed the provided population by the user more than the size of the largest family defined in the family patterns. Clearly, one family more or less in an entire society does not alter the results of the simulations for the problem for which this software is developed.

\begin{algorithm}[!t]
    \SetAlgoLined
    \KwData{Population size $M$, Family patterns $\{\varphi_1, \varphi_2, \ldots, \varphi_n\}$.}
    \KwResult{A structured society consisting of almost $M$ individuals instantiated from the~\texttt{Person} class and distributed to families.}

    \Begin{
    Society = [$\;$]

    \While{ $i < M$}{
        $j\leftarrow$ generate a sample from $\textrm{Multinomial}(\pi_1, \pi_2, \ldots, \pi_k)$

        $\phi\leftarrow$ generate the set of family members from the pattern $\varphi_j$
        
        Society.append($\phi$)
        
        $i\leftarrow i+|\phi|$
    }
    }
    \KwRet{Society}
    \caption{Generating the population}
    \end{algorithm}

\subsubsection{Community}
\label{sec:community}
The class \texttt{Community} consists of a set of \texttt{Person} objects with a shared interest that makes them interact closely. The class \texttt{Family} is the simplest class inherited from \texttt{Community}. The concept of community in this software covers a wide range of real-world communities. It can be as small as two or three friends talking to each other, or it can refer to larger entities such as everyone who walks in the streets of a city. Each \texttt{Community} object consists of multiple \texttt{SubCommunity} objects each of which contains a subset of the members of the enclosing \texttt{Community} that belong to that subcommunity.

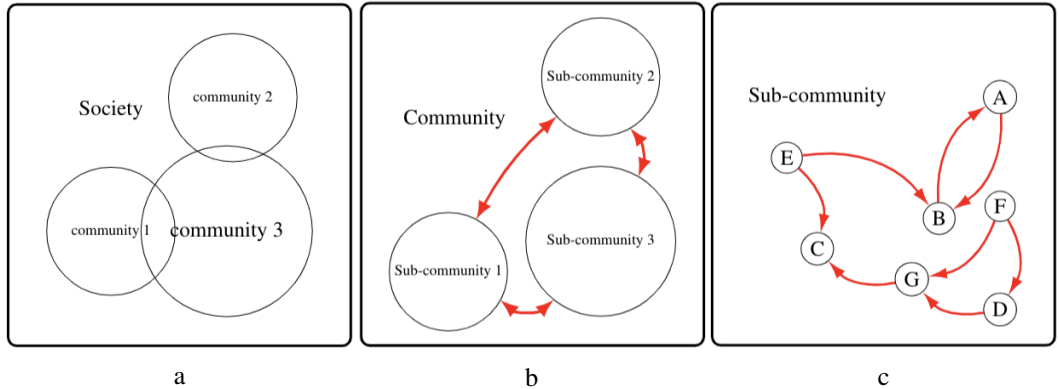
\begin{figure}[!t]
    \begin{subfigure}{0.33\columnwidth}
	\centering
        \begin{tikzpicture}
            \node (img)  {\includegraphics[width=\columnwidth]{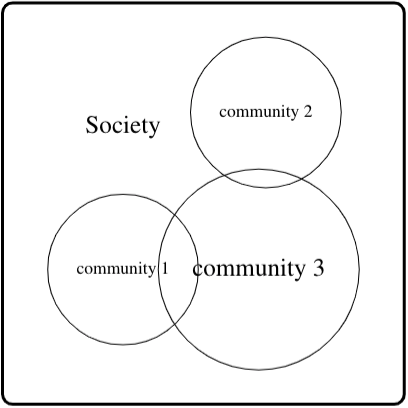}};
            \node[{shape=circle, text=black, minimum size=0.1em}] (v) at  (0,-2.7) {a};
        \end{tikzpicture}
	\end{subfigure}
    \begin{subfigure}{0.33\columnwidth}
	\centering
        \begin{tikzpicture}
            \node (img)  {\includegraphics[width=\columnwidth]{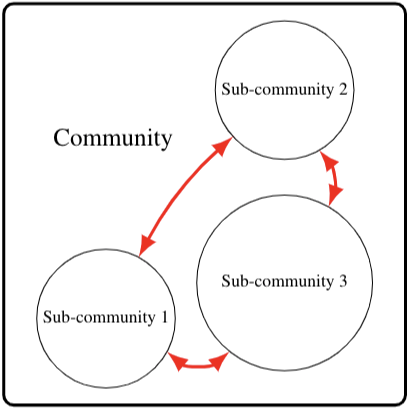}};
            \node[{shape=circle, text=black, minimum size=0.1em}] (v) at  (0,-2.7) {b};
        \end{tikzpicture}
	\end{subfigure}
	\hfill
    \begin{subfigure}{0.33\columnwidth}
	\centering
        \begin{tikzpicture}
            \node (img)  {\includegraphics[width=\columnwidth]{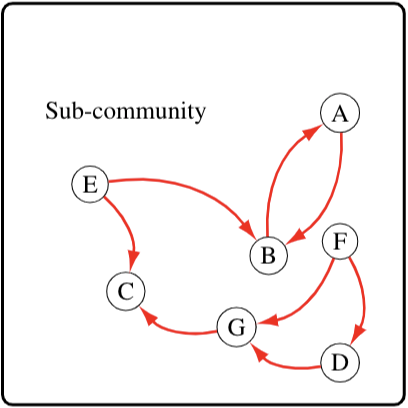}};
            \node[{shape=circle, text=black, minimum size=0.1em}] (v) at  (0,-2.72) {c};
        \end{tikzpicture}
	\end{subfigure}
    \caption{(a) In the first phase, all individuals of the society are created based on the population size and the set of provided family patterns as explained in~\Cref{sec:population_generation}. (b) The abstract hierarchical structure of the communities and subcommunities of the society is created. Observe that communities may be intersecting as an individual may be a member of various communities such as family, school, restaurant, etc. Within a community, there are two types of interactions shown as red arrows. One type of interactions is inter-subcommunity, such as the interactions among teachers and students in a school. (c) Another type of interactions, shown with red arrows, is within a subcommunity. For example, the way students interact with each other in a school.}
    
	\label{fig:community_overview}
\end{figure}

As an example, every school is an instance of the \texttt{Community} class. The concept of the school contains two main roles, students and teachers, each of which is a \texttt{SubCommunity} object of the school community. An overview of this hierarchical structure is depicted in~\Cref{fig:community_overview} and its benefits are explained below.

\paragraph{The advantages of a hierarchical structure. } The members of a community are assigned to subcommunities based on their shared role. In the following, we provide a list of reasons that explains the logic behind such division:

\begin{enumerate}
    \item The members of every subcommunity in a given community have special attributes. Hence, the persons of a society who are instantiated from a class~\texttt{Person} need to pass through different filters to be assigned to each subcommunity.
    \item Each subcommunity has a special pattern of internal interactions. As a result, separating them allows us to capture their influence on the spread of the infection more realistically. For example, in a school, teachers have different kinds of internal interactions compared to internal interactions among students.
    \item Each subcommunity may have its own daily time schedule even though all belong to the same community. For instance, in a restaurant that is an instance of a community, the time that a cashier spends in the restaurant is different (much longer) than the time that a customer spends there. Hence, the risk of getting infected in the restaurant is much higher for the cashier compared to the customer.
\end{enumerate}
	
\subsubsection{Community assignment}
\label{sec:community_assignment}
Once the needed \texttt{Community} and \texttt{SubCommunity} classes for a target society are constructed, the individuals who are instantiated from the~\texttt{Person} class in~\Cref{sec:population_generation} take their role by being assigned to the instances of \texttt{SubCommunities}. The challenge is that the assignment process is not trivial in the sense that we can not fill the subcommunities from top to bottom by an ordered list of individuals. Each subcommunity accepts people whose attributes belong to a certain range. For example, the subcommunity of students in a particular school accepts individuals whose \texttt{age} attribute is less than those that are acceptable to the subcommunity of teachers. 

Each subcommunity has its own set of special admissible attributes. A trivial assignment process would be an exhaustive search over the entire population to find the individuals whose attributes match those of the target subcommunity. In addition to the time intractability of this approach, the individuals may race for positions in some subcommunities while other subcommunities do not receive sufficient individuals. Hence, we developed a~\emph{stochastic filtering} approach inspired by importance sampling where the importance score is determined by how fit an individual's attributes are for a specific subcommunity. Therefore, it is helpful to view each subcommunity as a probabilistic filter that passes its matched attributes with higher probability. Individuals are assigned to their roles in the society by passing through a number of these stochastic filters. To decide whether an individual can be accepted to a subcommunity, the unnormalized density of the joint attributes is computed as a fitness score. 

Consider the subcommunity $S$ and the individual $p$. Assume $S$ has the $L$ admissible types of attributes and their corresponding probability densities $\{f_1, f_2, \ldots, f_L\}$. Suppose the individual $p$ has the set of attributes $\{\alpha_1, \alpha_2, \ldots, \alpha_L\}$ matching with the types of attributes that are admissible to $S$. Thus, the fitness score of $p$ for the subcommunity $S$ is calculated by $\prod_{l=1}^L f_l(\alpha_l)$. Notice that this score is not a probability, and computing its normalizing constant is intractable. However, this is not a problem because only the relative values matter. The scores are computed for all individuals in the society, and those with the highest scores are assigned to each subcommunity.

Among all attributes, the profession of an individual requires special treatment, as discussed below.

\paragraph{Special case of profession assignment. } Among the attributes of an individual, the profession needs special treatment. Every member of the society can be given only one profession. Hence, once the profession is assigned to an individual, she cannot be given another profession. Hence, the subcommunities that model professions (e.g., teachers, students, cashiers, bus drivers, etc.) must become blocking against her. This effect is modeled by multiplying the fitness score by $(1- 1_\text{has profession}\infty)$.

\subsubsection{Connectivity graph}
\label{sec:connectivity_graph}
The backbone of \ours is a connectivity graph that captures the interactions among the individuals of the society. Let $G(V, E)$ be the connectivity graph with the set of nodes $V$ and the set of edges $E$. In the following, we explain how this graph is created.

\begin{enumerate}
    \item Every individual is represented by a node of the graph (See~\Cref{fig:community_overview2}(a)).
    \item Due to the tight connections among the family members, a family is modeled by a complete and directed graph (See~\Cref{fig:community_overview2}(b)).

    \item A community defines the pattern of connections within and between its subcommunities. Assume the community $C$ has the set of $J$ subcommunities $\{S_1, S_2, \ldots, S_J\}$ and a $J\times J$ connectivity matrix $(c_{ij})$ with $1\leq i \leq j \leq J$. The entry $c_{ij}$, called~\emph{connectivity density}, represents the probability of the existence of an edge from an individual in the subcommunity $i$ to an individual in the subcommunity $j$. Formally speaking, Let $X_{a\rightarrow b}$ be an indicator random variable denoting whether there is a directed edge from node $a$ to node $b$. Then $X_{a\rightarrow b}$ follows a Bernoulli distribution with parameter $c_{ij}$ if $a \in S_i$ and $b \in S_j$. The connectivity density $c_{ij}$ itself comes from a Beta
    distribution. That is,
    
    \begin{equation}
        X_{a\rightarrow b}\sim \textrm{Bernoulli}(c_{ij}), \quad \text{for  } a\in S_i, b\in S_j,\textrm{ and } 1\leq i\leq j \leq J
        \label{eq:edges_across_subcommunties}
    \end{equation}

    where $c_{ij}\sim \textrm{Beta}(\alpha_{ij}, \beta_{ij})$ with $\alpha_{ij}$ the shape and $\beta_{ij}$ the scale parameter. See~\Cref{fig:community_overview2}(c) for an overview of the created edges. The Erdős–Rényi model of generating random graphs~\cite{erdHos1960evolution} is used. Notice, that the above edge creation process does not differentiate between edges within a subcommunity and among subcommunities. However, the connections are expected to be denser within a subcommunity, that is, $c_{ij}\ll c_{ii}$ for $i\neq j$. Because $c_{ij}$ itself is a sample from Beta($\alpha_{ij}$, $\beta_{ij}$), the parameters, $\alpha_{ij}$ and $\beta_{ij}$, of the corresponding beta distribution are chosen such that the probability mass is concentrated around $1$ for $i=j$ and around smaller values for $i\neq j$. The detailed pseudocode is given in~\Cref{alg:graph}.

\begin{figure}[!t]
    \begin{subfigure}{0.33\columnwidth}
    \label{fig:community_overview_a}
	\centering
        \begin{tikzpicture}
            \node (img)  {\includegraphics[width=\columnwidth]{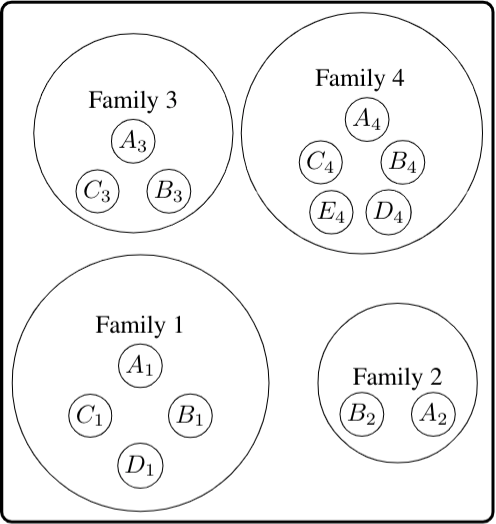}};
            \node[{shape=circle, text=black, minimum size=0.1em}] (v) at  (0,-2.7) {a};
        \end{tikzpicture}
	\end{subfigure}
    \begin{subfigure}{0.33\columnwidth}
    \label{fig:community_overview_b}
	\centering
        \begin{tikzpicture}
            \node (img)  {\includegraphics[width=\columnwidth]{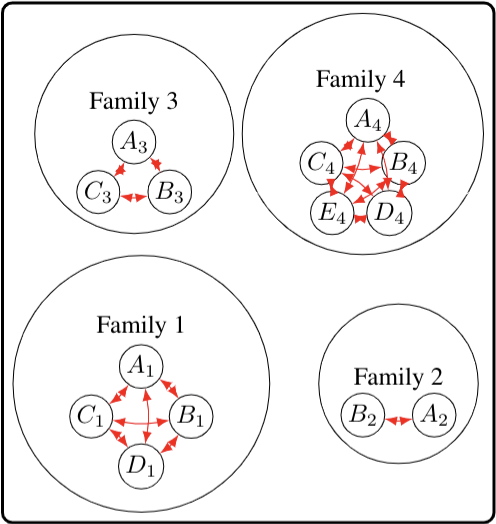}};
            \node[{shape=circle, text=black, minimum size=0.1em}] (v) at  (0,-2.7) {b};
        \end{tikzpicture}
	\end{subfigure}
	\hfill
    \begin{subfigure}{0.33\columnwidth}
        \label{fig:community_overview_c}
	\centering
        \begin{tikzpicture}
            \node (img)  {\includegraphics[width=1.04\columnwidth]{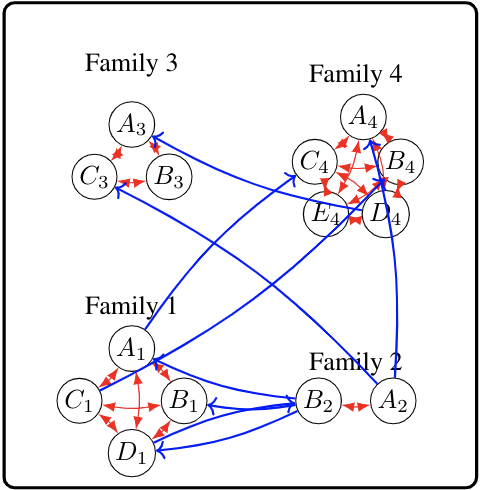}};
            \node[{shape=circle, text=black, minimum size=0.1em}] (v) at  (0,-2.72) {c};
        \end{tikzpicture}
	\end{subfigure}
    \caption{a) A node is added to the graph for every individual to generate the population. b) Due to the tight interactions within a family, a family's subgraph is a complete directed graph (red edges). c) The interactions across communities and subcommunities are represented by the blue edges, which are created according to~\Cref{eq:edges_across_subcommunties}.}
    \label{fig:community_overview2}
\end{figure}
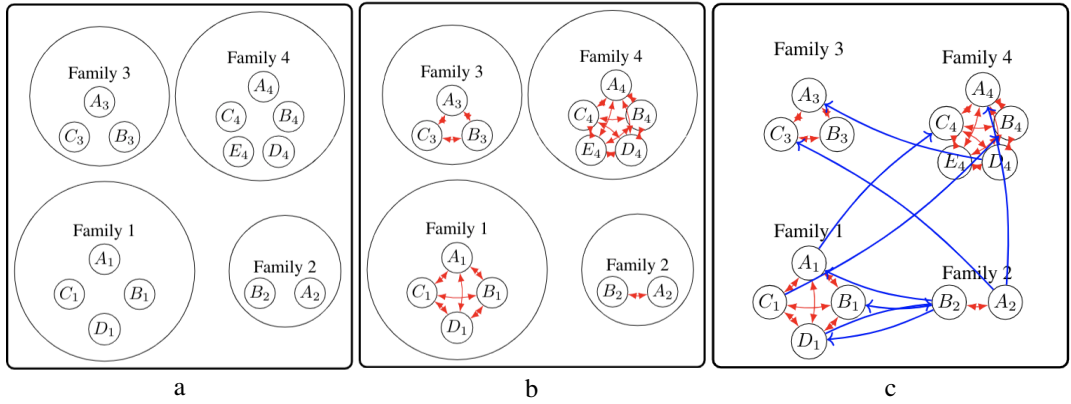

\end{enumerate}

\begin{algorithm}[!t]
\SetAlgoLined
\KwData{People $\{p_i\}$, families $\{f_i\}$, communities $\{C_i\}$ and their subcommunities $\{S_k^i\}$.}
\KwResult{The connectivity graph $G(V, E)$ with the set of nodes $V$ and the set of edges $E$.} 
\Begin{

$V \longleftarrow [\;]$

$E \longleftarrow [\;]$

\lForEach{person $p_i$}{add a vertex $v_i$ to $V$}
\tcp{creating within-family interactions}
 \For{family $f_i$}
 {
    \For{bidirected edge $e$ in the complete subgraph $K_{\text{size}(f_i)}$}{
    add $e$ to $E$\;
    }
 }
 \For{community $C_i$}{
    \For{sub-communities $S_j^i$ and. $S_k^i$ in $C_i$}{
$c_{jk} \longleftarrow \text{generate a random variable from } \Gamma_{jk} \text{ distribution}$\;
        $p \longleftarrow c_{jk}$\;
        \For{sub-communities $S_j^i$ and $S_k^i$ in $C_i$}{
            \For{person $a$ in $S_j^i$}{
                \For{person $b$ in $S_k^i$}{
                    $X_{a\rightarrow b} \longleftarrow \text{flip a biased coin with head probability } p$\;
                    \If{$X_{a\rightarrow b} = 1$}{
                    \tcp{creating inter-subcommunity interactions}
                    add directed edge $a\rightarrow b$ to $E$\; 
                    }
                }
            }
        }
    }
 }        
}
\KwRet{$G(V, E)$}
\caption{Creating the connectivity graph}
\label{alg:graph}
\end{algorithm}

\subsection{Propagation of the infection}
\label{sec:disease_propagation}

After the population is created in \Cref{sec:population_generation_procedure} and its structure is determined in \Cref{sec:connectivity_graph} to model the potential interactions among every pair of individuals of the society, this section describes how \ours models the propagation of a generic disease in the population.

\subsubsection{Disease transmission between two individuals}
\label{sec:disease_transmission}
The probability of the disease transmission from an individual to another depends on both the parameters of the disease and the attributes of the individuals. The following three parameters are critical in modeling the disease propagation.

\begin{itemize}
    \item Immunity ($\upsilon$): A real-valued parameter $v\in[0, 1]$ that shows how immune an individual is against being infected. For example, being infected once or being vaccinated increases this number towards the upper limit. Similar to other parameters of the model, immunity is also a random variable with an arbitrary distribution. A natural choice would be a beta distribution, i.e., $\upsilon \sim \textrm{Beta}(\alpha_\upsilon, \beta_\upsilon)$ whose parameters $\alpha_\upsilon, \beta_\upsilon$ needs to be determined according the the characteristics of the disease. Note that \ours comes with a versatile family of distributions that can be used for any model parameter, including immunity, if they are more suitable for a certain scenario.
    
    \item Infection rate ($r$): A real-valued parameter $r\in[0, 1]$ that models how easily an infection transmits. It is determined by either how fast a specific infection transmits in a society with no control measure or how robust the society is against the infection by observing the control measures such as wearing a mask, using hand sanitizer, etc. The infection rate is also a random variable for which we assume a Beta distribution here, i.e., $ r \sim \textrm{Beta}(\alpha_r, \beta_r)$.

    \item Transmission potential ($\gamma$): A real-valued parameter $\gamma\in[0, 1]$ that models the possibility for the transmission of the disease between two individuals based on the type of interaction they have. Hence, this parameter is determined by the connectivity graph under the population. The individuals who meet regularly (e.g., being the members of the same family) have strong connectivity and hence stronger potential for transmitting the infection. This parameter is also a random variable with Gamma distribution whose hyper-parameters are functions of the connectivity strength, i.e., $ \gamma^{\text{edge}} \sim \textrm{Beta}(\alpha_\gamma, \beta_\gamma)$.
\end{itemize}

Given the above influential variables in the disease transmission, the probability of the transmission of the infection from an infected individual (sender) to another individual (receiver) is calculated by:
\begin{equation}
        P(\textrm{sender} \xrightarrow[]{\text{Transmit}}\text{receiver}) = \upsilon_{\text{receiver}} \times r_{\text{sender}} \times r_{\text{receiver}} \times \gamma_{\text{edge}}.
\end{equation}
At each interaction between an infected and uninfected individuals, the above probability is calculated and kept as a threshold $p_{\textrm{thresh}}$. Then, a sample is generated from a uniform distribution $\zeta\sim \text{U}[0, 1]$ and a disease transmission event occurs if $\zeta\leq p_{\textrm{thresh}}$.

\subsubsection{The dynamics of infection in a patient} When an individual gets infected, the period of the disease $\tau$ and the probability of death $p_\textrm{death}$ is calculated based on the specified properties of the disease and the attributes of the individual. More specifically, the disease period is a random variable whose distribution is calculated from the real-world experimental data. The same holds for death probability (see the examples in~\Cref{sec:disease_properties_setting}). It can be seen that the death probability usually depends on the health condition and age. High ages and poor health conditions increase the likelihood of fatality when other factors are alike.
The diseased individual contributes to the propagation of the infection up to time $\tau$. At time $\tau$, the outcome is decided as death by probability $p_\textrm{death}$ or recovery by probability $1-p_\textrm{death}$. If the patient recovers, her attributes will be updated according to the characteristics of the infection and her attributes before getting infected. For example, a temporary immunity may be gained as a result of surviving the infection once.

\subsection{Time management}
A challenging issue when simulating a physical phenomenon is the immense computational resources needed to approximate the continuous evolution of the system in time. As a result, a reliable simulation becomes quickly intractable even for fairly low-dimensional systems. However, not all temporal details of the environment are relevant to the target application. When the goal is to simulate how an infectious disease propagates through a population, the only relevant events are those in which there is a potential in transmitting the disease.

Simulators often implement the evolution of time by either an event-based or clock-based method. In event-based methods, a queue of events is formed and ordered by the time-of-occurrence of them. In clock-based methods, any change in the system occurs at the pulses of a running clock. We propose a novel mixture event/clock-driven method to bring together the benefits of both worlds. The queue of events is formed similar to the event-based methods, but the pulses of a background clock determine which events are effective in the outcome. By changing the frequency of the background clock, the details of the timeline can be traded off with the computational demand. The constituent components of the time management module of \ours are explained in~\Cref{sec:events,sec:initialize_simulator,sec:standardizing_disease_transmission_prob} below.

\subsubsection{Timer}
\label{sec:timer}
The timer object is a pointer to a specific position of the time axis during the simulation. This pointer keeps moving forward as the simulation progresses.

\subsubsection{Events}
\label{sec:events}
\begin{figure}[t]
	\begin{subfigure}{0.4\columnwidth}
	\centering
		\begin{tikzpicture}
		    \node[{shape=circle, text=black, minimum size=0.1em}] (v) at  (-0.33,0) {(a)};
            \begin{axis}[
                height=2cm, width = 7.5cm,
                xmin=0, xmax=21,
                axis x line=bottom,
                hide y axis,    
                ymin=0,ymax=5,
                xlabel={Time (h)},
                scatter/classes={%
                    a={mark=o,draw=black}}
                ]
            
            \addplot[scatter,only marks,
                mark=otimes*,
                mark size = 3pt,
                fill = yellow,
                scatter src=explicit symbolic]
            table {
            3 0 
            6 0 
                };
            \addplot[scatter,only marks,
                mark=*,
                mark size = 3pt,
                fill = yellow,
                scatter src=explicit symbolic]
            table {
            9 0 
            12 0 
            15 0 
            18 0 
                };
            \addplot[scatter,only marks,
                mark=triangle*,
                mark size = 3pt,
                fill = black,
                scatter src=explicit symbolic]
            table {
            9 0
                };
            \end{axis}

        \end{tikzpicture}
        \caption{}
        \label{fig:time_transition_before}
	\end{subfigure}
	\hfill
		\begin{subfigure}{0.45\columnwidth}
		
	\centering
		\begin{tikzpicture}
		    \node[{shape=circle, text=black, minimum size=0.1em}] (v) at  (-0.33,0) {(b)};
            \begin{axis}[
                name=boundary,
                height=2cm, width = 7.5cm,
                xmin=0, xmax=21,
                axis x line=bottom,
                hide y axis,    
                ymin=0,ymax=5,
                xlabel={Time (h)},
                scatter/classes={%
                    a={mark=o,draw=black}}
                ]
            
            \addplot[scatter,only marks,
                mark=otimes*,
                mark size = 3pt,
                fill = yellow,
                scatter src=explicit symbolic]
            table {
            3 0 
            6 0 
            9 0 
                };
            \addplot[scatter,only marks,
                mark=*,
                mark size = 3pt,
                fill = yellow,
                scatter src=explicit symbolic]
            table {
            12 0 
            15 0 
            18 0 
                };
            \addplot[scatter,only marks,
                mark=triangle*,
                mark size = 3pt,
                fill = black,
                scatter src=explicit symbolic]
            table {
            12 0
                };
            \end{axis}
            
            \node[draw,fill=white,inner sep=2pt,below left=2em] at (boundary.south east) {\small
                \begin{tabular}{cc}
                    \begin{tikzpicture} \node [shape=circle, draw, fill=yellow, minimum size=0.7em] {}; \end{tikzpicture} & Event
                \end{tabular}};

        \end{tikzpicture}
        \caption{}
        \label{fig:time_transition_after}

	\end{subfigure}
	\caption{Yellow circles on the timeline axis indicate events. The crossed circles represent the events that are already executed. The filled triangle represents the current simulation time. After a task is executed, the simulation time jumps to the next event in the queue. An exemplary transition is shown by moving from from~\Cref{fig:time_transition_before} to~\Cref{fig:time_transition_after}.} 
	\label{fig:timer}
\end{figure}
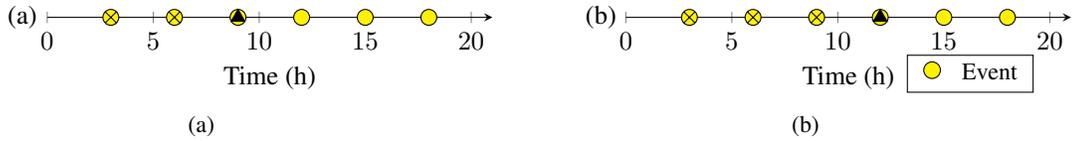

An event refers to any alternation of the simulation setting, i.e., individuals' states, connectivity graph, virus spread, disease properties, etc. Every event is an instance of a class called~\texttt{Event} with the following two properties: the \texttt{Activation Time} (time-of-occurrence) and the \texttt{Activation Task}. When the simulator's timer reaches the activation time, the associated task to that event (defined by the activation task) is executed. A series of events are kept in a queue and are executed in the order of their activation times (See~\Cref{fig:timer}). To prevent the events from racing for execution, each event is given a distinct priority index as a tiebreaker in case two events happen to have the same activation time. The following events are included in the current version (V1.0) of Pyfectious. The events with higher priorities come earlier in the list: \{Incubation Event, Infection Event, Plan Day Event, Transition Event, Virus Spread Event\}. Each of these events is explained in the following. 

\paragraph{Transition event}
Each transition event is associated with a certain individual, and it is triggered when that individual changes its location from one subcommunity to another. The change of location is implemented by changing the weights of the edges connected to an individual in the connectivity graph.

\paragraph{Plan-Day event}
The daily dynamics of society consist of the motion of the people and their interactions based on the role they play in society. Hence, individuals' daily schedule in a city is roughly determined by their attributes and their role. In \ours, the daily schedule for every individual is determined by the activation of an event at the beginning of each day. The Plan-Day event is a sequential random variable that takes the attributes and associated communities / subcommunities to an individual and generates a sample from the schedule suited to her. Looking more closely into the implementation, every daily plan is an ordered sequence of events whose start time and duration are sampled from specified probability distributions. The hyper-parameters of these distributions are determined by the attributes of the individual and the subcommunities that are related to a certain event. To prevent the overlapping between the time intervals of the events, a time priority index is assigned to each event to resolve potential conflicts. For example, mandatory events, such as \emph{going to work} and \emph{going to school} have a higher priority compared with optional events such as \emph{going to restaurant}.

\paragraph{Incubation event} After a disease transmission occurs, an incubation event is added to the event queue containing the end time of the incubation period. This marks the period in which the disease is still dormant in the body. When this event is activated at the end of the incubation period, the state of the respective individual will be updated to infected.

\paragraph{Infection event} After the incubation period, the respective individual's health condition is updated to infected, and an infection event is added to the event queue. This event's activation time marks the end time of the duration of the disease as a function of the individual's attributes. When this event is activated at the end of the disease period, the outcome of the disease is decided, and the individual's condition is updated to either~\emph{dead} or~\emph{recovered}.

\paragraph{Virus spread event}
The running clock of the simulator is converted to a series of virus spread events. The period of the clock works as temporal snapshots on which a virus transmission can occur. Hence, it trades off the needed computational resources with the temporal resolution of the simulator. To save computational time, the temporal resolution can be chosen long enough that assures the spread of the infection in the city does not change drastically within that period. For example, for respiratory diseases such as COVID-19, the fastest transmission way takes two individuals to get near each other. Hence, the temporal resolution can be set accordingly.

\subsubsection{Initializing the simulator}
\label{sec:initialize_simulator}
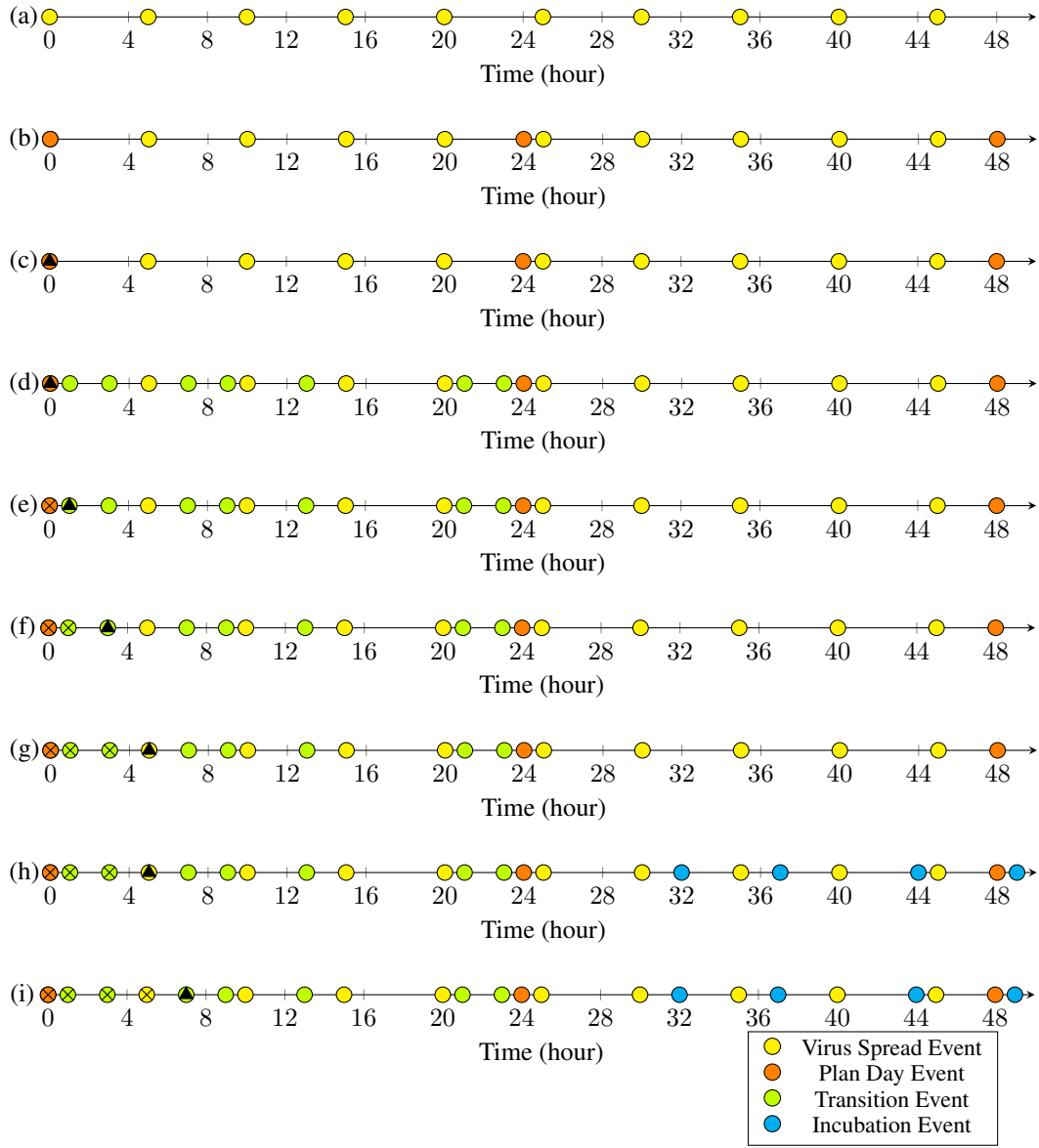
\begin{figure}[!t]
\captionsetup[subfigure]{labelformat=empty, aboveskip=-3pt,belowskip=-3pt}
	\begin{subfigure}{0.99\columnwidth}
	\centering
		\begin{tikzpicture}
		    \node[{shape=circle, text=black, minimum size=0.1em}] (v) at  (-0.35,0) {(a)};
            \begin{axis}[
                width=15cm,height=2cm,
                xmin=0, xmax=50,
                axis x line=bottom,
                hide y axis,    
                ymin=0,ymax=5,
                xtick distance=4,
                xlabel={Time (hour)},
                scatter/classes={%
                    a={mark=o,draw=black}}
                ]
            
            \addplot[scatter,only marks,
                mark=*,
                mark size = 3pt,
                fill = yellow,
                scatter src=explicit symbolic]
            table {
            0 0
            5 0 
            10 0
            15 0
            20 0
            25 0
            30 0
            35 0
            40 0
            45 0
                };
            \end{axis}

        \end{tikzpicture}
        \caption{}
        \label{fig:execution_a}
	\end{subfigure}
	
	\begin{subfigure}{0.99\columnwidth}
	\centering
		\begin{tikzpicture}
		    \node[{shape=circle, text=black, minimum size=0.1em}] (v) at  (-0.35,0) {(b)};
            \begin{axis}[
                width=15cm,height=2cm,
                xmin=0, xmax=50,
                axis x line=bottom,
                hide y axis,    
                ymin=0,ymax=5,
                xtick distance=4,
                xlabel={Time (hour)},
                scatter/classes={%
                    a={mark=o,draw=black}}
                ]
            
            \addplot[scatter,only marks,
                mark=*,
                mark size = 3pt,
                fill = yellow,
                scatter src=explicit symbolic]
            table {
            0 0
            5 0 
            10 0
            15 0
            20 0
            25 0
            30 0
            35 0
            40 0
            45 0
                };
            
            \addplot[scatter,only marks,
                mark=*,
                mark size = 3pt,
                fill = orange,
                scatter src=explicit symbolic]
            table {
            0 0
            24 0 
            48 0
                };
                
            \end{axis}

        \end{tikzpicture}
        \caption{}
        \label{fig:execution_b}
	\end{subfigure}
	
	\begin{subfigure}{0.99\columnwidth}
	\centering
		\begin{tikzpicture}
		    \node[{shape=circle, text=black, minimum size=0.1em}] (v) at  (-0.35,0) {(c)};
            \begin{axis}[
                width=15cm,height=2cm,
                xmin=0, xmax=50,
                axis x line=bottom,
                hide y axis,    
                ymin=0,ymax=5,
                xtick distance=4,
                xlabel={Time (hour)},
                scatter/classes={%
                    a={mark=o,draw=black}}
                ]
            
            \addplot[scatter,only marks,
                mark=*,
                mark size = 3pt,
                fill = yellow,
                scatter src=explicit symbolic]
            table {
            0 0
            5 0 
            10 0
            15 0
            20 0
            25 0
            30 0
            35 0
            40 0
            45 0
                };
            
            \addplot[scatter,only marks,
                mark=*,
                mark size = 3pt,
                fill = orange,
                scatter src=explicit symbolic]
            table {
            0 0
            24 0 
            48 0
                };
              
            \addplot[scatter,only marks,
                mark=triangle*,
                mark size = 3pt,
                fill = black,
                scatter src=explicit symbolic]
            table {
            0 0
                };
                
            \end{axis}

        \end{tikzpicture}
        \caption{}
        \label{fig:execution_c}
	\end{subfigure}
	
	\begin{subfigure}{0.99\columnwidth}
	\centering
		\begin{tikzpicture}
		    \node[{shape=circle, text=black, minimum size=0.1em}] (v) at  (-0.35,0) {(d)};
            \begin{axis}[
                width=15cm,height=2cm,
                xmin=0, xmax=50,
                axis x line=bottom,
                hide y axis,    
                ymin=0,ymax=5,
                xtick distance=4,
                xlabel={Time (hour)},
                scatter/classes={%
                    a={mark=o,draw=black}}
                ]
            
            \addplot[scatter,only marks,
                mark=*,
                mark size = 3pt,
                fill = yellow,
                scatter src=explicit symbolic]
            table {
            0 0
            5 0 
            10 0
            15 0
            20 0
            25 0
            30 0
            35 0
            40 0
            45 0
                };
            
            \addplot[scatter,only marks,
                mark=*,
                mark size = 3pt,
                fill = orange,
                scatter src=explicit symbolic]
            table {
            0 0
            24 0 
            48 0
                };
              
            \addplot[scatter,only marks,
                mark=triangle*,
                mark size = 3pt,
                fill = black,
                scatter src=explicit symbolic]
            table {
            0 0
                };
                
            \addplot[scatter,only marks,
                mark=*,
                mark size = 3pt,
                fill = lime,
                scatter src=explicit symbolic]
            table {
            1 0
            3 0 
            7 0
            9 0
            21 0
            23 0
            13 0
                };
                
            \end{axis}

        \end{tikzpicture}
        \caption{}
        \label{fig:execution_d}
	\end{subfigure}

	\begin{subfigure}{0.99\columnwidth}
	\centering
		\begin{tikzpicture}
		    \node[{shape=circle, text=black, minimum size=0.1em}] (v) at  (-0.35,0) {(e)};
            \begin{axis}[
                width=15cm,height=2cm,
                xmin=0, xmax=50,
                axis x line=bottom,
                hide y axis,    
                ymin=0,ymax=5,
                xtick distance=4,
                xlabel={Time (hour)},
                scatter/classes={%
                    a={mark=o,draw=black}}
                ]
            
            \addplot[scatter,only marks,
                mark=*,
                mark size = 3pt,
                fill = yellow,
                scatter src=explicit symbolic]
            table {
            0 0
            5 0 
            10 0
            15 0
            20 0
            25 0
            30 0
            35 0
            40 0
            45 0
                };
            
            \addplot[scatter,only marks,
                mark=*,
                mark size = 3pt,
                fill = orange,
                scatter src=explicit symbolic]
            table {
            24 0 
            48 0
                };
                
            \addplot[scatter,only marks,
                mark=otimes*,
                mark size = 3pt,
                fill = orange,
                scatter src=explicit symbolic]
            table {
            0 0
                };
                
            \addplot[scatter,only marks,
                mark=*,
                mark size = 3pt,
                fill = lime,
                scatter src=explicit symbolic]
            table {
            1 0
            3 0 
            7 0
            9 0
            21 0
            23 0
            13 0
                };
                
            \addplot[scatter,only marks,
                mark=triangle*,
                mark size = 3pt,
                fill = black,
                scatter src=explicit symbolic]
            table {
            1 0
                };
                
            \end{axis}

        \end{tikzpicture}
        \caption{}
        \label{fig:execution_e}
	\end{subfigure}
	
	\begin{subfigure}{0.99\columnwidth}
	\centering
		\begin{tikzpicture}
		    \node[{shape=circle, text=black, minimum size=0.1em}] (v) at  (-0.35,0) {(f)};
            \begin{axis}[
                width=15cm,height=2cm,
                xmin=0, xmax=50,
                axis x line=bottom,
                hide y axis,    
                ymin=0,ymax=5,
                xtick distance=4,
                xlabel={Time (hour)},
                scatter/classes={%
                    a={mark=o,draw=black}}
                ]
            
            \addplot[scatter,only marks,
                mark=*,
                mark size = 3pt,
                fill = yellow,
                scatter src=explicit symbolic]
            table {
            0 0
            5 0 
            10 0
            15 0
            20 0
            25 0
            30 0
            35 0
            40 0
            45 0
                };
            
            \addplot[scatter,only marks,
                mark=*,
                mark size = 3pt,
                fill = orange,
                scatter src=explicit symbolic]
            table {
            24 0 
            48 0
                };
                
            \addplot[scatter,only marks,
                mark=otimes*,
                mark size = 3pt,
                fill = orange,
                scatter src=explicit symbolic]
            table {
            0 0
                };
                
            \addplot[scatter,only marks,
                mark=*,
                mark size = 3pt,
                fill = lime,
                scatter src=explicit symbolic]
            table {
            3 0 
            7 0
            9 0
            21 0
            23 0
            13 0
                };
            
            \addplot[scatter,only marks,
                mark=otimes*,
                mark size = 3pt,
                fill = lime,
                scatter src=explicit symbolic]
            table {
            1 0
                };
                
            \addplot[scatter,only marks,
                mark=triangle*,
                mark size = 3pt,
                fill = black,
                scatter src=explicit symbolic]
            table {
            3 0
                };
                
            \end{axis}
            
        \end{tikzpicture}
        \caption{}
        \label{fig:execution_f}
	\end{subfigure}
	
	\begin{subfigure}{0.99\columnwidth}
	\centering
		\begin{tikzpicture}
		    \node[{shape=circle, text=black, minimum size=0.1em}] (v) at  (-0.35,0) {(g)};
            \begin{axis}[
                width=15cm,height=2cm,
                xmin=0, xmax=50,
                axis x line=bottom,
                hide y axis,    
                ymin=0,ymax=5,
                xtick distance=4,
                xlabel={Time (hour)},
                scatter/classes={%
                    a={mark=o,draw=black}}
                ]
            
            \addplot[scatter,only marks,
                mark=*,
                mark size = 3pt,
                fill = yellow,
                scatter src=explicit symbolic]
            table {
            0 0
            5 0 
            10 0
            15 0
            20 0
            25 0
            30 0
            35 0
            40 0
            45 0
                };
            
            \addplot[scatter,only marks,
                mark=*,
                mark size = 3pt,
                fill = orange,
                scatter src=explicit symbolic]
            table {
            24 0 
            48 0
                };
                
            \addplot[scatter,only marks,
                mark=otimes*,
                mark size = 3pt,
                fill = orange,
                scatter src=explicit symbolic]
            table {
            0 0
                };
                
            \addplot[scatter,only marks,
                mark=*,
                mark size = 3pt,
                fill = lime,
                scatter src=explicit symbolic]
            table {
            7 0
            9 0
            21 0
            23 0
            13 0
                };
            
            \addplot[scatter,only marks,
                mark=otimes*,
                mark size = 3pt,
                fill = lime,
                scatter src=explicit symbolic]
            table {
            1 0
            3 0
                };
                
            \addplot[scatter,only marks,
                mark=triangle*,
                mark size = 3pt,
                fill = black,
                scatter src=explicit symbolic]
            table {
            5 0
                };
                
            \end{axis}

        \end{tikzpicture}
        \caption{}
        \label{fig:execution_g}
	\end{subfigure}

	\begin{subfigure}{0.99\columnwidth}
	\centering
		\begin{tikzpicture}
		    \node[{shape=circle, text=black, minimum size=0.1em}] (v) at  (-0.35,0) {(h)};
            \begin{axis}[
                width=15cm,height=2cm,
                xmin=0, xmax=50,
                axis x line=bottom,
                hide y axis,    
                ymin=0,ymax=5,
                xtick distance=4,
                xlabel={Time (hour)},
                scatter/classes={%
                    a={mark=o,draw=black}}
                ]
            
            \addplot[scatter,only marks,
                mark=*,
                mark size = 3pt,
                fill = yellow,
                scatter src=explicit symbolic]
            table {
            0 0
            5 0 
            10 0
            15 0
            20 0
            25 0
            30 0
            35 0
            40 0
            45 0
                };
            
            \addplot[scatter,only marks,
                mark=*,
                mark size = 3pt,
                fill = orange,
                scatter src=explicit symbolic]
            table {
            24 0 
            48 0
                };
                
            \addplot[scatter,only marks,
                mark=otimes*,
                mark size = 3pt,
                fill = orange,
                scatter src=explicit symbolic]
            table {
            0 0
                };
                
            \addplot[scatter,only marks,
                mark=*,
                mark size = 3pt,
                fill = lime,
                scatter src=explicit symbolic]
            table {
            7 0
            9 0
            21 0
            23 0
            13 0
                };
            
            \addplot[scatter,only marks,
                mark=otimes*,
                mark size = 3pt,
                fill = lime,
                scatter src=explicit symbolic]
            table {
            1 0
            3 0
                };
                
            \addplot[scatter,only marks,
                mark=triangle*,
                mark size = 3pt,
                fill = black,
                scatter src=explicit symbolic]
            table {
            5 0
                };
                
            \addplot[scatter,only marks,
                mark=*,
                mark size = 3pt,
                fill = cyan,
                scatter src=explicit symbolic]
            table {
            32 0
            37 0
            44 0
            49 0
                };
                
            \end{axis}

        \end{tikzpicture}
        \caption{}
        \label{fig:execution_h}
	\end{subfigure}
	
	\begin{subfigure}{0.99\columnwidth}
	\centering
		\begin{tikzpicture}
		    \node[{shape=circle, text=black, minimum size=0.1em}] (v) at  (-0.35,0) {(i)};
            \begin{axis}[
                name=boundary,
                width=15cm,height=2cm,
                xmin=0, xmax=50,
                axis x line=bottom,
                hide y axis,    
                ymin=0,ymax=5,
                xtick distance=4,
                xlabel={Time (hour)},
                scatter/classes={%
                    a={mark=o,draw=black}}
                ]
            
            \addplot[scatter,only marks,
                mark=*,
                mark size = 3pt,
                fill = yellow,
                scatter src=explicit symbolic]
            table {
            10 0
            15 0
            20 0
            25 0
            30 0
            35 0
            40 0
            45 0
                }; \label{pgfplots:virus}
                
            \addplot[scatter,only marks,
                mark=otimes*,
                mark size = 3pt,
                fill = yellow,
                scatter src=explicit symbolic]
            table {
            0 0
            5 0 
                };
            
            \addplot[scatter,only marks,
                mark=*,
                mark size = 3pt,
                fill = orange,
                scatter src=explicit symbolic]
            table {
            24 0 
            48 0
                };
                
            \addplot[scatter,only marks,
                mark=otimes*,
                mark size = 3pt,
                fill = orange,
                scatter src=explicit symbolic]
            table {
            0 0
                };
                
            \addplot[scatter,only marks,
                mark=*,
                mark size = 3pt,
                fill = lime,
                scatter src=explicit symbolic]
            table {
            7 0
            9 0
            21 0
            23 0
            13 0
                };
            
            \addplot[scatter,only marks,
                mark=otimes*,
                mark size = 3pt,
                fill = lime,
                scatter src=explicit symbolic]
            table {
            1 0
            3 0
                };
                
            \addplot[scatter,only marks,
                mark=*,
                mark size = 3pt,
                fill = cyan,
                scatter src=explicit symbolic]
            table {
            32 0
            37 0
            44 0
            49 0
                };
                
            \addplot[scatter,only marks,
                mark=triangle*,
                mark size = 3pt,
                fill = black,
                scatter src=explicit symbolic]
            table {
            7 0
                };
            \end{axis}
            
            \node[draw,fill=white,inner sep=2pt,below left=2em] at (boundary.south east) {\small
            \begin{tabular}{ccl}
                \begin{tikzpicture} \node [shape=circle, draw, fill=yellow, minimum size=0.7em] {}; \end{tikzpicture} & Virus Spread Event \\
                \begin{tikzpicture} \node [shape=circle, draw, fill=orange, minimum size=0.7em] {}; \end{tikzpicture} & Plan Day Event \\
                \begin{tikzpicture} \node [shape=circle, draw, fill=lime, minimum size=0.7em] {}; \end{tikzpicture} & Transition Event \\
                 \begin{tikzpicture} \node [shape=circle, draw, fill=cyan, minimum size=0.7em] {}; \end{tikzpicture} & Incubation Event
            \end{tabular}};

        \end{tikzpicture}
        \caption{}
        \label{fig:execution_i}
	\end{subfigure}
	
	\caption{In this figure, a toy example of the execution process for two days is depicted. Events are indicated by yellow circles on the horizontal axis that represents the timeline. The current time of the simulation is shown by the filled triangle. The crossed circles represent those events that are already executed. After the execution of each event, the simulator time jumps forward to the nearest event in the queue. Each row of this figure shows one step of the simulation time. a) The simulator is initialized. The clock period is set to 5 hours based on which the virus spread events are placed. b) The plan day events are placed at the beginning of each day. These events are supposed to schedule the individuals' daily lives based on their roles in society. c) The timer is set at the start of the simulation time. d) During the execution of the plan day events, transition events are added to the event queue. These events will change the location of the individuals. e) The timer takes a step forward. f, g) The transition events are executed that change the locations of the individuals. h) The infection is being spread by interactions among individuals. New infections create new incubation events, which are added to the queue. i) The simulation moves on with the same rules.
	} 
	\label{fig:execution}
\end{figure}

To initiate the simulation, an empty queue of events is created. Then, the following two tasks are carried out to populate the queue with \texttt{Events} to be run.
\begin{enumerate} 	
    \item The simulator clock period is set, and the virus spread events are added to the queue. In~\Cref{fig:execution_a}, the clock pulses every 5 hours, and the virus spread events are placed on consecutive 5-hour intervals starting from hour 0.
    \item The Plan-Day events are placed at the beginning of each day. They create the daily schedule for every individual and fill in the time progression queue with the events that make up the daily plan (See Figure \Cref{fig:execution_b} and \Cref{fig:execution_d} for illustration.)

\end{enumerate}

The progress of a simple case of the simulation for two days (48 hours) is depicted in \Cref{fig:execution}. In~\Cref{fig:execution_c}, the plan day event is activated as the first event of the day. As a result, the transition events are added to the timeline as is shown in~\Cref{fig:execution_d}. The transition events are activated in~\Cref{fig:execution_e,fig:execution_f} that leads to changing the location of the respective individuals. A virus spread event is executed in~\Cref{fig:execution_g,fig:execution_h} and the virus transmission occurs depending on the individual's location and the connectivity graph. If a new individual gets infected as a result of a virus spread event, the infection end events are added as shown in~\Cref{fig:execution_h}. This process continues, and the added events to the timeline get executed in order until the end of the simulation time.

\subsubsection{Standardizing the probability of disease transmission}
\label{sec:standardizing_disease_transmission_prob}
We propose a multiresolution simulator to make the simulation possible on machines with weaker computational resources. A crucial point in the multiresolution simulator is to make sure the outcome is consistent regardless of the employed resolution (the period of the clock). 

\begin{figure}[!t]
	\begin{subfigure}{0.99\columnwidth}
	\centering
		\begin{tikzpicture}
		    \node[{shape=circle, text=black, minimum size=0.1em}] (v) at  (-0.35,0) {(a)};
            \begin{axis}[
                width=14.5cm,height=2cm,
                xmin=0, xmax=80,
                axis x line=bottom,
                hide y axis,    
                ymin=0,ymax=5,
                xtick={0, 9, 18, 27, 36, 45, 54, 67, 80},
                xticklabels={0, $T_1$, $2T_1$, $3T_1$, $4T_1$, $5T_1$, $6T_1$, $\dots k T_1 \dots$, $t$},
                extra x ticks={0, 9, 18, 27, 36, 45, 54, 67, 80},
                extra x tick style={
                    ,grid=major
                    ,ticklabel pos=top
                },
                extra x tick labels={$1$, ${(1-p_1)}^1$, ${(1-p_1)}^2$, ${(1-p_1)}^3$, ${(1-p_1)}^4$, ${(1-p_1)}^5$, ${(1-p_1)}^6$, ${(1-p_1)}^k$, ${(1-p_1)}^{\lfloor t/T_1 \rfloor}$},
                xlabel={Time (hour)},
                scatter/classes={%
                    a={mark=o,draw=black}}
                ]
            
            \addplot[scatter,only marks,
                mark=*,
                mark size = 3pt,
                fill = yellow,
                scatter src=explicit symbolic]
            table {
            0 0
            9 0 
            18 0
            27 0
            36 0
            45 0
            54 0
            67 0
                };
            \end{axis}

        \end{tikzpicture}
	\end{subfigure}
	
	\begin{subfigure}{0.99\columnwidth}
	\centering
		\begin{tikzpicture}
		    \node[{shape=circle, text=black, minimum size=0.1em}] (v) at  (-0.35,0) {(b)};
            \begin{axis}[
                width=14.5cm,height=2cm,
                xmin=0, xmax=80,
                axis x line=bottom,
                hide y axis,    
                ymin=0,ymax=5,
                xtick={0, 17, 34, 51, 67, 80},
                xticklabels={0, $T_2$, $2T_2$, $3T_2$, $\dots k T_2 \dots$, $t$},
                extra x ticks={0, 17, 34, 51, 67, 80},
                extra x tick style={
                    ,grid=major
                    ,ticklabel pos=top
                },
                 extra x tick labels={$1$, ${(1-p_2)}^1$, ${(1-p_2)}^2$, ${(1-p_2)}^3$, ${(1-p_2)}^k$, ${(1-p_2)}^{\lfloor t/T_2 \rfloor}$},
                xlabel={Time (hour)},
                scatter/classes={%
                    a={mark=o,draw=black}}
                ]
            
            \addplot[scatter,only marks,
                mark=*,
                mark size = 3pt,
                fill = yellow,
                scatter src=explicit symbolic]
            table {
            0 0
            17 0 
            34 0
            51 0
            67 0
                };
            \end{axis}
            
            
            \node[draw,fill=white,inner sep=2pt, below left = 22pt] at (boundary.south east) {\small
                \begin{tabular}{cc}
                    \begin{tikzpicture} \node [shape=circle, draw, fill=yellow, minimum size=0.7em] {}; \end{tikzpicture} & Simulator's Clock
                \end{tabular}};

        \end{tikzpicture}
	\end{subfigure}
	
	\caption{Simulator's clock is translated to Virus Spread Events indicated by yellow circles on the timeline axis. In (a) and (b) two characteristically similar simulators' timelines are displayed that only differ in their clock periods ($T_1$ for (a) and $T_2$ for (b)). If the probability of transmission in a single clock trigger is $p$, the probability of disease transmission for the whole time is a geometric distribution with parameters equals to $(p, \lfloor t/T \rfloor)$, where $T$ is the clock period. The probability of disease not being transmitted is shown above the axes for a simple interaction between two individuals. In order to have consistent results for both similar simulators with different resolutions, the mentioned probabilities should be equal to each other for the simulators. In other words, ${(1-p_1)}^{\lfloor t/T_1 \rfloor} = {(1-p_2)}^{\lfloor t/T_2 \rfloor}.$} 
	\label{fig:standard_probability}
\end{figure}
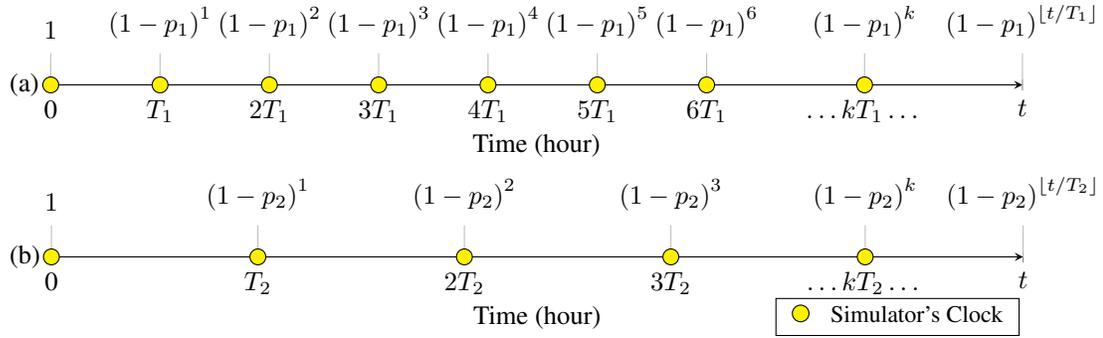

We argue that consistency is achieved if the probability of the virus transmission between two individuals is invariant with respect to the clock's period. Equivalently, it is sufficient to show that the probability of the virus not being transmitted remains invariant to the period of the clock. Hence, we equate the probability of non-transmission under clock periods $T_1$ and $T_2$ before time $t$ as

\begin{equation}
    {(1-p_1)}^{\lfloor t/T_1 \rfloor} = {(1-p_2)}^{\lfloor t/T_2 \rfloor},
\end{equation}
where $p_1$ and $p_2$ are the probabilities of a single virus transmission. Hence, as can be seen in~\Cref{fig:standard_probability}, the virus transmission probability under a target clock period can be derived as a function of the transmission probability under a different clock period and the ratio of clock periods. That is
\begin{equation}
    p_2 = 1 - {(1 - p_1)}^{\frac{\lfloor t/T_1 \rfloor}{\lfloor t/T_2 \rfloor}}.
\end{equation}

\subsection{Even management}
To emulate the real-world condition, \ours is equipped with event-based measurement and control modules. These modules are triggered by the occurrence of certain events along the timeline.

\subsubsection{Conditions}
\label{sec:condition}
A condition object is instantiated from the class~\texttt{Condition} and acts as a watchdog that triggers when a certain event occurs. Hence, the purpose of a condition to notify the simulator about a predetermined event. The triggering event is a property of the condition instance. For example, one condition triggers when a specific date arrives. Another condition triggers when the number of infected people surpasses a specified threshold. 

Every condition object has two standard methods. One determines whether the condition should be triggered while the other method checks if the condition has served its purposes, i.e., whether the simulator still needs the condition.

Each type of condition may have its own attributes that are implemented on demand. A list of conditions implemented in the current version (V1.0) of Pyfectious is explained below. They can be extended fairly easily by the user to support customized conditions.

\begin{itemize}
    \item \texttt{Time Point Condition} notifies the simulator when the simulation time reaches a specified point in the timeline. Having the deadline as its parameter, the condition compares the current time with the simulation time on each pulse of the simulator's clock.
    \item \texttt{Time Period Condition} operates based on a given period. Starting from the beginning of the simulation, the condition notifies the simulator periodically whenever the simulator timer is divisible by the mentioned period. By increasing this divisor period, we can reduce the simulation's time resolution to reduce the computational burden at the cost of missing some short events that occur in an interval between two pulses. 

    \item \texttt{Statistical Ratio Condition} is an example of time-independent conditions that are defined by three items: a threshold ratio, a mathematical operator, and a pair of statistics from the population. For example, let the threshold ratio be $0.2$, the mathematical operator be division and the pair of statistics be the number of deaths, and the number of active infected cases. Hence, this condition is triggered when more than $20\%$ of the infected people die.
    
    \item \texttt{Statistical Family Condition} and \texttt{Statistical Role Condition} are other time-independent conditions. They follow the same logic as the \texttt{Statistical Ratio Condition} except that the given statistics that trigger the condition are taken from a specific family, role, subcommunity, or community. For example, an instance of this condition can get triggered if the number of infected students in a school surpasses a threshold. This allows interventions on specific roles within a particular community instead of treating all members of the same role alike. For example, suppose the infection is spread in a specific school instead of shutting every school in the society. In that case, \ours allows investigating the outcome of quarantining the subcommunities (teachers or students) of that particular school.
\end{itemize}

\subsection{Commands}
\label{sec:commands}
A command object is an instance of a class named \texttt{Command}, and its function is to intervene on the other objects in a running simulation. Every command corresponds to a real-world action that changes one or more attributes in individuals, communities, subcommunities, or the edges of the connectivity graph. These actions allow the implementation of a wide range of quarantine and restriction policies. Since the command objects are designed to mimic the policies issued by the health authorities to contain the infection, they cannot change the parameters that such policies in the real world cannot alter. For example, a command can shut down a school but cannot change the inherent attributes of the disease or fixed attributes of individuals such as age and background health condition. To elaborate more, the currently implemented commands in \ours are explained below.
\begin{itemize}
    \item \texttt{Quarantine Single Person or Quarantine Multiple People} force an individuals or a group of individuals to stay at home. The quarantine remains effective until another command lifts it. In the \texttt{Quarantine Multiple People} command, the quarantined people can be any subset of the population and do not need to belong to the same community or subcommunity. For example, if the infection is detected in the schools of a certain neighborhood of the city, a command can be issued to quarantine $50\%$ of the students and teachers of that neighborhood's schools
    \item \texttt{Quarantine Infected People} puts the currently infected people in quarantine. This command operates with the idealistic assumption that there is no inaccuracy in detecting the infected people.
    \item \texttt{Quarantine Single Family} and \texttt{Quarantine Multiple Families} put people living in the same residence in quarantine.
    \item \texttt{Quarantine Infected People Noisy} resembles \texttt{Quarantine Infected People} command but emulates some inaccuracy in the detection of the infected people. It also models another stochasticity in the application of the quarantine policy to better mimic real-world scenarios.
    \item \texttt{Restrict Certain Roles} imposes restrictions on one or more roles in the communities of the society. It has a parameter called \texttt{Restriction Ratio} that shows what ratio of the people with the target role must obey the restriction. For example, it can enforce that only $30\%$ of the students of a school can be present on-site, and the other $70\%$ must remain at home for a specified period.
    \item \texttt{Quarantine Single Community} or \texttt{Quarantine Multiple Communities} are relatively coarse-grained commands that shut down a specific community such as a particular school or restaurant.
    \item \texttt{Quarantine Community Type} is a coarse-grained command that shuts down all communities of the same type. For example, shutting down all the schools or restaurants of society are examples of this \texttt{Command}.
    \item \texttt{Change Immunity Level} is capable of increasing or decreasing the immunity level of a given set of individuals. Consequently, this command enables us to investigate the situations in which the immunity level of a specific group of individuals changes during the pandemic. In particular, as the vaccination increases the immunity level, the effect of vaccinating specific groups on changing the course of the pandemic can be helpful in designing the distribution of vaccines when the resources are limited.
    \item \texttt{Change Infection Rate} emulates the effect of the policies on individuals' behavior that affect how infectious a disease is. For example, in respiratory diseases transmitted via droplets while sneezing, coughing, or talking, governmental policies such as mandatory mask-wearing in certain locations result in decreasing the infection rate in those locations.
\end{itemize}

\subsection{Data logging with observers}
In reality, the virus spread information is logged with limited spatial and temporal coverage and resolution. Meaning that the information of only a subset of the population and at a sparse set of time points is logged and available to the policymakers. To emulate this real-world condition and at the same time to reduce the computational and memory usage, \ours is equipped with a class named~\emph{Observer} whose instances are meant to emulate the limitations of the real-world measurements. Similar to the command objects, the observer objects are also invoked by the activation of a condition object (see~\Cref{sec:condition}). For instance, to collect the simulation data at regular intervals, a \texttt{Time Period Condition} is created to activate an observer that measures the health condition of a specific group of individuals. Every observer stores the data related to individuals, families, communities, and subcommunities into a database. \ours is equipped with a comprehensive interface to get detailed reports form this database during or after the simulation is carried out for a desired duration.

\section{Experiments}
\label{sec:experiments}
The concepts, novelties and implementation details of \ours were presented in-depth in the previous sections. To showcase the wide range of applications this simulator can be used for, here we present a few examples by conducting illustrative experiments. In \Cref{sec:general_experiment_setup}, the general settings of the experiments are described and the results are discussed in \Cref{sec:evaluation_and_assessment_results}.

\subsection{General experimental setup}
\label{sec:general_experiment_setup}
In order to assess the simulation using a real-world scenario, the simulator requires a sample population structure, defining the primary properties of the population and the attributes of the disease that is planned to spread through the population. This data is provided to the software by two configuration files, one containing the population settings and the other containing the disease attributes. Prior to running the simulation engine, the configuration files need to be prepared.

For our experiments, a configuration file for the population generator is designed based on the structure of a small town's population. Likewise, a configuration file is prepared based on the known attributes of COVID-19 as an exemplary infectious disease. Before the simulation starts, the software parses these configuration files and constructs the population based on the retrieved information.

This section is dedicated to explaining the properties involved in the configuration files, along with a concise justification of the design procedure.

\subsubsection{Population generator}
\label{sec:population_generator_setting}
Designing a representative population structure is critical for the simulation to generate reasonable results that bear a resemblance to the available real-world statistics of the pandemic. Thus, the configuration file for the population generator must be designed carefully and based on realistic assumptions. We use the information provided by reputable population census and statistics centers. In particular, we use two primary sources of information to adjust the structure of the generated population. First, the United States Census Bureau's tables and data of the US population \cite{us2020census}. Second, the Eurostat \cite{eurostat2020census} data collection that provides information on the demographics of a number of European countries.

\begin{table}[]

\caption{ Set of family patterns and the probability of their occurrence (M: male, F: female, \{\}: A family gender pattern).  Any other arbitrary family pattern can be easily defined in \ours. For brevity, the age and the health condition of the family members are excluded from this table. They are sampled from a truncated normal distribution for every family member.}

\centering
\begin{tabular}{@{}ccc@{}}
\hline
Family Size & Genders                & Probability \\ \hline
2           & \{M, F\}                    & 0.21        \\ \hline
3           & \{M, F, M\} , \{M, F, F\}       & 0.3         \\ \hline
4           & \{M, F, M, F\} , \{M, M, M, F\} & 0.19        \\ \hline
1           & \{F\} , \{M\}                   & 0.126       \\ \hline
5           & \{M, F, M, F, F\}           & 0.124       \\ \hline
6           & \{M, M, M, F, F, F\}        & 0.05        \\ \hline
\end{tabular}
\label{tab:family_patterns}
\end{table}

The population generator configuration file comprises the following items. 
\begin{enumerate}
    \item \texttt{Population Size}: This attribute determines the total size of the population and is set to 20,000 in our experiments, which is roughly the average population of a small town.
    \item \texttt{Family Patterns}: A list of family patterns and the probability of their existence in society are required to disperse the population among the families. Furthermore, as discussed in \Cref{sec:population_generative_model}, a family pattern enables the simulation to assign individual attributes, for instance, gender and health condition, to the people. Six family patterns are designed and reported in \Cref{tab:family_patterns} for our intended experiments.
    \item \texttt{Community Types}: As illustrated in \Cref{sec:population_generative_model}, after distributing the individuals of the population to families based on the provided family patterns, the individuals are also assigned to communities that are the closest match to their personal attributes and the attributes of the family they belong to. The conducted experiments in this section are based on the communities briefly described in \Cref{tab:communities_information}. 
    \item \texttt{Distance Function}: A function that determines if two individuals are in close contact. Here, the \emph{Euclidean distance} is used as a measure of proximity between two individuals. Other measures can be employed depending on the real-world circumstances.
\end{enumerate}

\begin{table}[!t]
\centering
\caption{The design of the communities and subcommunities for an exemplary city (society).} 

\begin{tabular}{@{}ccccc@{}}
\toprule
\begin{tabular}[c]{@{}c@{}}Community Type\\ Name\end{tabular} &
\begin{tabular}[c]{@{}c@{}}Number of\\ Communities\end{tabular} &
\begin{tabular}[c]{@{}c@{}}Sub-Community\\ Types\end{tabular} &
\begin{tabular}[c]{@{}c@{}}Transmission \\ Potential\end{tabular} &
\begin{tabular}[c]{@{}c@{}}Connectivity \\ Matrix\end{tabular} \\ \midrule
School &
  40 &
  \begin{tabular}[c]{@{}c@{}}Teacher \\ Student\end{tabular} &
  Very High &
  High Density \\ \midrule
\begin{tabular}[c]{@{}c@{}}Workspace \\ (Medium)\end{tabular} &
  800 &
  \begin{tabular}[c]{@{}c@{}}Worker\\ Potential Client\end{tabular} &
  Very High &
  Medium Density \\ \midrule
\begin{tabular}[c]{@{}c@{}}Workspace \\ (Large)\end{tabular} &
  50 &
  \begin{tabular}[c]{@{}c@{}}Worker\\ Potential Client\end{tabular} &
  Moderate to High &
  Medium Sparsity \\ \midrule
Gym &
  50 &
  \begin{tabular}[c]{@{}c@{}}Trainer\\ Client\end{tabular} &
  Moderate to High &
  Medium Density \\ \midrule
Public Transportation &
  10 &
  \begin{tabular}[c]{@{}c@{}}Commuter\\ Staff\end{tabular} &
  Low to Moderate &
  High Sparsity \\ \midrule
Restaurant &
  80 &
  \begin{tabular}[c]{@{}c@{}}Staff\\ Costumer\end{tabular} &
  Moderate to High &
  Medium Density \\ \midrule
Cinema &
  15 &
  \begin{tabular}[c]{@{}c@{}}Staff\\ Costumer\end{tabular} &
  Low to Moderate &
  Medium Sparsity \\ \midrule
Mall &
  5 &
  \begin{tabular}[c]{@{}c@{}}Staff\\ Costumer\end{tabular} &
  Very Low &
  High Sparsity \\ \bottomrule
\end{tabular}
\label{tab:communities_information}
\end{table}

\subsubsection{Disease properties}
\label{sec:disease_properties_setting}
The disease configuration file contains the fundamental attributes of an infectious disease. Here, we explain these attributes and the reasoning behind the selected value for each of them.

\begin{enumerate}
    \item \texttt{Infection Rate}:

    Sampled from a uniform distribution whose support is determined based on real-world reports. We adjusted the support of the distribution, i.e. $\textrm{Uniform}[0.1, 0.6]$, in the experiments based on the data available from Wuhan \cite{ChinaData}, where there was less restrictive measures at the beginning. 

    \item \texttt{Immunity}: Based on the previous studies on COVID-19, e.g., \cite{tay2020trinity}, no prior immunity has been detected in most of the studied cases. Therefore, the immunity against infection for the first time must be sampled from a distribution with a support relatively close to zero. However, once a person recovers from the disease, the possibility of reinfection in the short term is infinitesimal. The immunity parameter is sampled from a distribution that is designed  based on this assumption: A person has a negligible immunity against the virus in the first infection, and the immunity is exponentially raised upon each infection.
    \item \texttt{Incubation Period}:  We consider a truncated normal distribution (mean = 5.44, 95\%-CI\footnote{Confidence Interval} = (4.98, 5.99), STD = 2.3) for the the duration of the incubation period based on the reported values in the data collected by \cite{shan2020epidemiological} and more notably the study conducted by \cite{li2020early}.
    \item \texttt{Disease Period}:

    This parameter is sampled from a truncated normal distribution (mean = 6.25, 95\%-CI = (5.09, 7.51), STD = 3.72) based on the values reported in \cite{backer2020incubation,li2020early}.

    \item \texttt{Death Probability}: The value of death probability, also known as mortality rate, is available through the
    hospital's statistics for patients who are diagnosed with COVID-19. Here, the death probability is sampled from a truncated normal distribution (mean = 1.4\%, 95\%-CI = (1.1\%, 1.6\%), STD = 2.4) as reported in \cite{hauser2020estimation, riou2020pattern}.
\end{enumerate}

\subsection{Evaluation and assessment of the results}
\label{sec:evaluation_and_assessment_results}
To illustrate the simulator's potential and applications, we have conducted several experiments, ranging from exploring the outcome of changing the disease parameters to imposing restrictions to control the disease's spread. The conducted experiments are based on the configurations described in \Cref{sec:disease_properties_setting,sec:population_generator_setting}.

We deployed our experiments on a cluster distributed on a multitude of computational nodes to accelerate using parallel execution. Every particular curve that we obtained through our experiments is produced by 32 simultaneous executions combined in a single graph. We use the moving average, with a window size of 2, to reduce the effect of high-frequency oscillations and improve the visibility of the trend. The variations are typically caused by the chosen sampling rate that is 7 hours in our experiments.
Comparing our observations in \Cref{fig:normal_execution_plus_average} with the statistics published by the authorities, for instance, the US COVID-19 statistics \cite{cdc2020graphs}, we conclude that the observed oscillations are not unexpected and do not affect the trend of the simulation graphs.

To deliberately evaluate the experiments, they are divided into five categories, each explained in the subsequent sections.

\subsubsection{Performance and sanity checks}
In this section, we present primary experiments related to the soundness and performance of the simulation software \ours.

\begin{figure}[!t]
    \begin{subfigure}{0.5\columnwidth}
    	\centering
    	\includegraphics[width=1\columnwidth]{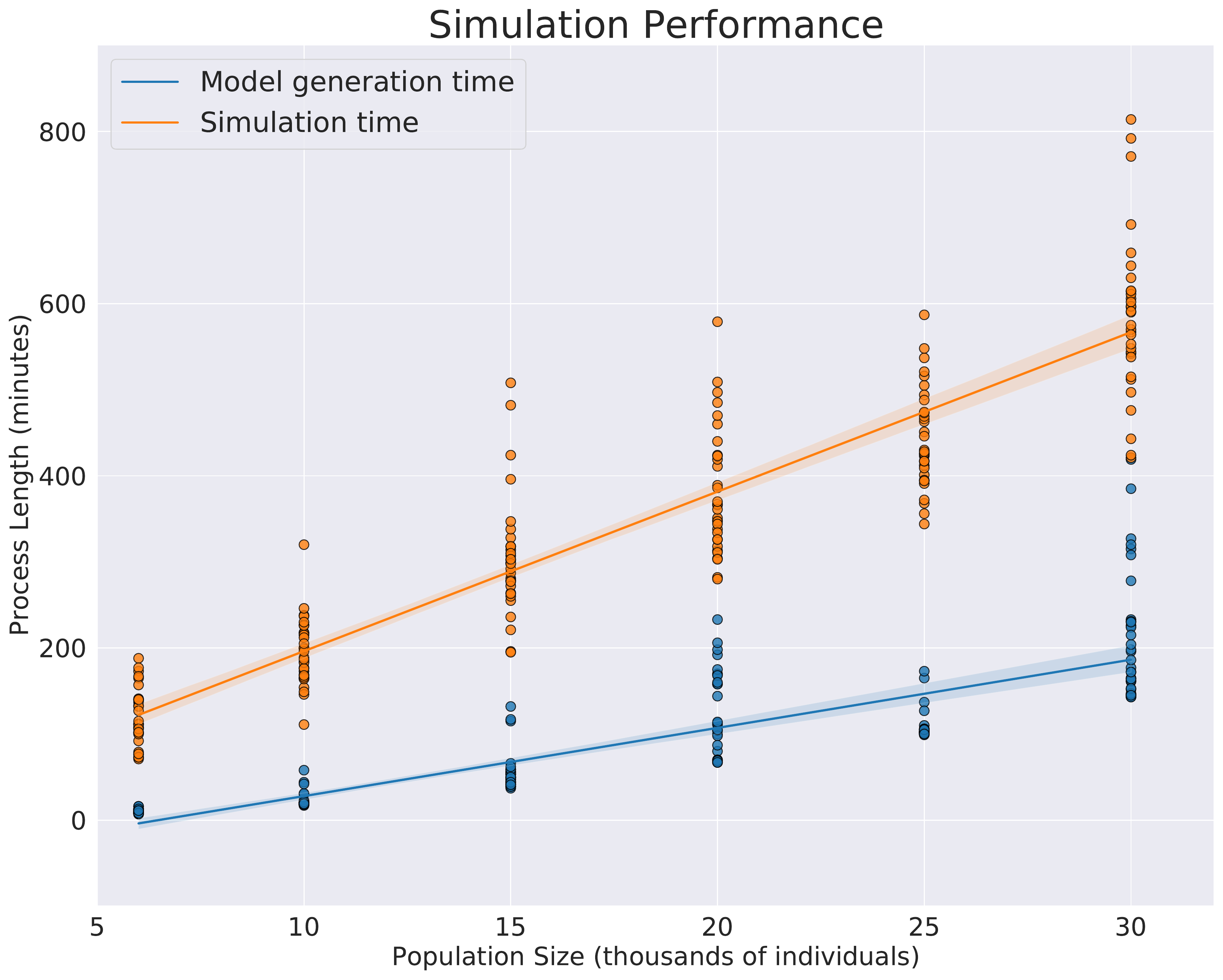}
        \caption{}
        \label{fig:simulator_performance}
	\end{subfigure}
	\hfill
    \begin{subfigure}{0.5\columnwidth}
	    \centering
        \includegraphics[width=1\columnwidth]{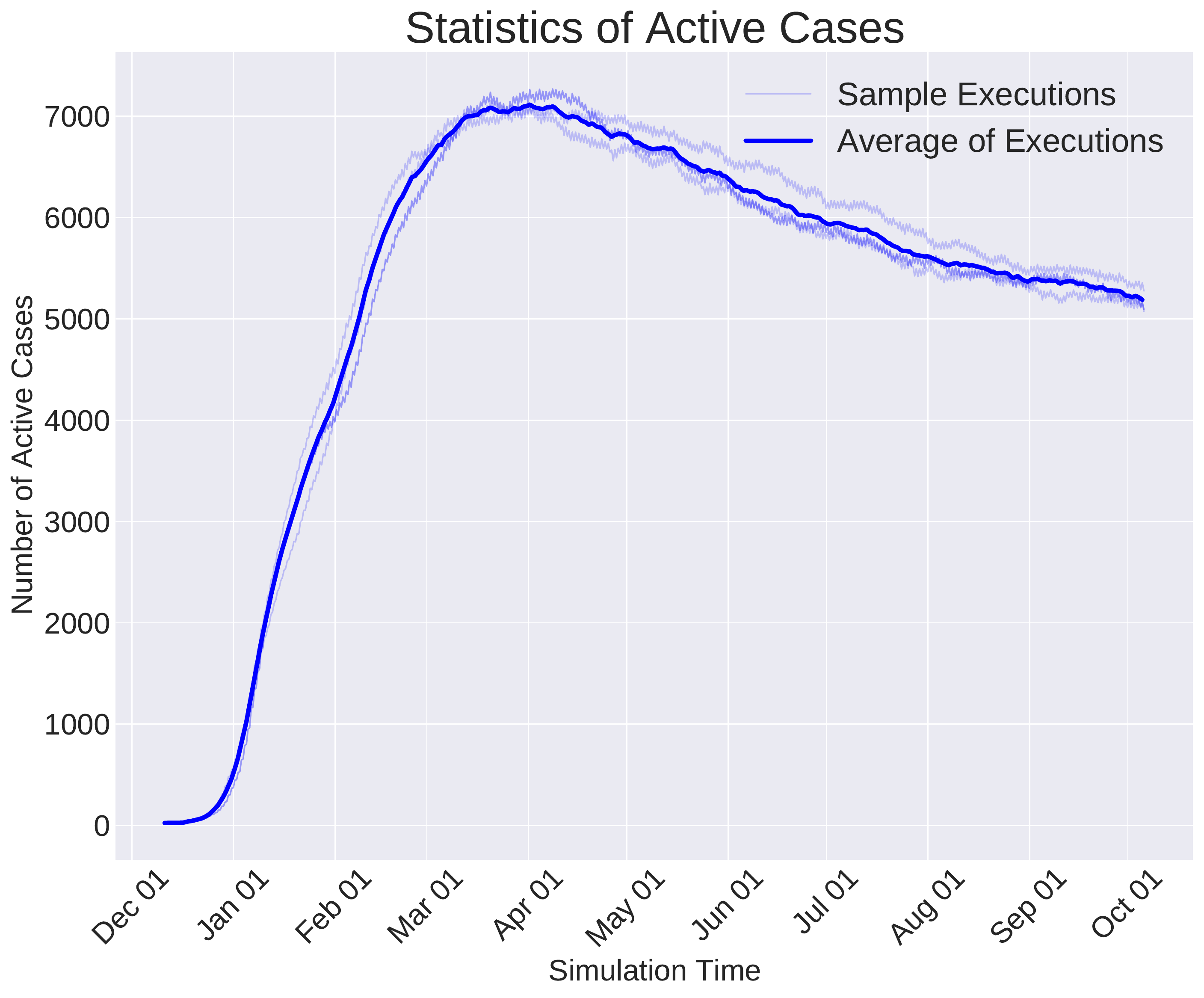}
        \caption{}
        \label{fig:normal_execution_plus_average}
	\end{subfigure}
	\caption{a) The time it takes to generate the population and to simulate the propagation of the disease for 48 hours is plotted for six cities with population sizes from 6k to 30k. The horizontal axis represents the population size, and the vertical axis represents the total process time. The experiment for every size of the population is repeated multiple times (each vertically aligned dot corresponds to an experiment) to achieve confidence, and the straight line indicates the trend. b) The number of active cases versus time is shown for a sample city with a population size of 20k. To emphasize the probabilistic nature of \ours, the same experiment is repeated multiple times, and the effect of this randomness is seen by observing slightly different trajectories. The blue curve is the moving average (window size of 2) of all executions to show the trend.} 
	\label{fig:simulation_performance_normal_executions_average}
\end{figure}

\paragraph{Performance measures.} The first experiment in this section, shown in \Cref{fig:simulator_performance}, aims at measuring the required time for the model generation and simulation phases. Since the generated models may be saved and reused when running the simulation more than once, its required time can be amortized when simulating the same model multiple times or with different disease attributes. Evidently, the experiment proves a linear relationship between the population size and both the simulation and model generation times. This result is promising and shows the scalability of \ours for simulating larger cities with more complex population structures.

\begin{figure}[!t]
    \begin{subfigure}{0.5\columnwidth}
    	\centering
        \includegraphics[width=1\columnwidth]{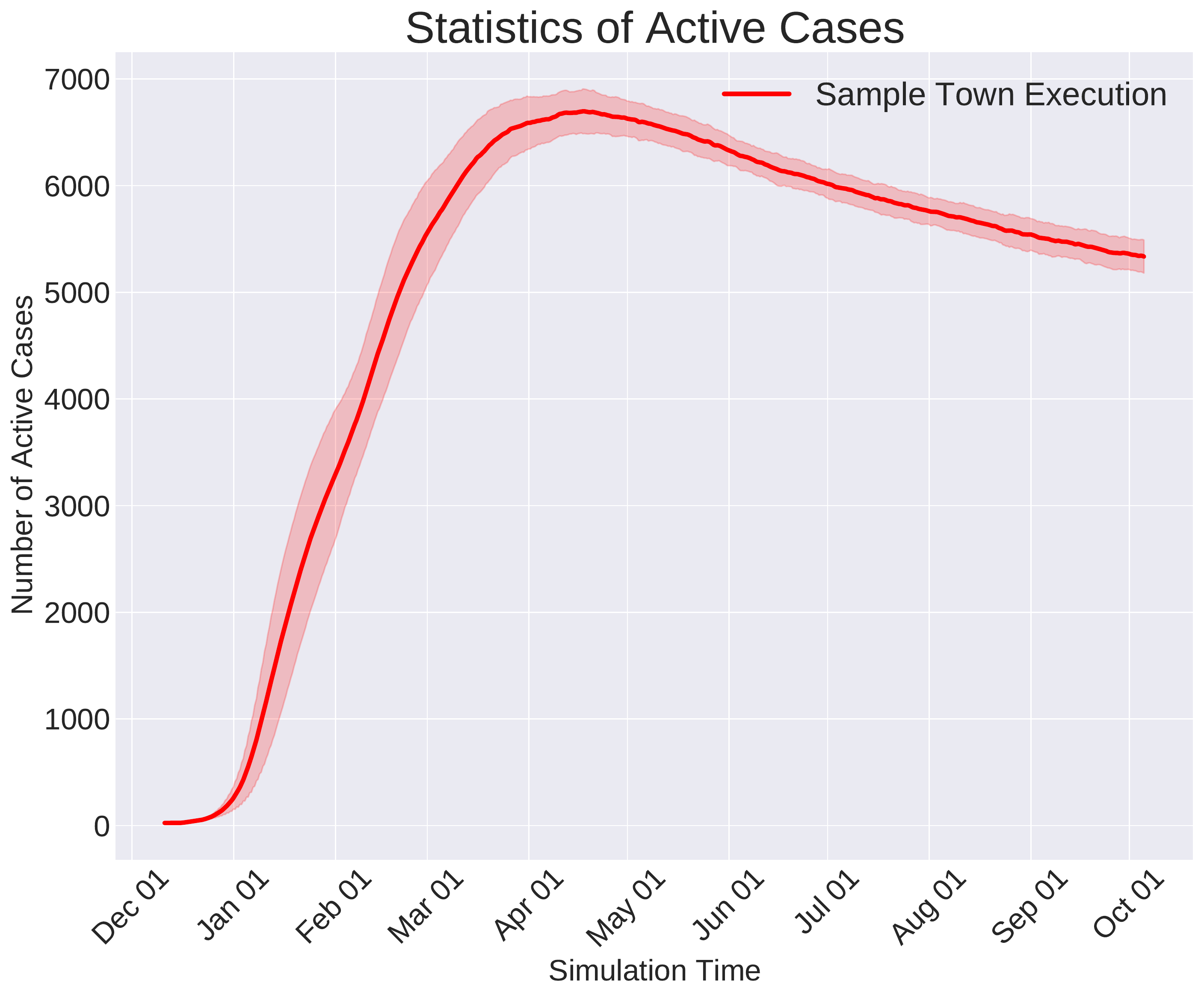}
        \caption{}
        \label{fig:normal_execution_plus_error_band}
	\end{subfigure}
	\hfill
    \begin{subfigure}{0.5\columnwidth}
	    \centering
        \includegraphics[width=1\columnwidth]{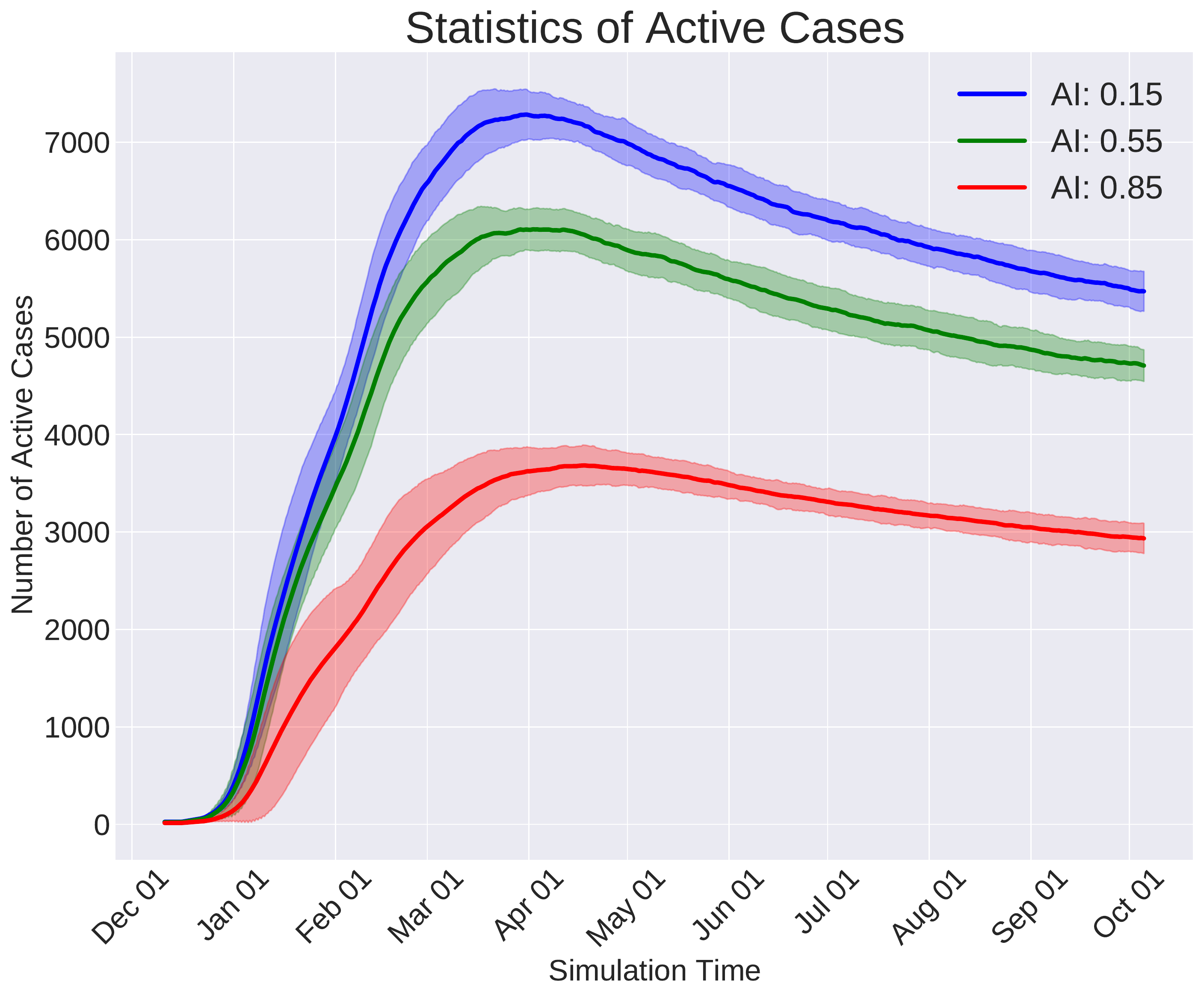}
        \caption{}
        \label{fig:immunity_effect_on_infected}
	\end{subfigure}
	\caption{a) The simulation is executed without any control measure, and the number of infected individuals is plotted versus time for the period of 10 months. The halo around the solid curve is the confidence interval obtained by multiple runs. At each round, the parameters of the population and the disease are re-sampled from the specified distributions. b) The curves of the number of active cases versus time are plotted for different immunity rates. The immunity rate is sampled from three uniform distributions with different mean values. As can be seen, that more significant immunity rates give rise to flatter curves. Note that AI initials in the legend stand for Average Immunity, which is the mean of the uniform distribution from which each curve's immunity rate is sampled.} 
	\label{fig:normal_executions_errorbands_immunity_effect}
\end{figure}

\paragraph{Sanity checks.} In the experiment whose result is shown in \Cref{fig:normal_execution_plus_average,fig:normal_execution_plus_error_band}, 
the graphs of the number of active cases versus time are reported for six executions. The oscillations caused by sudden changes in the active cases' statistics are observable, especially near the curve's global maximum, where there is the largest number of active cases. As mentioned before, the moving average of these curves is shown to illustrate the trend better. Notice that a slight difference between the produced trajectories in \Cref{fig:normal_execution_plus_average} is expected due to the probabilistic nature of \ours at multiple levels. Numerous sampled trajectories similar to those shown in \Cref{fig:normal_execution_plus_average} form a halo around the average trajectory as in \Cref{fig:normal_execution_plus_error_band} to obtain confidence in the results. Every experiment in this paper is executed multiple times to obtain such confidence intervals and be robust against randomness artifacts.

\paragraph{Discussion on herd immunity.} Although no control measure is applied during the simulation in \Cref{fig:normal_execution_plus_error_band}, the number of active cases declines after reaching a certain level. This incident, called~\emph{herd immunity}~\cite{anderson1985vaccination}, is aligned with the course of infectious diseases in the real world. It shows a reduction in the number of active cases after a certain fraction of the population has been infected by the virus and developed a level of immunity. We implemented an immunity model based on the available real-world information. Inspired by~\cite{duggan2021novel}, our model incorporates a minor chance of reinfection (about 1\%), which significantly drops after the second infection. 
Aligned with the real-world experiences, a decline occurs in the number of cases after a sufficient portion of the population (about 70-80\% in our simulation) recovers from the disease.

\subsubsection{Changing the attributes of the disease}
To investigate the effect of changing the disease properties, this section covers a comprehensive study on how changing disease attributes affect its spread through a structured population.

\paragraph{Immunity variations. }To assess the influence of immunity distribution on the results, we organize three sets of experiments, depicted in \Cref{fig:immunity_effect_on_infected}. Each curve shows the evolution of the number of active cases where the immunity rate is sampled from a uniform distribution with a specified range. In the lowest immunity level, a larger population contracts the disease as the population has the least resistance against the infection. Notably, increasing the immunity level to intermediate and high results in lowering the peak of the active cases' curve. For instance, if a fraction of the population is vaccinated, the immunity rises within that group, and the trajectory of active cases reorients from the blue curve to the red or the green one. This incident is often referred to as "flattening the curve."

\begin{figure}[!t]
    \begin{subfigure}{0.5\columnwidth}
    	\centering
        \includegraphics[width=1\columnwidth]{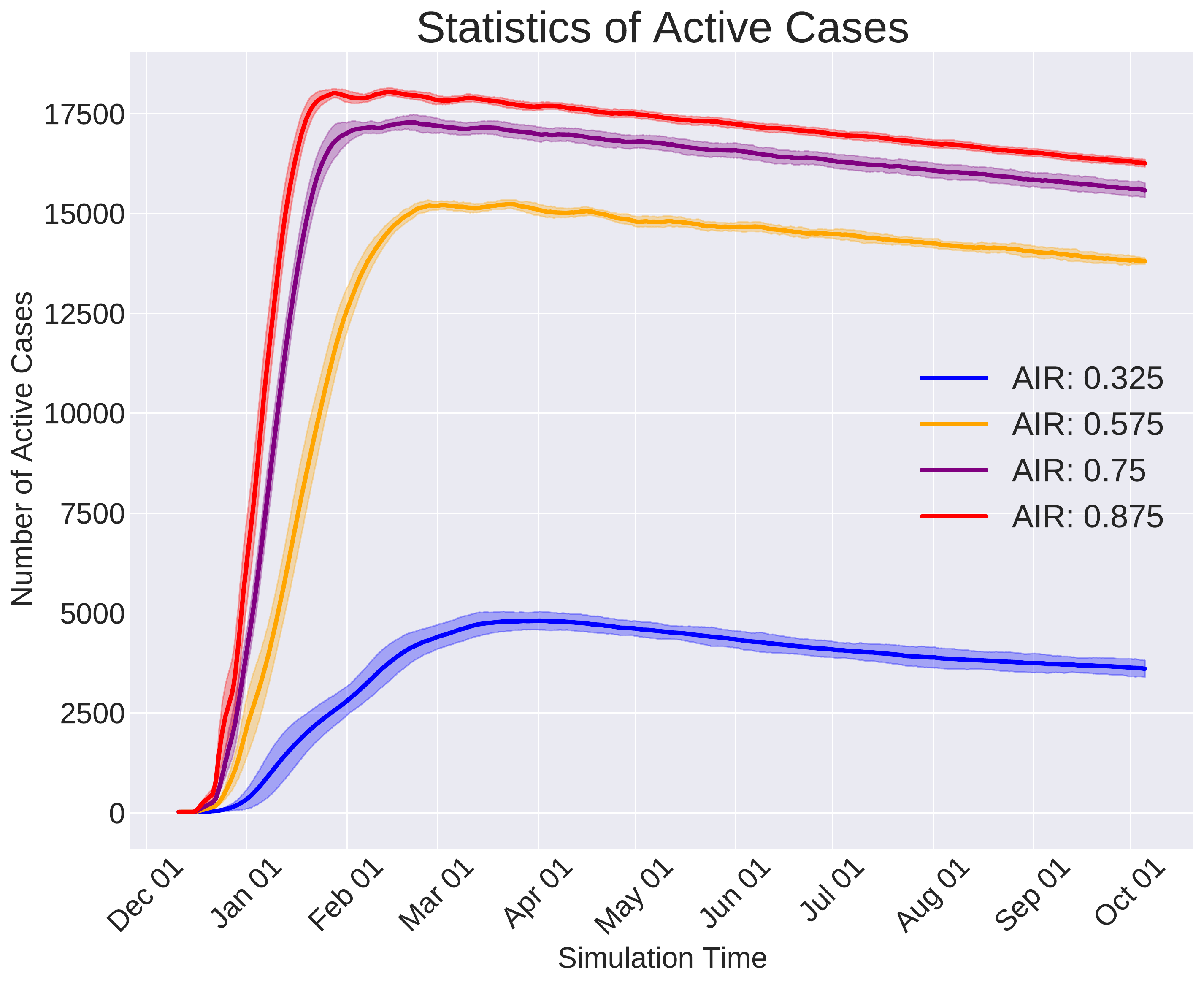}
        \caption{}
        \label{fig:infectious_effect_on_infected}
	\end{subfigure}
	\hfill
    \begin{subfigure}{0.5\columnwidth}
    	\centering
        \includegraphics[width=1\columnwidth]{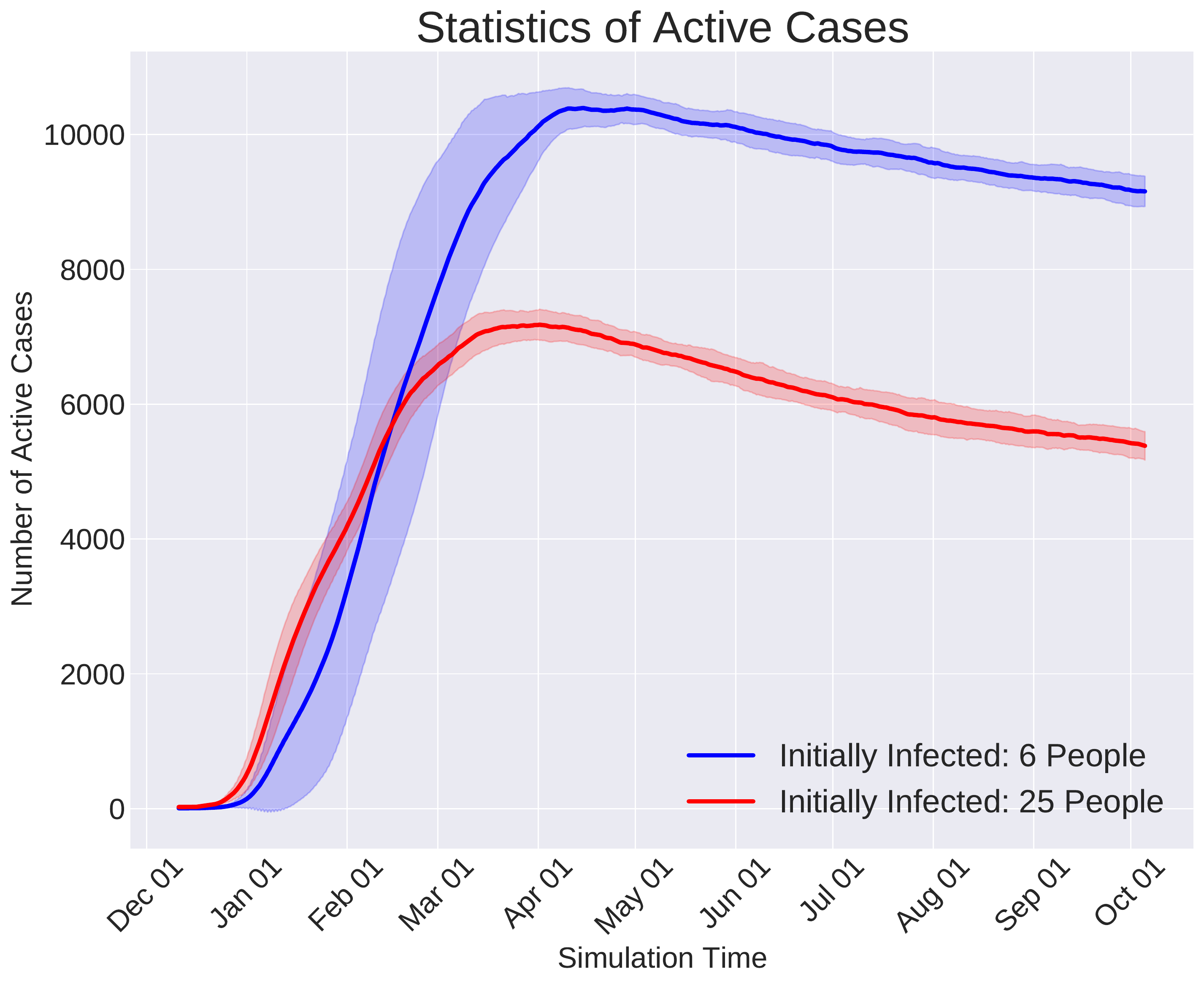}
        \caption{}
        \label{fig:reduce_initially_infecteds}
	\end{subfigure}
	\caption{a) The number of currently infected individuals is plotted versus time for different values of the infection rates. The infection rates are sampled from uniform distributions with different mean values. It is observed that a larger infection rate increases the slope of the curve that means a faster spread of the disease early after the advent of the outbreak. As a result, it takes less time for the number of active cases to reach its peak. Notice that AIR stands for the Average Infection Rate, which is the mean of the distribution from which the infection rate is sampled. b) The spread of the infection is shown versus time for different numbers of initial spreaders where the smaller set is chosen from communities such as large workspaces and schools that are suitable places for infecting many people. The confidence intervals are expectedly wider for smaller initially infected set because it results in some communities without an initial spreader and consequently a less homogeneous spread of the disease. The observation that the peak of the graph with a smaller initial set is higher than the one with a larger initial set emphasizes the hypothesis that some roles and places need special treatment early in an epidemic even though only a few of their individuals can be initially infected.}

	\label{fig:infectious_effect_reduce_initially_infected}
\end{figure}

\paragraph{Modifying the infection rate.} In this experiment, we investigate the effect of infection rate on the disease's spread in the population. To obtain confidence in the results, each experiment is executed 32 times with the infection rate that is sampled from a uniform distribution with a specified range. The curves in~\Cref{fig:infectious_effect_on_infected} indicate the results for five non-overlapping uniform distributions from which the infection rate is sampled. It can be seen that for the exceptionally high rate of infection (i.e., the infection rate is sampled from a uniform distribution whose support occupies larger values), the number of active cases increases with a significant slope before reaching the peak of the curve. As the area under the curve of active cases shows in~\Cref{fig:infectious_effect_on_infected}, the total number of infected people during the epidemic increases with an increase in the infection rate.

\begin{figure}[!t]
    \begin{subfigure}{0.5\columnwidth}
    	\centering
        \includegraphics[width=1\columnwidth]{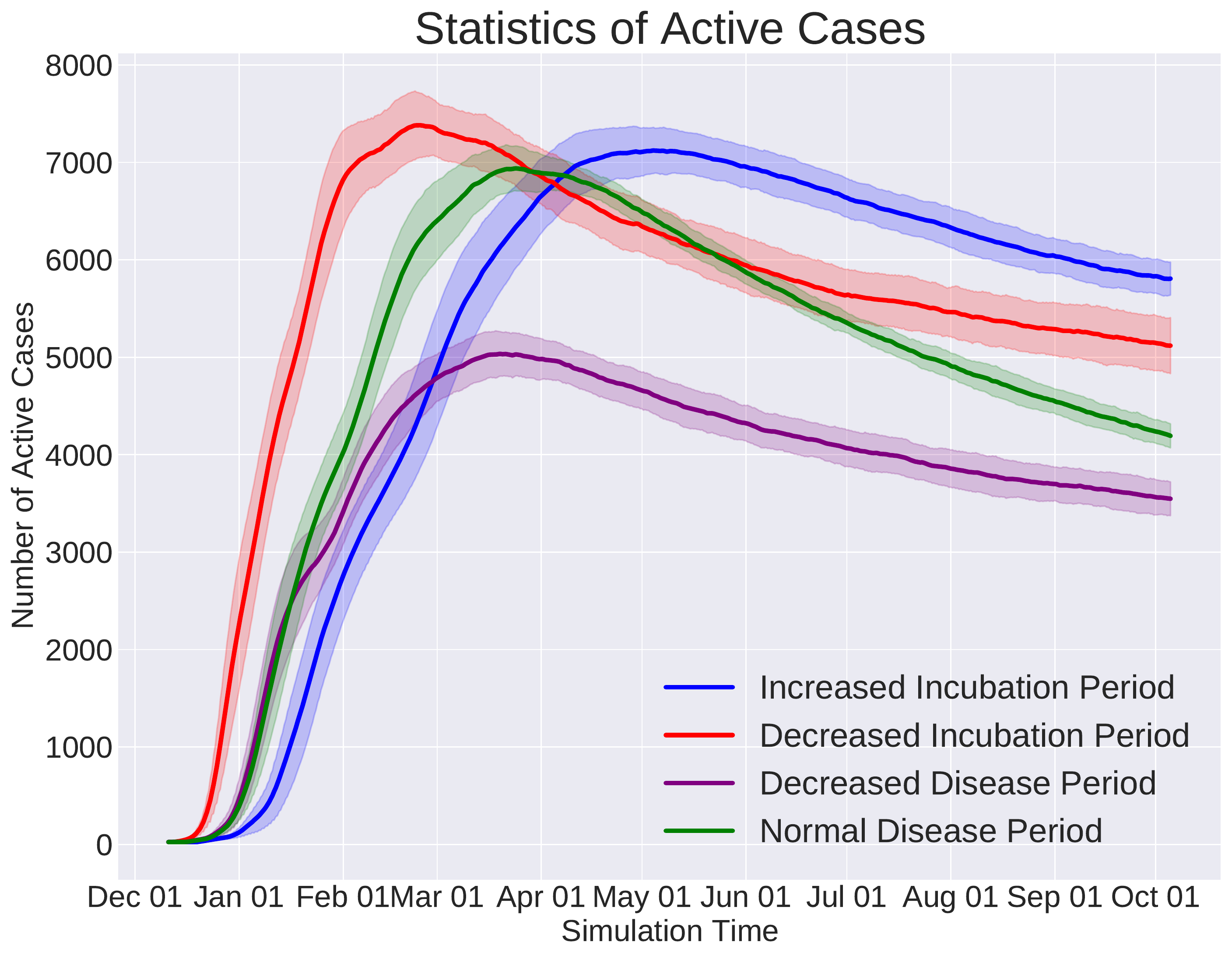}
        \caption{}
        \label{fig:disease_properties_variations}
	\end{subfigure}
	\hfill
    \begin{subfigure}{0.5\columnwidth}
    	\centering
        \includegraphics[width=1\columnwidth]{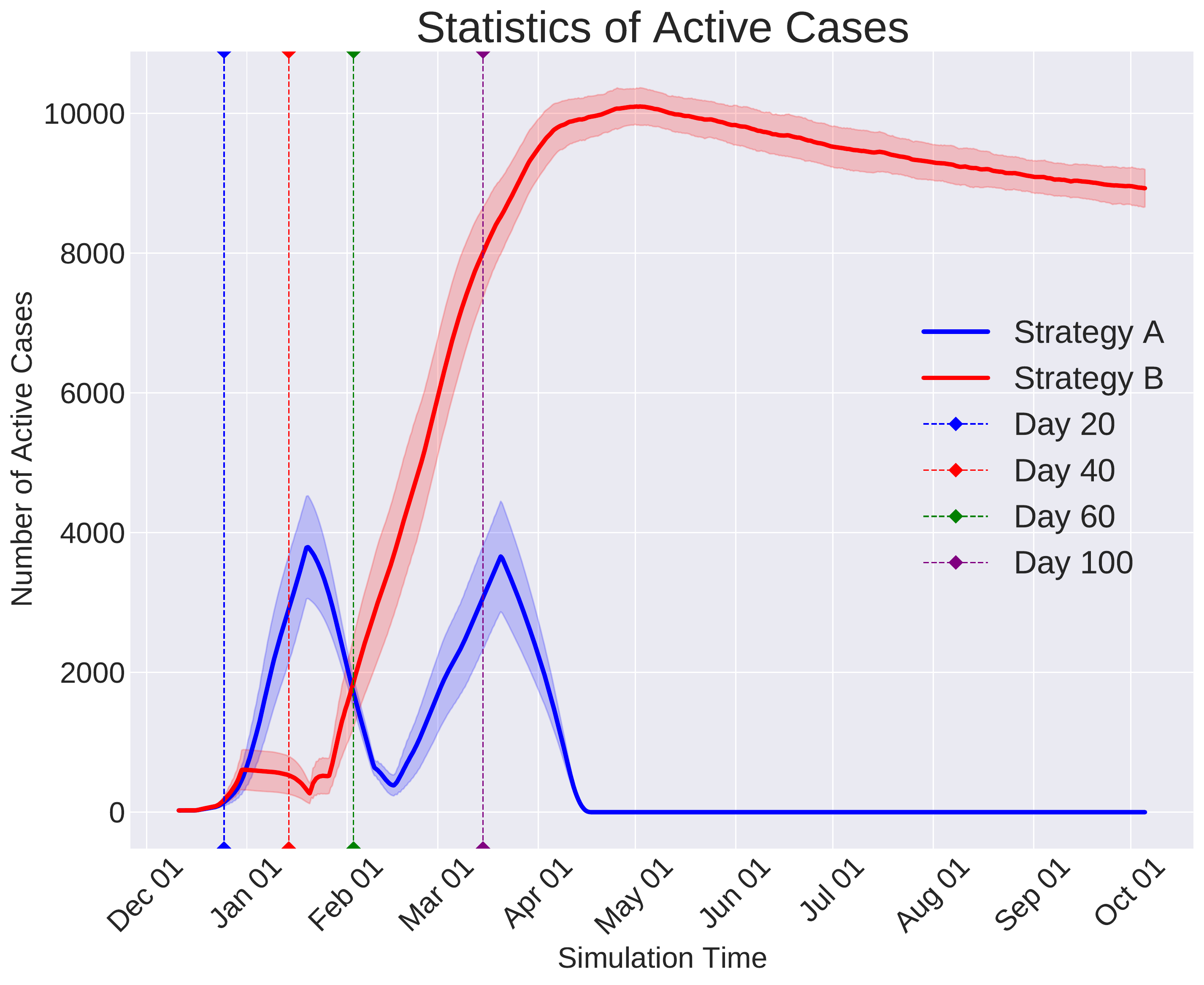}
        \caption{}
        \label{fig:quarantine_unquarantine_infecteds}
	\end{subfigure}
	\caption{a) The effect of the length of the incubation period (the period in which the infection is not detectable) and the disease period (the period in which the individual is infectious) is shown by changing these parameters of the disease. The curves correspond to a normal incubation and disease period, increased incubation period by 3.5 days, decreased incubation period by 3.5 days, and the decrease disease period by seven days. b) The outcome of two quarantine strategies. Strategy A: Enforce a quarantine 20 days after the outbreak and lift it 20 days later. Strategy B: Enforce a quarantine 40 days after the outbreak, lift it after 20 days and enforce it again after 40 days. The oscillatory curve is expected as the remaining active cases after the initial quarantine will be the initial spreaders for the next wave of the epidemic.
	}
	\label{fig:disease_periods_quarantine}
\end{figure}

\paragraph{Initially infected cases.} The first set of infected individuals in a population plays a crucial role in spreading the disease. Here, we check this effect by studying the course of the epidemic when the number of initially infected people is set to either $6$ or $25$ with the added information that the smaller set is chosen from the large workspaces and schools. Again, we run every experiment multiple times to ensure the results are not by accident. As \Cref{fig:reduce_initially_infecteds} shows, when only six infected people exist at the beginning of the simulation, the error bounds are wider compared to when there are $25$ initially infected people. This effect is expected because, for a more significant number of initially infected people that are randomly assigned to communities, most of the communities will have at least one infectious member. Hence, there will be minor variations across the executions of the simulation compared to the situation when a few individuals are chosen from a different set of communities at each execution. The other notable observation is that even though the blue curve initiates from fewer infected people, it shows a larger set of infected people eventually. The reason is that its initial set is intentionally chosen from communities, such as schools or large workspaces, making it easier to spread the disease throughout a large population in a short time.

\paragraph{Incubation and disease period.} As already discussed in \Cref{sec:disease_properties_setting}, incubation and disease period parameters define the temporal behavior of the infectious disease. As \Cref{fig:disease_properties_variations} shows, changing these parameters has a significant effect on the curve of active cases. For instance, increasing the incubation period creates a longer flat curve at the beginning of the outbreak. It can also be seen that an elongated incubation period shifts the curve forward in time with an almost equal peak of the number of active cases. Simultaneously, a shortened length of the disease period decreases the peak and the total cumulative number of active cases. This result is expected because a longer period of disease without a strict control measure to keep the infected individuals away from the others increases the chance of spreading the infection. 

\begin{figure}[!t]
    \begin{subfigure}{0.5\columnwidth}
	    \centering
        \includegraphics[width=1\columnwidth]{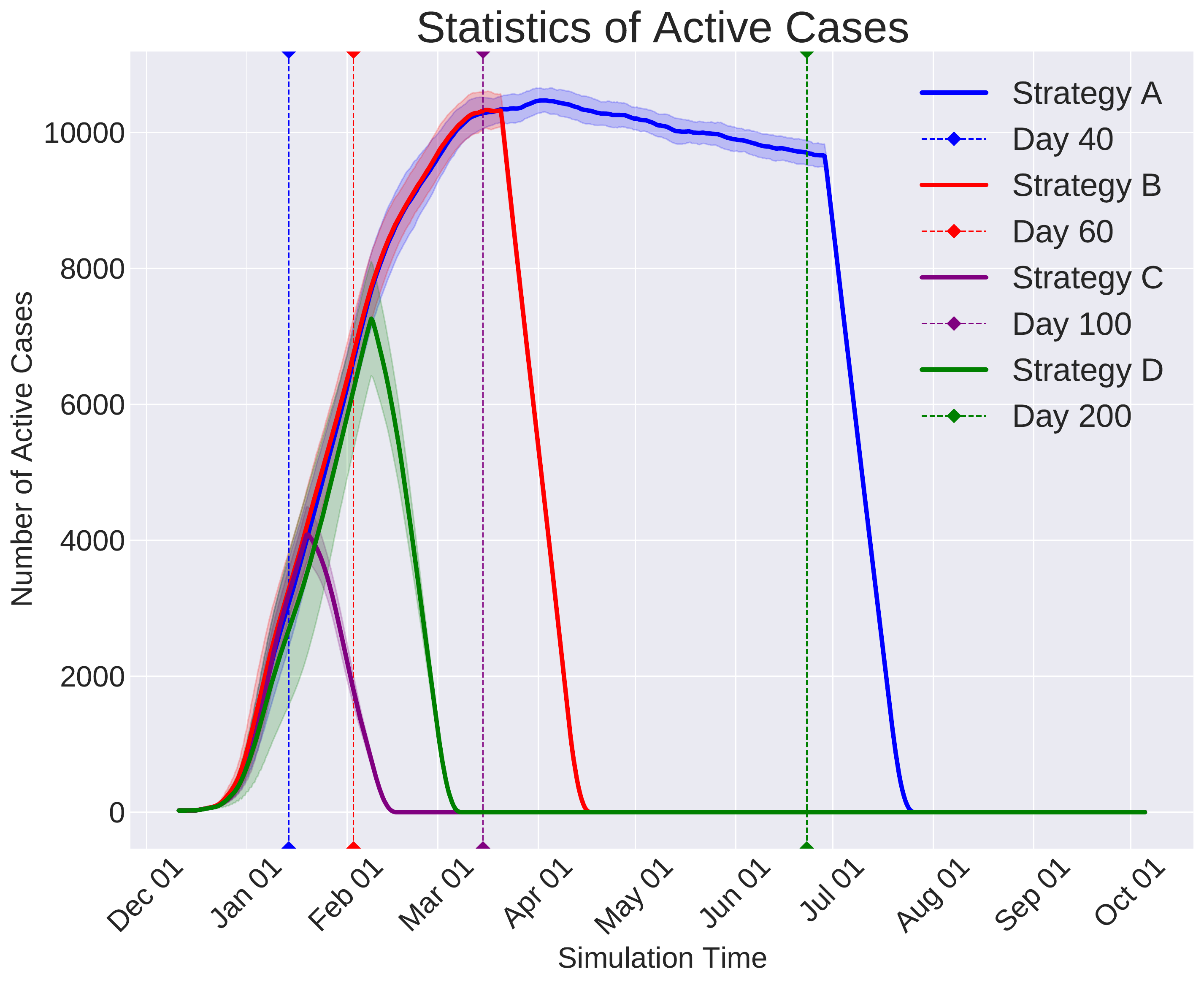}
        \caption{ }
        \label{fig:quarantine_infected_effect}
	\end{subfigure}
	\hfill
    \begin{subfigure}{0.5\columnwidth}
	    \centering
        \includegraphics[width=1\columnwidth]{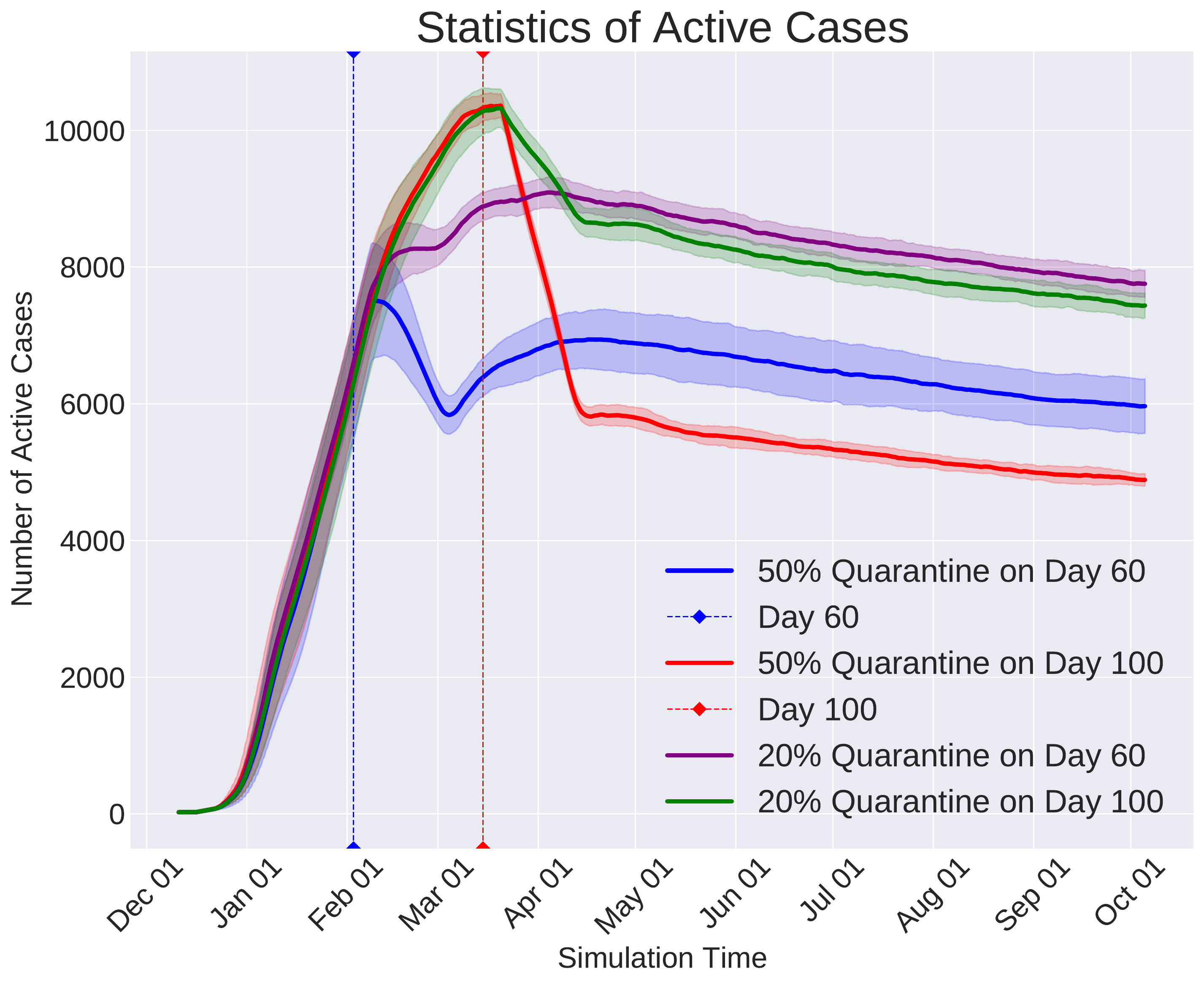}
        \caption{ }
        \label{fig:quarantine_partially_effect}
	\end{subfigure}
	\caption{a) This experiment focuses on enforcing universal quarantines (isolating every infected individual after detection) on a specific day after the outbreak. Here, the quarantines are applied both before and after the day when the curve of the active cases reaches its peak. Strategies A, B, C, and D enforce a quarantine at 40, 60, 100, and 200 days after the outbreak, respectively. b) This experiment studies the effect of partial quarantine where a specified ratio of currently infected individuals are isolated at a specified date (i.e., the control measure is triggered by a time point condition). The partial quarantine represents a real-world scenario where there is uncertainty in detecting the infected individuals that can be caused by numerous reasons such as inaccurate test kits or individuals with mild symptoms that do not visit hospitals or test facilities.}
	\label{fig:quarantine_infected_quarantine_infected_partially}
\end{figure}

\subsubsection{Applying control measures} As introduced in \Cref{sec:commands}, \ours offers a straightforward and flexible way to impose control measures during the course of the simulation to emulate real-world epidemic containment policies. The control policies can appear in numerous forms, including quarantining people or communities and reducing the number of people in some sectors of society. Here, we present a couple of experiments to illustrate the effect of control measures that are similar to those applied in the real world. Typically, there is a trade-off between the strength of the control measure and its outcome. The most strict measures, such as forcing everyone to stay at home and in isolation from other family members, stop the spread of the disease but entails enormous economic and societal costs. \ours allows us to investigate this trade-off by changing the strength of the control measures in an almost continuous way to find the optimal restrictive rules with a reasonable cost. In the following, some of the control measures inspired by real-world policies are tested.

\paragraph{Full quarantine. } This policy, whose result is shown in \Cref{fig:quarantine_infected_effect} refers to the most strict quarantine method isolating every discovered infected individual. Each curve of~\Cref{fig:quarantine_infected_effect} represents the effect of applying the strict full quarantine with different starting dates. As the quarantined infected people are no longer able to infect others, the complete isolation causes a sharp drop in the number of active cases until the epidemic eventually vanishes. As expected, it is clear that the full quarantine is most effective if it is applied as early as possible after the outbreak of the infection.

\paragraph{Enforce and remove a quarantine. } In \Cref{fig:quarantine_unquarantine_infecteds}, the effect of enforcing a quarantine early and removing it after some time is presented. As appears of the results, an early quarantine could be effective if placed and removed at particular time points (as in Strategy A). However, it could also fail to contract the virus if not appropriately planned (as in Strategy B), i.e., the start and termination time are not sufficient to control the spread, and the pandemic strikes back after removing the quarantine.

\paragraph{Partial quarantine. } As opposed to the full quarantine measures, in \Cref{fig:quarantine_partially_effect}, we study the effect of quarantine under, a more realistic assumption that the detection of the infected people is not absolute. For instance, a $50\%$ error indicates that the detection and testing mechanisms are cables of detecting only half of the infected population. Once infected people are detected, they are treated with total isolation as in the full quarantine policy. Nevertheless, complete isolation is still effective in controlling the epidemic even when the detection rate is significantly lower, e.g., only $20\%$ of infections are detected. 

\begin{figure}[!t]
    \begin{subfigure}{0.5\columnwidth}
	    \centering
        \includegraphics[width=1\columnwidth]{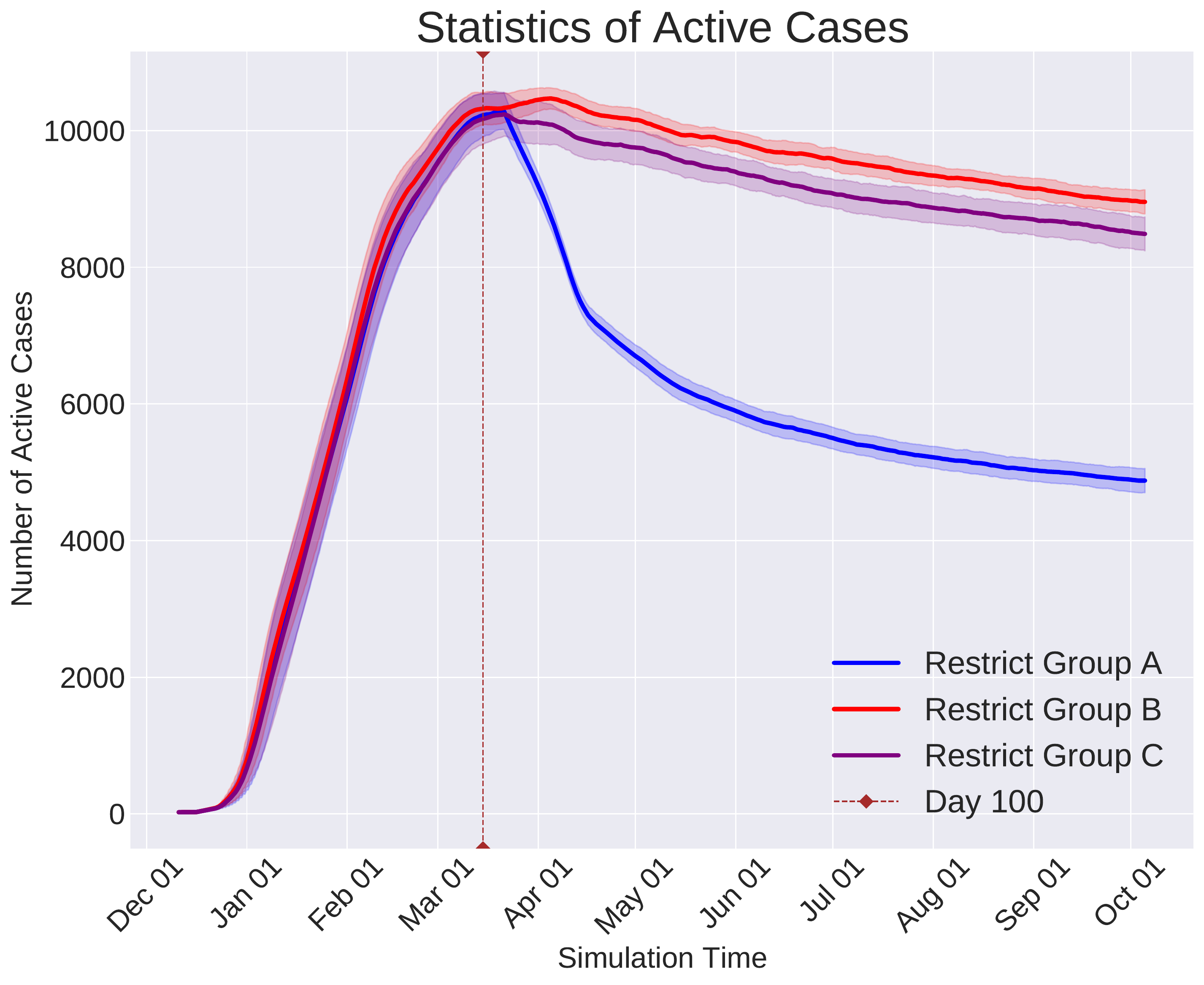}
        \caption{}
        \label{fig:quarantine_society_sectors}
	\end{subfigure}
	\hfill
    \begin{subfigure}{0.5\columnwidth}
	    \centering
        \includegraphics[width=1\columnwidth]{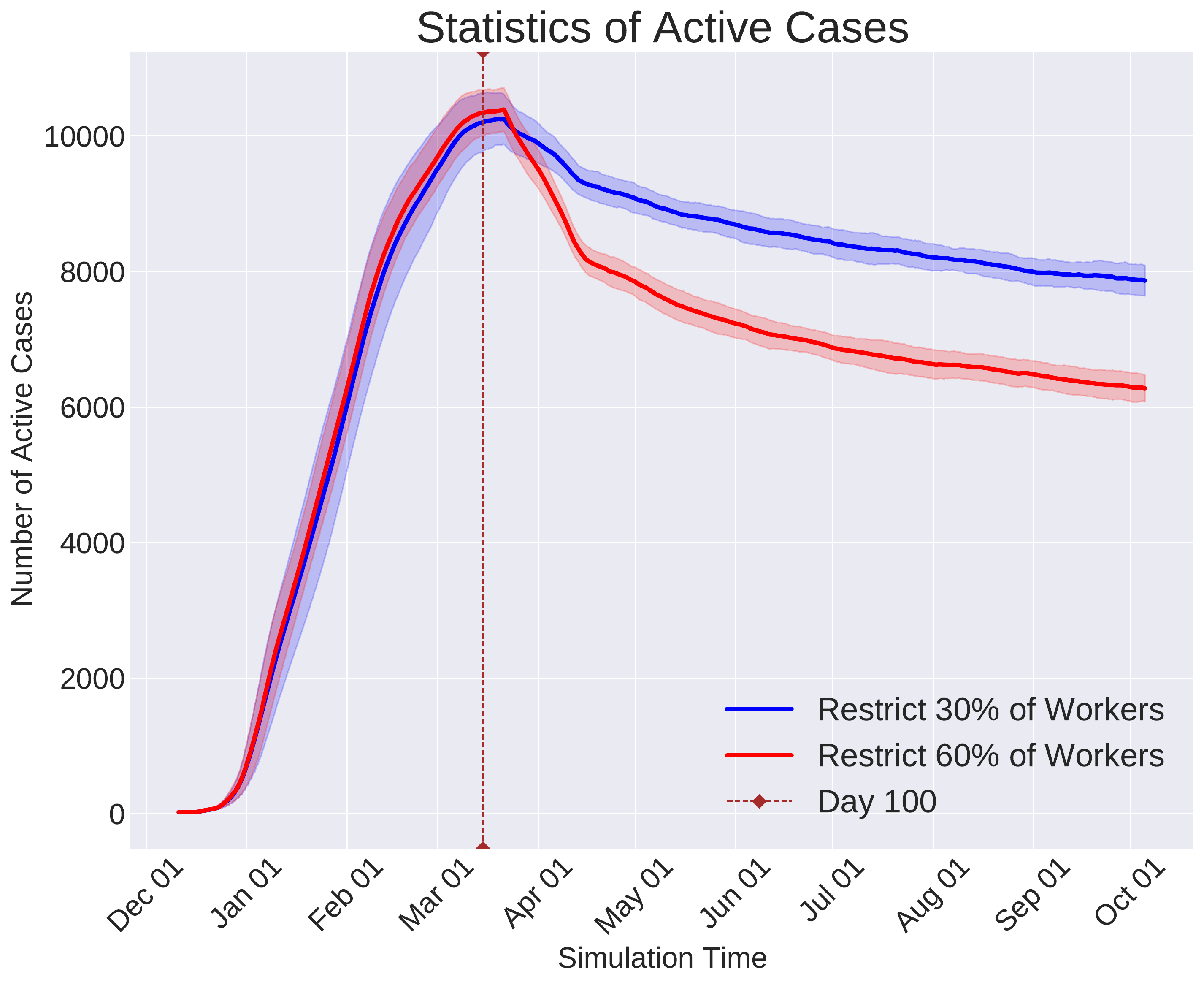}
        \caption{}
        \label{fig:quarantine_roles}
	\end{subfigure}
	\caption{a) This graph shows the effect of control policies that target specific sectors of society. Each curve corresponds to shutting down a different place. Group A includes all the workplaces of any size. Group B consists of gyms, restaurants, and cinemas. Group C includes more public places such as malls and public transportation. b) This graph shows the effectiveness of quarantining specific roles (subcommunities) in society. The curves show the spread of the infection when different ratios of the workers of any kind are quarantined. The effect is expectedly significant because the workers spend so much time in their workplaces every day, and many individuals often visit a workplace during working hours that makes it a suitable place for spreading the infection.} 
	\label{fig:quarantine_society_sector_quarantine_roles}
\end{figure}

\paragraph{Quarantining specific sectors. } This control measure restricts specific communities or sub-communities, such as gyms, public transportation, and schools. The result of shutting down some communities is shown in~\Cref{fig:quarantine_society_sectors}. We study the effect of closing gyms, restaurants, cinemas, malls, and public transportation. As a result, we observe that the most effective decision is to impose a general lock-down on workspaces. It can be seen that closing the public transportation and shopping malls has a negligible impact on flattening the epidemic curve. This seemingly counterintuitive observation is justified as people will end up in close contact with the infected individuals at their destinations regardless of how they get there. Moreover, due to the size of the considered city, the passengers spend a short period in public transport facilities that decreases the chance of getting infected.

\paragraph{Restricting specific roles. }This class of control measures, also known as working from home (WFH) policies, concerns restricting the physical presence of the employees of specified jobs whose physical presence is not absolutely necessary. It was described in~\Cref{sec:population_generation} that, in \ours and inspired by the real-world population structure, each community (e.g., schools) can have multiple roles (e.g., teachers, students, staffs) with their own special daily schedule. This control measure is especially effective and less costly because workspaces are critical hubs in the spread of the disease, and many jobs can be performed remotely thanks to the developed online communications in many areas. The results in~\Cref{fig:quarantine_roles} show a decline in the active cases a couple of days after the restrictions are enforced. It is observed that the ratio of the isolated employee plays an essential role in the outcome. In our experimented setting, a restriction that involves only $30\%$ of the working force seems ineffective compared to a situation where $60\%$ of the employees are working from home.

\subsubsection{Closed-loop policies} 
\label{section:automated_restrictions}
The experimented policies in the previous sections did not automatically react to the changes in the epidemic condition in the population. A more innovative policy should be able to adjust its commands when the conditions change. \ours is equipped with special objects that constantly monitor the population and fire a trigger signal when a pre-specified condition is met. As discussed in~\Cref{sec:condition}, the trigger signal of condition objects can be fed to either an observer object to record the statistics or a command object to issue a new restrictive rule or relieve the existing ones based on the current state of the epidemic. This closes the loop between command and observation and renders a closed-loop policy. To illustrate closed-loop policies, a simple controller is designed and is triggered when a condition is satisfied. Here, we define the condition as the moment when a ratio of two statistics from the population surpasses a specified threshold. These experiments also illustrate the substantial flexibility of the simulator to assess unlimited scenarios for epidemic control.

\begin{figure}[!t]
    \begin{subfigure}{0.5\columnwidth}
    	\centering
        \includegraphics[width=1\columnwidth]{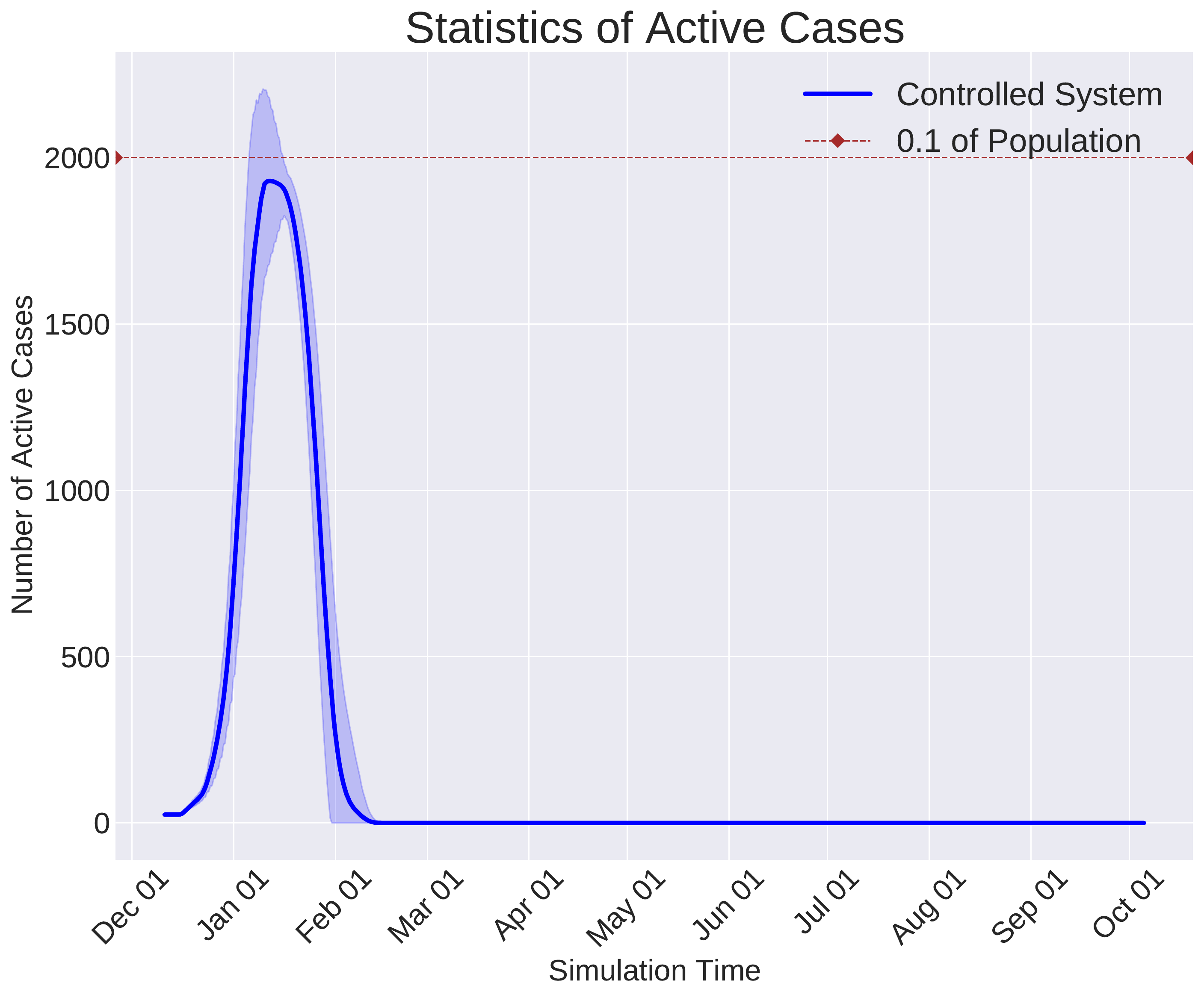}
        \caption{ }
        \label{fig:cutoff_control}
	\end{subfigure}
	\hfill
    \begin{subfigure}{0.5\columnwidth}
	    \centering
        \includegraphics[width=1\columnwidth]{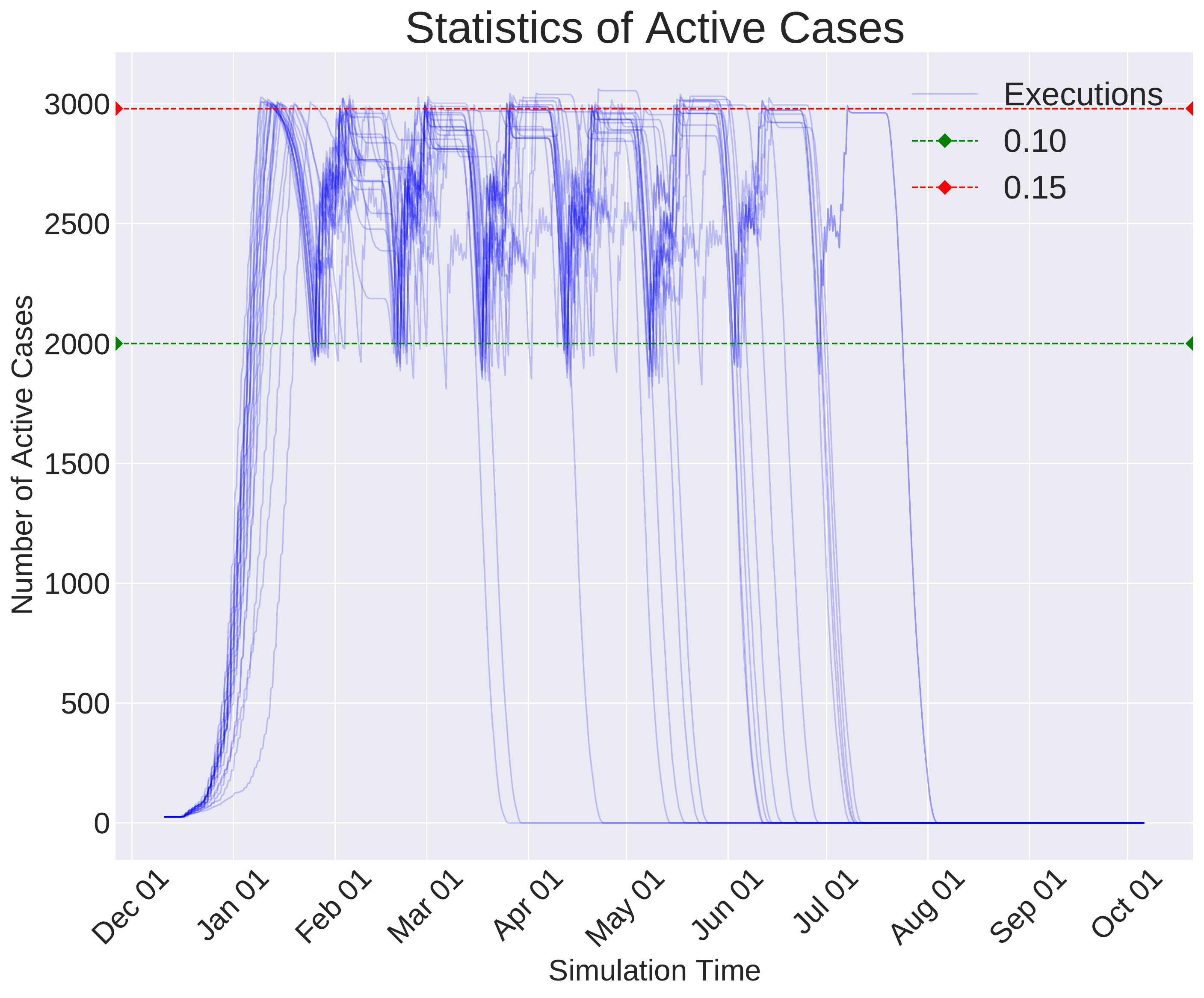}
        \caption{ }
        \label{fig:bang_bang_controller}
	\end{subfigure}
	\caption{a) In this experiment, a command is set to quarantine the infected individuals in the population when $10\%$ of the whole population is infected (a ratio condition triggers the control command). b) In this experiment, infected individuals are quarantined when more than 15\% of the population are infected, and the quarantine is lifted when the ratio of infected individuals drops below 10\%. In the terminology of control theory, this strategy is known as a~\emph{bang-bang} controller.}

	\label{fig:cutoff_control_bang_bang_controller}
\end{figure}

\paragraph{Simple cut-off mechanism.} Viewing the entire population as a dynamic system, this policy acts as a~\emph{step} controller~\cite{vidyasagar2002nonlinear} that gets activated when more than $10\%$ of the population are infected (see~\Cref{fig:cutoff_control}). The ratio threshold implies the tolerance of the policymakers and may be imposed by economic and societal factors.

\paragraph{Bang-bang controller.} This policy, called \emph{bang-bang} controller, switches between two defined states: enforce and release the quarantine. As depicted in \Cref{fig:bang_bang_controller}, this strategy keeps the ratio of active cases to the population size between 0.1 and 0.15 until the disease is no longer capable of spreading, i.e., herd immunity is reached. This scenario is specifically important since it reduces the burden on the healthcare systems by keeping the active cases below the capacity of the hospitals and, at the same time, controls the financial burden of a long-term comprehensive lockdown. Multiple executions are plotted in~\Cref{fig:bang_bang_controller} whose oscillations between two thresholds depict the actions of the bang-bang controller. It can be seen that based on the initial condition of the simulation and the initially infected individuals, some curves start dropping earlier than others. However, all dropping curves share the same slope from the point when herd immunity occurs and no new individual gets infected.

\subsubsection{Finding an optimal policy} 
\label{section:finding_an_optimal_policy}
As mentioned earlier in~\Cref{sec:introduction}, in addition to predicting the course of an epidemic under different individual-level control policies, the ultimate goal of \ours is to discover smart and detailed policies that might be hard for humans experts to find. The purpose of this experiment is to showcase this feature and find the most effective policy that minimizes the negative impacts of an epidemic of a specified disease on a specified population. In the real world, the outcome of the employed control measures in previous epidemics combined with the knowledge of the epidemiologists is used to devise control strategies to contain the spread of a novel infection. However, the complexity of the problem grows so quickly that the predicted outcome becomes unreliable even if there is a slight difference between the current and the previously experienced conditions. \ours as a detailed and high-performance simulator gives the possibility to test many proposed control strategies quickly and accurately. To find the best control measure, we offer an automatic method that cleverly searches in the space of fine-grained policies to approach the one that performs best in the specified population.

The experiment follows the following steps.

\begin{enumerate}
    \item Constructing the structure of the population of interest: We employ the structure as the previous experiments with the exception that the population size is one-tenth. Since the population structure, i.e., communities and family patterns, is unchanged, the results are still practical.
    
    \item Defining a cost function: We pick the maximum height of the curve of the active case as the cost function. Then, the goal will be finding the control measure during the entire course of the epidemic such that the number of active cases never gets so large at any time. This objective function assures that the health system will not saturate.
    
    \item Determining the optimization parameters: Every optimization is done concerning a set of parameters. As we search in the space of control measures, the control policy needs to be encoded in a few numerical parameters. In our experiment, the control measure consists of restricting students, workers, and customers. The ratio of the restricted fraction of each community is a controllable variable denoted by $\alpha, \beta$, and $\gamma$, respectively, for instance, in the case where $\alpha = 0.8$, only 20\% of students are permitted to attend the schools physically. Hence, searching for an optimal policy amounts to finding the best values for these parameters.
    
    \item Defining the economic constraints: Searching for the optimal control measure is essentially an optimal control problem where the control action always comes with some cost. Moreover, in order to avoid picking a trivial solution, we have to introduce a set of constraints on the restriction ratios. For instance, setting all restriction ratios to 1 is clearly the most effective and yet, the most expensive solution. We introduce some constraints to eliminate the trivial solutions. To do so, the sum of the restricted proportion of the roles is constrained as $\alpha + \beta + \gamma = 1.4.$. Moreover, the ratio of each role that can be isolated is also upper bounded as $0 < \alpha < 0.7$, $0 < \beta  < 0.7$, and $0 < \gamma  < 1$, implying that some roles cannot be completely remote. In the absence of these constraints, the trivial solution would be isolating the $100\%$ ratio of all three considered roles.
    
    \item Selecting an optimization algorithm: After defining the above-mentioned components of the problem, various non-gradient-based methods can be employed to carry out the optimization. It is clear that gradient-based methods cannot be used here because the gradients need to be back-propagated through the entire simulator that is not differentiable. Viewing the population and disease as the environment and control commands as the policies, the discovery of the optimal epidemic control policy can be seen as a reinforcement learning problem. We postpone further elaboration on this application of \ours to future work. Here, a simple probabilistic search method (Bayesian optimization) is employed that can be regarded as a special subset of reinforcement learning algorithms. We use Hyperopt \cite{bergstra2011algorithms} package to find the optimal set of restriction ratios based on the predetermined criteria. Hyperopt exploits the Tree of Parzen Estimator (TPE) algorithm, explained in \cite{bergstra2011algorithms}, to find the set of parameters corresponding to the most significant expected improvement at each iteration. 
\end{enumerate}

\begin{figure}[!t]
    \begin{subfigure}{0.5\columnwidth}
	    \centering
        \includegraphics[width=1\columnwidth]{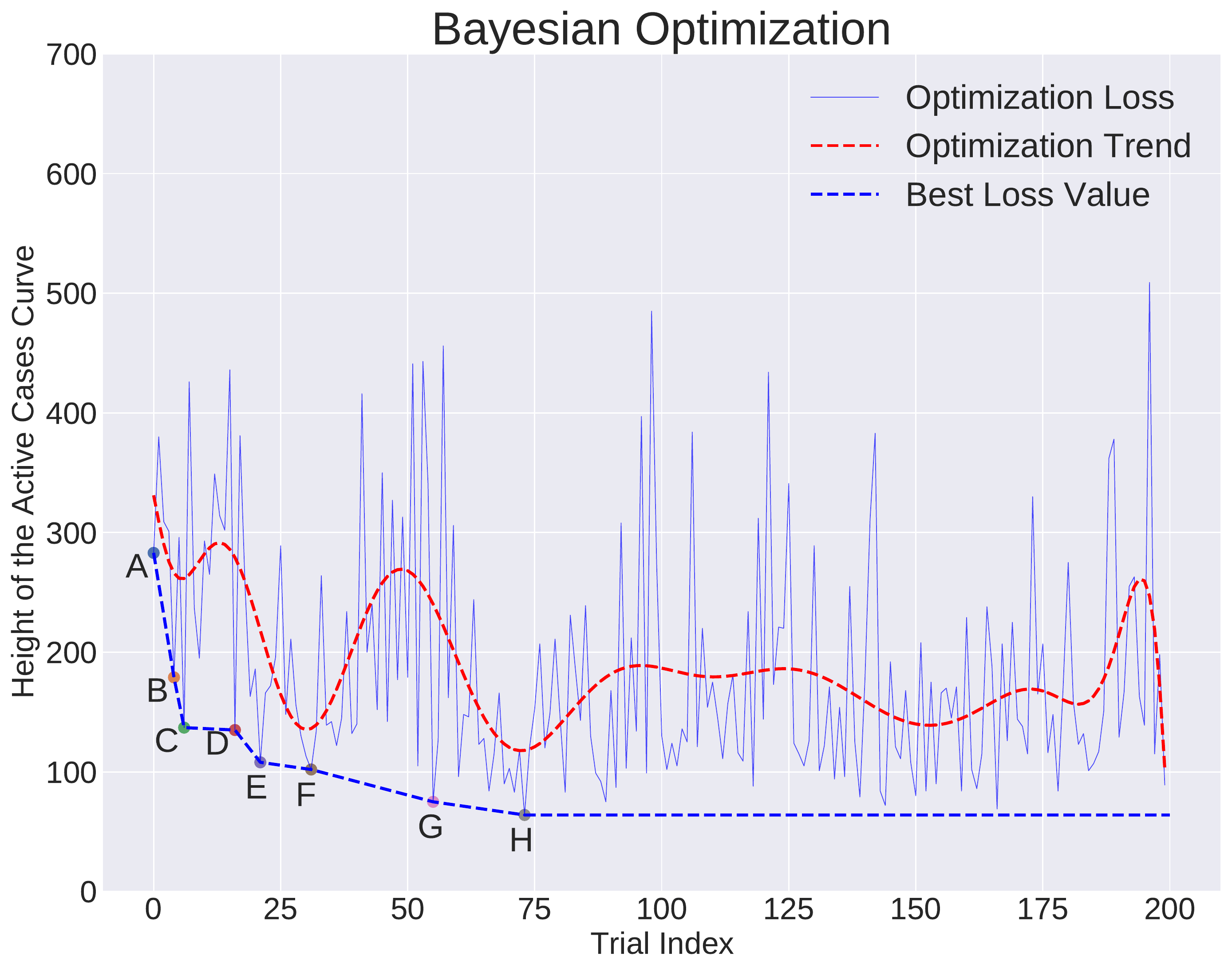}
        \caption{ }
        \label{fig:optimization_trials}
	\end{subfigure}
	\hfill
    \begin{subfigure}{0.5\columnwidth}
	    \centering
        \includegraphics[width=1\columnwidth]{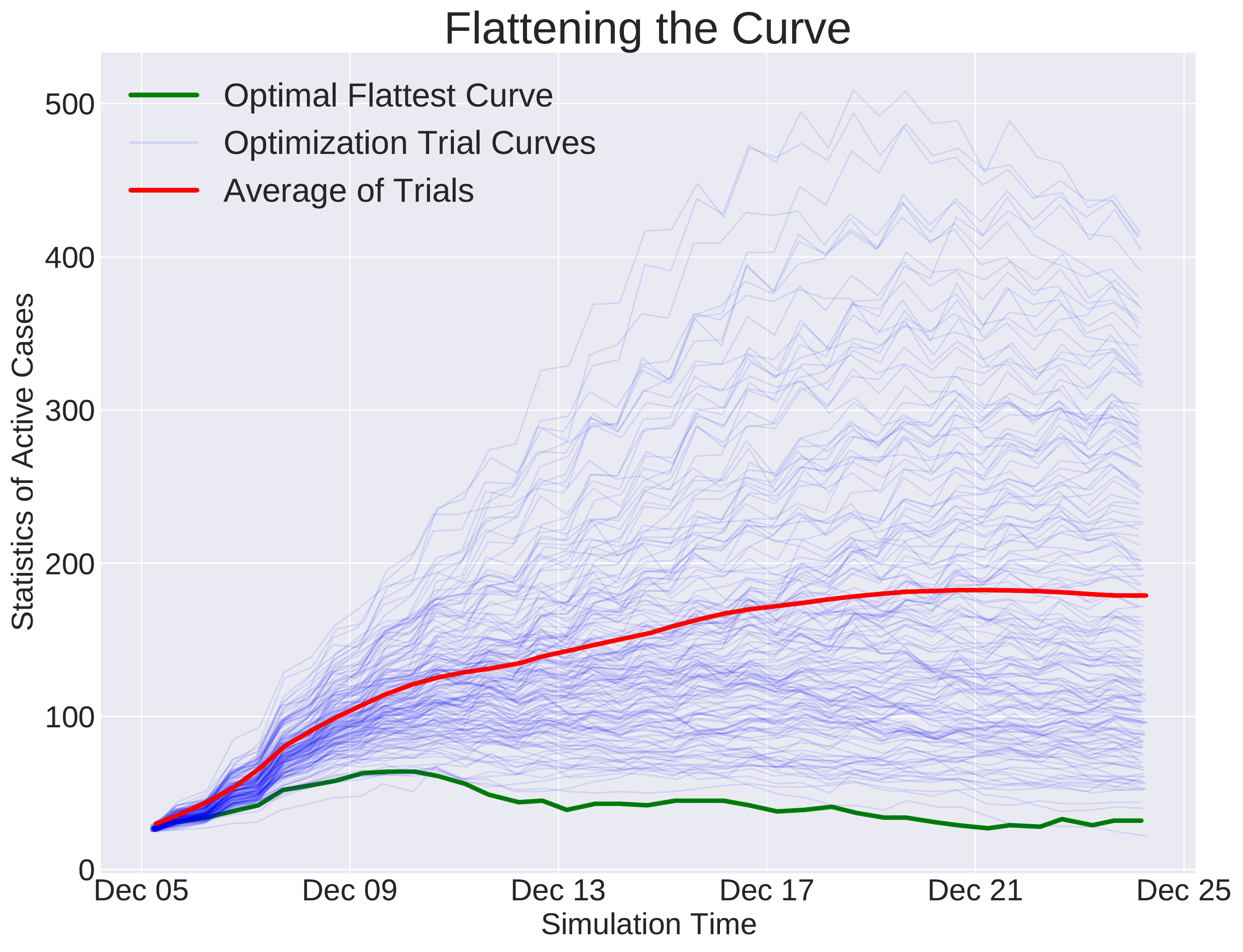}
        \caption{ }
        \label{fig:optimization_curves}
	\end{subfigure}
	\caption{a) The agent aims to minimize the loss function defined as the peak of the active cases. The optimization variables are the ratio of three roles that must be quarantined, and the ratios are constrained to be bounded from above and sum up to a constant value. The upper bound constraints are placed to take into account the cost of shutting down the economy and the trivial solution that is quarantining all individuals. The graph shows the result for 200 trials. The blue dashed line is the lower envelope of the cost produced by the discovered solution at every trial. Each point from A to H corresponds to the minimum cost up until that trial. The discovered policy associated with each of these points can be seen in~\Cref{tab:optimization_summary}. b) The curves that show the number of active cases versus time for each round of the optimization are plotted in this figure. These are actually the curves we need to flatten to protect the healthcare system against overloading. It can be seen that the discovered strategy with the least cost corresponds to the flattest curve. (The population size is reduced by a scale of 10 to boost the computation time.)}
	\label{fig:optimized_policy_procedure}
\end{figure}

\begin{table}[!t]
\centering
\caption{A summary of the discovered control strategies during the rounds of the optimization process. The first column is the index of the discovered quarantine strategy. Each strategy consists of three ratios that show the portion of the size of each group of \{students, workers, customers\} to put under quarantine. The second column indicates the iteration at which the associated strategy is found, and the rightmost column shows the value of the cost function for that strategy. This table only incorporates the iterations at which the agent improves the strategy. It can be seen that, at earlier rounds, the agent picks a large ratio for students, and only in later rounds, it realizes the critical effect of quarantining workers.}
\label{tab:optimization_summary}
\begin{tabular}{@{}cccccc@{}}
\toprule

Optimization State & Iteration & Students & Workers & Customers & Max Infected People \\ \hline
A                           & 0                  & 0.4               & 0.3              & 0.7                & 283                          \\ \hline
B                           & 4                  & 0.6               & 0.3              & 0.5                & 179                          \\ \hline
C                           & 6                  & 0.6               & 0.2              & 0.6                & 137                          \\ \hline
D                           & 16                 & 0.5               & 0.6              & 0.3                & 135                          \\ \hline
E                           & 21                 & 0.6               & 0.4              & 0.4                & 108                          \\ \hline
F                           & 31                 & 0.7               & 0.1              & 0.6                & 102                          \\ \hline
G                           & 55                 & 0.7               & 0.5              & 0.2                & 75                           \\ \hline \rowcolor{LightGreen}
H                           & 73                 & 0.6               & 0.6              & 0.2                & 64                          
\end{tabular}
\end{table}

\paragraph{Discussion on optimization results.} The course of the optimization process is shown in \Cref{fig:optimized_policy_procedure}. The found optimal restriction rule is to enforce 60\% of students and workers to stay home. The restriction is more severe for the customers of unnecessary activities where only 20\% of their population are allowed to be physically present in their communities. The found policy is aligned with our experience of the real-world scenarios, where schools and workspaces have the highest risk for spreading the infection since people are in close contact for a relatively long period per day. More importantly, based on the constraints mentioned in item 4 of~\Cref{section:finding_an_optimal_policy}, maximally 70\% of students and workers can be restricted while this ratio is unbounded for customers. As seen in~\Cref{tab:optimization_summary}, it turns out that, although workers and students are considered the groups with the highest spreading potential, imposing the maximum possible restriction on these groups is not the most effective policy. Instead, and in light of the economic constraint $\alpha+\beta+\gamma=1.4$, the algorithm learns to restrict 60\% of the population of students and workers and leave the other 20\% of the restriction budget for customers. This solution is relatively counter-intuitive, which emphasizes the benefit of \ours to find control strategies that might be difficult to find by human experts.

\section{Conclusion}
\label{sec:conclusion}
We have introduced \ours, an agent-based individual-level simulation software built upon sophisticated statistical models capable of high-granularity simulation of an epidemic disease in a structured population. The modularity and hierarchical structure allow the simulator to quickly adapt to various sizes of the population such as cities, countries, continents, or even worldwide. Thanks to several algorithmic and implementational novelties such as multiresolution timelines, \ours can be used on machines with a wide range of computational resources. The control and monitoring modules are designed to facilitate implementing real-world inspired testing and quarantining at various scales from subsets of the population to every individual. These features altogether make \ours a full-fledged environment to search for the most effective policy that controls the spread of the epidemic with minimum economic side-effects. As the next step, we explore general-purpose reinforcement learning algorithms, especially those that combine an effective representation learning with Monte Carlo Tree Search, such as AlphaZero, to learn the policies that might be impossible for human experts to find due to the immense complexity of the problem.



\bibliography{main} 

\begin{thebibliography}{10}

\bibitem{specktor2020coronavirus}
Brandon Specktor.
\newblock Coronavirus: What is' flattening the curve,'and will it work.
\newblock {\em Live Science}, 2020.

\bibitem{ferguson2020report}
Neil Ferguson, Daniel Laydon, Gemma Nedjati~Gilani, Natsuko Imai, Kylie
  Ainslie, Marc Baguelin, Sangeeta Bhatia, Adhiratha Boonyasiri, ZULMA
  Cucunuba~Perez, Gina Cuomo-Dannenburg, et~al.
\newblock Report 9: Impact of non-pharmaceutical interventions (npis) to reduce
  covid19 mortality and healthcare demand.
\newblock 2020.

\bibitem{tseng2012immunization}
Chien-Te Tseng, Elena Sbrana, Naoko Iwata-Yoshikawa, Patrick~C Newman, Tania
  Garron, Robert~L Atmar, Clarence~J Peters, and Robert~B Couch.
\newblock Immunization with sars coronavirus vaccines leads to pulmonary
  immunopathology on challenge with the sars virus.
\newblock {\em PloS one}, 7(4), 2012.

\bibitem{chinazzi2020effect}
Matteo Chinazzi, Jessica~T Davis, Marco Ajelli, Corrado Gioannini, Maria
  Litvinova, Stefano Merler, Ana~Pastore y~Piontti, Kunpeng Mu, Luca Rossi,
  Kaiyuan Sun, et~al.
\newblock The effect of travel restrictions on the spread of the 2019 novel
  coronavirus (covid-19) outbreak.
\newblock {\em Science}, 368(6489):395--400, 2020.

\bibitem{viner2020school}
Russell~M Viner, Simon~J Russell, Helen Croker, Jessica Packer, Joseph Ward,
  Claire Stansfield, Oliver Mytton, Chris Bonell, and Robert Booy.
\newblock School closure and management practices during coronavirus outbreaks
  including covid-19: a rapid systematic review.
\newblock {\em The Lancet Child \& Adolescent Health}, 2020.

\bibitem{leung2020respiratory}
Nancy~HL Leung, Daniel~KW Chu, Eunice~YC Shiu, Kwok-Hung Chan, James~J
  McDevitt, Benien~JP Hau, Hui-Ling Yen, Yuguo Li, Dennis~KM Ip, JS~Malik
  Peiris, et~al.
\newblock Respiratory virus shedding in exhaled breath and efficacy of face
  masks.
\newblock {\em Nature medicine}, 26(5):676--680, 2020.

\bibitem{wang2020four}
Tianbing Wang, Yanqiu Wu, Johnson Yiu-Nam Lau, Yingqi Yu, Liyu Liu, Jing Li,
  Kang Zhang, Weiwei Tong, and Baoguo Jiang.
\newblock A four-compartment model for the covid-19 infection-implications on
  infection kinetics, control measures and lockdown exit strategies.
\newblock {\em Precision Clinical Medicine}, 2020.

\bibitem{silver2017mastering}
David Silver, Thomas Hubert, Julian Schrittwieser, Ioannis Antonoglou, Matthew
  Lai, Arthur Guez, Marc Lanctot, Laurent Sifre, Dharshan Kumaran, Thore
  Graepel, et~al.
\newblock Mastering chess and shogi by self-play with a general reinforcement
  learning algorithm.
\newblock {\em arXiv preprint arXiv:1712.01815}, 2017.

\bibitem{block2020social}
Per Block, Marion Hoffman, Isabel~J Raabe, Jennifer~Beam Dowd, Charles Rahal,
  Ridhi Kashyap, and Melinda~C Mills.
\newblock Social network-based distancing strategies to flatten the covid-19
  curve in a post-lockdown world.
\newblock {\em Nature Human Behaviour}, pages 1--9, 2020.

\bibitem{prem2020effect}
Kiesha Prem, Yang Liu, Timothy~W Russell, Adam~J Kucharski, Rosalind~M Eggo,
  Nicholas Davies, Stefan Flasche, Samuel Clifford, Carl~AB Pearson, James~D
  Munday, et~al.
\newblock The effect of control strategies to reduce social mixing on outcomes
  of the covid-19 epidemic in wuhan, china: a modelling study.
\newblock {\em The Lancet Public Health}, 2020.

\bibitem{hanley2006object}
Brian Hanley.
\newblock An object simulation model for modeling hypothetical disease
  epidemics--epiflex.
\newblock {\em Theoretical Biology and Medical Modelling}, 3(1):32, 2006.

\bibitem{mniszewski2008episims}
Susan~M Mniszewski, Sara~Y Del~Valle, Phillip~D Stroud, Jane~M Riese, and
  Stephen~J Sydoriak.
\newblock Episims simulation of a multi-component strategy for pandemic
  influenza.
\newblock In {\em Proceedings of the 2008 Spring simulation multiconference},
  pages 556--563, 2008.

\bibitem{stroud2007spatial}
Phillip Stroud, Sara Del~Valle, Stephen Sydoriak, Jane Riese, and Susan
  Mniszewski.
\newblock Spatial dynamics of pandemic influenza in a massive artificial
  society.
\newblock {\em Journal of Artificial Societies and Social Simulation}, 10(4):9,
  2007.

\bibitem{jenness2018epimodel}
Samuel~M Jenness, Steven~M Goodreau, and Martina Morris.
\newblock Epimodel: an r package for mathematical modeling of infectious
  disease over networks.
\newblock {\em Journal of statistical software}, 84, 2018.

\bibitem{hoertel2020stochastic}
Nicolas Hoertel, Martin Blachier, Carlos Blanco, Mark Olfson, Marc Massetti,
  Marina~S{\'a}nchez Rico, Fr{\'e}d{\'e}ric Limosin, and Henri Leleu.
\newblock A stochastic agent-based model of the sars-cov-2 epidemic in france.
\newblock {\em Nature Medicine}, 26(9):1417--1421, 2020.

\bibitem{weissman2020locally}
Gary~E Weissman, Andrew Crane-Droesch, Corey Chivers, ThaiBinh Luong, Asaf
  Hanish, Michael~Z Levy, Jason Lubken, Michael Becker, Michael~E Draugelis,
  George~L Anesi, et~al.
\newblock Locally informed simulation to predict hospital capacity needs during
  the covid-19 pandemic.
\newblock {\em Annals of internal medicine}, 2020.

\bibitem{carcione2020simulation}
Jos{\'e}~M Carcione, Juan~E Santos, Claudio Bagaini, and Jing Ba.
\newblock A simulation of a covid-19 epidemic based on a deterministic seir
  model.
\newblock {\em arXiv preprint arXiv:2004.03575}, 2020.

\bibitem{fang2020many}
Zhiming Fang, Zhongyi Huang, Xiaolian Li, Jun Zhang, Wei Lv, Lei Zhuang,
  Xingpeng Xu, and Nan Huang.
\newblock How many infections of covid-19 there will be in the" diamond
  princess"-predicted by a virus transmission model based on the simulation of
  crowd flow.
\newblock {\em arXiv preprint arXiv:2002.10616}, 2020.

\bibitem{nunez2020epidemiology}
Santiago N{\'u}{\~n}ez-Corrales and Eric Jakobsson.
\newblock The epidemiology workbench: a tool for communities to strategize in
  response to covid-19 and other infectious diseases.
\newblock {\em medRxiv}, 2020.

\bibitem{d2020restart}
Marco D'Orazio, Gabriele Bernardini, and Enrico Quagliarini.
\newblock How to restart? an agent-based simulation model towards the
  definition of strategies for covid-19" second phase" in public buildings.
\newblock {\em arXiv preprint arXiv:2004.12927}, 2020.

\bibitem{akbarpour2020socioeconomic}
Mohammad Akbarpour, Cody Cook, Aude Marzuoli, Simon Mongey, Abhishek Nagaraj,
  Matteo Saccarola, Pietro Tebaldi, Shoshana Vasserman, and Hanbin Yang.
\newblock Socioeconomic network heterogeneity and pandemic policy response.
\newblock Technical report, National Bureau of Economic Research, 2020.

\bibitem{lorch2020quantifying}
Lars Lorch, H~Kremer, W~Trouleau, S~Tsirtsis, A~Szanto, B~Sch{\"o}lkopf, and
  M~Gomez-Rodriguez.
\newblock Quantifying the effects of contact tracing, testing, and containment
  measures in the presence of infection hotspots.
\newblock 2020.

\bibitem{erdHos1960evolution}
Paul Erd{\H{o}}s and Alfr{\'e}d R{\'e}nyi.
\newblock On the evolution of random graphs.
\newblock {\em Publ. Math. Inst. Hung. Acad. Sci}, 5(1):17--60, 1960.

\bibitem{us2020census}
The United States~Census Bureau.
\newblock Families and population data, 2020.
\newblock \url{https://data.census.gov/}, (last accessed: 12.09.2020).

\bibitem{eurostat2020census}
Eurostat by~European~Commission.
\newblock Population and social conditions, 2020.
\newblock \url{https://ec.europa.eu/eurostat/web/cities/data/databas}, (last
  accessed: 12.09.2020).

\bibitem{ChinaData}
Lai~Lin Thomala.
\newblock Coronavirus covid-19 in china - statistics \& facts, 2021.
\newblock
  \url{https://www.statista.com/topics/5898/novel-coronavirus-covid-19-in-china/},
  (last accessed: 23.03.2021).

\bibitem{tay2020trinity}
Matthew~Zirui Tay, Chek~Meng Poh, Laurent R{\'e}nia, Paul~A MacAry, and Lisa~FP
  Ng.
\newblock The trinity of covid-19: immunity, inflammation and intervention.
\newblock {\em Nature Reviews Immunology}, pages 1--12, 2020.

\bibitem{shan2020epidemiological}
Shan shan Wu, Pan pan Sun, Rui ling Li, Liang Zhao, Yan li~Wang, Li~fang Jiang,
  and Jin~Bo Deng.
\newblock Epidemiological development of novel coronavirus pneumonia in china
  and its forecast.
\newblock {\em medRxiv}, 2020.

\bibitem{li2020early}
Qun Li, Xuhua Guan, Peng Wu, Xiaoye Wang, Lei Zhou, Yeqing Tong, Ruiqi Ren,
  Kathy~SM Leung, Eric~HY Lau, Jessica~Y Wong, et~al.
\newblock Early transmission dynamics in wuhan, china, of novel
  coronavirus--infected pneumonia.
\newblock {\em New England Journal of Medicine}, 2020.

\bibitem{backer2020incubation}
Jantien~A Backer, Don Klinkenberg, and Jacco Wallinga.
\newblock Incubation period of 2019 novel coronavirus (2019-ncov) infections
  among travellers from wuhan, china, 20--28 january 2020.
\newblock {\em Eurosurveillance}, 25(5):2000062, 2020.

\bibitem{hauser2020estimation}
Anthony Hauser, Michel~J Counotte, Charles~C Margossian, Garyfallos
  Konstantinoudis, Nicola Low, Christian~L Althaus, and Julien Riou.
\newblock Estimation of sars-cov-2 mortality during the early stages of an
  epidemic: A modeling study in hubei, china, and six regions in europe.
\newblock {\em PLoS medicine}, 17(7):e1003189, 2020.

\bibitem{riou2020pattern}
Julien Riou and Christian~L Althaus.
\newblock Pattern of early human-to-human transmission of wuhan 2019 novel
  coronavirus (2019-ncov), december 2019 to january 2020.
\newblock {\em Eurosurveillance}, 25(4):2000058, 2020.

\bibitem{cdc2020graphs}
The United States~Centers for Disease~Control and Prevention.
\newblock Cdc covid data tracker, 2020.
\newblock
  \url{https://covid.cdc.gov/covid-data-tracker/#trends_dailytrendscases},
  (last accessed: 12.09.2020).

\bibitem{anderson1985vaccination}
Roy~M Anderson and Robert~M May.
\newblock Vaccination and herd immunity to infectious diseases.
\newblock {\em Nature}, 318(6044):323--329, 1985.

\bibitem{duggan2021novel}
Nicole~M Duggan, Stephanie~M Ludy, Bryant~C Shannon, Andrew~T Reisner, and
  Susan~R Wilcox.
\newblock Is novel coronavirus 2019 reinfection possible? interpreting dynamic
  sars-cov-2 test results.
\newblock {\em The American Journal of Emergency Medicine}, 39:256--e1, 2021.

\bibitem{vidyasagar2002nonlinear}
Mathukumalli Vidyasagar.
\newblock {\em Nonlinear systems analysis}.
\newblock SIAM, 2002.

\bibitem{bergstra2011algorithms}
James~S Bergstra, R{\'e}mi Bardenet, Yoshua Bengio, and Bal{\'a}zs K{\'e}gl.
\newblock Algorithms for hyper-parameter optimization.
\newblock pages 2546--2554, 2011.

\end{thebibliography}
\bibliographystyle{unsrt}
\clearpage

\section*{Supplementary Materials}
\label{sec:supplementary_material}
In this section, some supplementary materials are presented to clarify the capabilities and applications of the simulator's library, introduce a step-by-step guide to run the simulator, and provide further experiments conducted with different criteria than what was already discussed in \Cref{sec:experiments}.

The information provided in the next three sections aims at guiding developers to customize and deploy the simulator. To illustrate how the simulator interface works, we divide the tutorials into three major parts, each focused on a general aspect of the simulator's operation. 

\begin{enumerate}
    \item Manual simulation: Designing a simulation from scratch.
    \item Configured simulation: Deploying a configured simulation.
    \item Sanity check: Assessment of the simulator results.
\end{enumerate}

The simulator's programming tutorials presented in the first two parts can be significantly helpful for someone starting to work with the simulator for the first time, even with a basic knowledge of the Python programming language. Furthermore, the code snippets presented in this section are already accessible on GitHub by \href{https://github.com/amehrjou/Pyfectious/tree/master/example}{\color{blue}{https://github.com/amehrjou/Pyfectious/tree/master/example}}. 


In the last section, we present complementary results of our experiments, created by a different setting and smaller duration than the one introduced earlier in \Cref{sec:experiments}.

\hypertarget{manual-simulation}{%
\section{Manual simulation}\label{manual-simulation}}

This example is the best place to understand the simulator's software interface and a
comprehensive guide to designing and executing desired simulations with \ours. Please follow each section and carefully read the instructions about customizing the simulation to fit any specific settings and requirements.

\hypertarget{import-necessary-libraries}{%
\subsection{Import the required
libraries}\label{import-necessary-libraries}}

The source libraries have to be included in
the environment to start a simulation.

\begin{lstlisting}[language=Python]
import sys, os

# All the required lib files are located in the source folder
sys.path.insert(1, os.path.join(os.pardir, 'src'))
\end{lstlisting}

\hypertarget{build-a-test-environment}{%
\subsection{Build a test environment}\label{build-a-test-environment}}

This section starts from scratch and builds up all the necessary elements of a simulation environment. Moreover, these settings can also be saved in JSON format for later use cases. In the following steps, we demonstrate the process of
building a simple simulation configuration.

Note that this process is just for understanding the fundamental software concepts; therefore, in reality, there is no need to start from scratch, and one can use the tutorial presented in \Cref{configured-simulation} to run their customized simulation.

\hypertarget{family-patterns-dictionary}{%
\subsubsection{Family patterns
dictionary}\label{family-patterns-dictionary}}

The family pattern is the first object required to build the population generator class. Below is the procedure to construct a family pattern dictionary. The family pattern dictionary resembles the general pattern of the families in the simulation.

\hypertarget{location}{%
\paragraph{Location}\label{location}}

Creating a sample location distribution can be as easy as importing the
Test module. Alternatively, customized distributions
may be developed with the help of modules in distributions.py.

\begin{lstlisting}[language=Python]
from population_generator import Test
location_distribution = Test.get_location_distribution()
\end{lstlisting}

\hypertarget{age}{%
\paragraph{Age}\label{age}}

Age distributions can be created using the distribution classes implemented in
distributions.py, like the following code snippet. Accordingly, build the age distributions
list to gather all the distributions in one place. A more advanced distribution can be developed using the interface provided in the distributions.py.

\begin{lstlisting}[language=Python]
from distributions import Truncated_Normal_Distribution

# Adults age distribution
age_distribution_2 = Truncated_Normal_Distribution({"lower_bound": 30, "upper_bound": 50, "mean": 40, "std": 5})
age_distribution_1 = Truncated_Normal_Distribution({"lower_bound": 20, "upper_bound": 40, "mean": 30, "std": 5})

# Children age distribution
age_distribution_3 = Truncated_Normal_Distribution({"lower_bound": 5, "upper_bound": 15, "mean": 10, "std": 3})
age_distribution_4 = Truncated_Normal_Distribution({"lower_bound": 0, "upper_bound": 5, "mean": 2.5, "std": 2})

age_distributions = [age_distribution_1, age_distribution_2]
\end{lstlisting}
\hypertarget{health-condition}{%
\paragraph{Health condition}\label{health-condition}}

Health condition distribution is more or less determined in the same way as age
distribution. However, a person's health condition is modeled by a number
between 0 and 1, where one means the person has no history of significant health problems. 

\begin{lstlisting}[language=Python]
# Normal distribution is used here to describe the population's health condition
health_condition_distribution_1 = Truncated_Normal_Distribution({"lower_bound": 0, "upper_bound": 1,
                                                                 "mean": 0.5, "std": 0.1})
health_condition_distribution_2 = Truncated_Normal_Distribution({"lower_bound": 0, "upper_bound": 1,
                                                                 "mean": 0.5, "std": 0.1})
# A list of all the distributions must be assembled
health_condition_distributions = [health_condition_distribution_1, health_condition_distribution_2]
\end{lstlisting}

\hypertarget{family-pattern}{%
\paragraph{Family pattern}\label{family-pattern}}

Now we have almost all the required fields to generate a family pattern.
We also need to create a gender list consisting of all the family
members, respectively. Two instances of creating a family pattern are mentioned in the following code
snippet to demonstrate the object's flexibility in modeling any patterns.

\begin{lstlisting}[language=Python]
# Size of the family
number_of_members = 2

# Create a genders list based on the number of members
genders = [0, 1]

# Build the family pattern object
from population_generator import Family_Pattern
family_pattern_1 = Family_Pattern(number_of_members,
                                    age_distributions,
                                    health_condition_distributions,
                                    genders,
                                    location_distribution)

# Family size is increased here
number_of_members = 4

# For a four member family, age and health condition distributions most be of length 4, respectively
age_distributions = [age_distribution_1, age_distribution_2, age_distribution_3, age_distribution_4]

health_condition_distributions = [health_condition_distribution_1, health_condition_distribution_2, health_condition_distribution_1, health_condition_distribution_2]

# Here we have male and female equally
genders = [0, 1, 1, 0]

# Build the family pattern object
family_pattern_2 = Family_Pattern(4, age_distributions, health_condition_distributions, genders, location_distribution)
\end{lstlisting}

\hypertarget{probability-dictionary}{%
\paragraph{Probability dictionary}\label{probability-dictionary}}

Last but not least, the job here is to build a family probability
dictionary. This structure represents the presence probability of each
family pattern in society. Naturally, the accumulative sum of the probabilities must be equal to 1.

\begin{lstlisting}[language=Python]
# note that the sum of probabilities must be 1
family_pattern_probability_dict = {family_pattern_1: 0.4, family_pattern_2: 0.6}
\end{lstlisting}

\hypertarget{community-types}{%
\subsubsection{Community types}\label{community-types}}

A community type object represents the overall structure of a specific community, e.g.,
a school, in the simulation environment. Each community type consists of a
list of sub-community types, for instance, teacher, student, etc., and a
sub-community connectivity dictionary, representing the interactions
between sub-communities as a graph. Name and location distribution are
also other parts of the structure.

\hypertarget{sub-community-types}{%
\paragraph{Sub-community types}\label{sub-community-types}}

Sub-community types represent a smaller community, generally attached to
a particular community type role, e.g., student and teachers. To build a
sub-community type, the procedure indicated in the following code
snippet must be followed.

\begin{lstlisting}[language=Python]
# Generate an age distribution for this sub-community
age_distribution = Truncated_Normal_Distribution({"lower_bound": 5, "upper_bound": 15, "mean": 10, "std": 1})

# Create a time cycle distribution, when people are present in this community
from distributions import Uniform_Whole_Week_Time_Cycle_Distribution
time_cycle_distribution = Uniform_Whole_Week_Time_Cycle_Distribution({
    "start": {
        "lower_bound": 300,
        "upper_bound": 301
    },
    "length": {
        "lower_bound": 250,
        "upper_bound": 500
    }
})

# Generate a community type role object
from population_generator import Community_Type_Role
student_type = Community_Type_Role(age_distribution,
                                    Test.get_gender_distribution(),
                                    time_cycle_distribution,
                                    True,
                                    1)

# Build the number of members distribution
from distributions import UniformSet_Distribution
number_of_members_distribution = UniformSet_Distribution({
    "probability_dict": {
        30: 1,
    }
})

# Generate the sub community type using the Role, members and a connectivity distribution
from population_generator import Sub_Community_Type
sub_community_type_1 = Sub_Community_Type(student_type,
                                          "Student",
                                          number_of_members_distribution,
                                          Test.get_connectivity_distribution(),
                                          Test.get_transmission_potential_distribution())
\end{lstlisting}

Another sub-community type is generated below.

\begin{lstlisting}[language=Python]

age_distribution = Truncated_Normal_Distribution({"lower_bound": 20, "upper_bound": 40, "mean": 30, "std": 5})
teacher_type = Community_Type_Role(age_distribution,
                                   Test.get_gender_distribution(),
                                   time_cycle_distribution,
                                   True,
                                   1)
number_of_members_distribution = UniformSet_Distribution({
    "probability_dict": {
        5: 1,
    }
})
sub_community_type_2 = Sub_Community_Type(teacher_type,
                                          "Teacher",
                                          number_of_members_distribution,
                                          Test.get_connectivity_distribution(),
                                          Test.get_transmission_potential_distribution())
\end{lstlisting}
\hypertarget{build-a-community-type}{%
\paragraph{Build a community type}\label{build-a-community-type}}
The community type object can now be created by putting together the sub-community types
and connectivity distributions. It is noteworthy that connectivity dictionary explains the level of mutual contacts between sub-communities by employing a probabilistic distribution. Here, we use a prepared distribution that has already been implemented in the Test class. 

The transmission potential explains the possibility of virus transmission between the individuals of a community. For instance, infectious diseases can spread more quickly in a closed environment such as a classroom. 

\begin{lstlisting}[language=Python]
from population_generator import Community_Type

# Build the community types list
subcommunity_types_list = [sub_community_type_1, sub_community_type_2]

# Build the connectivity dictionary for sub-communities
sub_community_connectivity_dict = {0: {1:Test.get_connectivity_distribution()},
                                       1: {0:Test.get_connectivity_distribution()}}
# Prepare a transmission potential dictionary
transmission_potential_dict =
                {0: {1:Test.get_transmission_potential_distribution()},
                 1: {0:Test.get_transmission_potential_distribution()}}
                 
# Build the community type object
school_type = Community_Type(sub_community_types=subcommunity_types_list,
                    name="School",
                    number_of_communities=4,
                    sub_community_connectivity_dict=sub_community_connectivity_dict,
                    location_distribution=Test.get_location_distribution(),
                    transmission_potential_dict=transmission_potential_dict)
\end{lstlisting}

\hypertarget{population-generator}{%
\subsubsection{Population generator}\label{population-generator}}

Having the family pattern dictionary and community types, a population
generator may be created in the following way. This class contains all
the necessary information to generate a sample population in the
simulation environment.

\begin{lstlisting}[language=Python]
from population_generator import Population_Generator

# Build the Population Generator class
population_generator = Population_Generator(population_size=500,
                    family_pattern_probability_dict=family_pattern_probability_dict,
                    community_types=[school_type, school_type])
\end{lstlisting}

Moreover, the following command examines how the population
is spread among the families and communities. More importantly, The generate\_population method is necessary in order to build an operational set of parameters later used during the simulation procedure. For large populations, it is also possible to run
this task using the python multiprocessing library by just setting
is\_parallel to True.

\begin{lstlisting}[language=Python]
people, graph, families, communities = population_generator.generate_population(is_parallel=False)
\end{lstlisting}

\hypertarget{disease-properties}{%
\subsubsection{Disease properties}\label{disease-properties}}

The disease properties class represents the major specifications related
to the spread of the target infectious disease. Any disease specifications may be
applied here, e.g., attributes related to COVID-19, MERS, and SARS behavior.

\begin{lstlisting}[language=Python]
from distributions import Uniform_Disease_Property_Distribution
from time_handle import Time

# infection rate
infection_rate_distribution = Uniform_Disease_Property_Distribution({
    "lower_bound": 0.6,
    "upper_bound": 0.8
})

# Immunity rate
immunity_distribution = Uniform_Disease_Property_Distribution({
    "lower_bound": 0.05,
    "upper_bound": 0.15
})

# Disease period
disease_period_distribution = Uniform_Disease_Property_Distribution({
    "lower_bound": Time.convert_day_to_minutes(2),
    "upper_bound": Time.convert_day_to_minutes(5)
})

# Incubation period
incubation_period_distribution = Uniform_Disease_Property_Distribution({
    "lower_bound": Time.convert_day_to_minutes(7),
    "upper_bound": Time.convert_day_to_minutes(16)
})

# Death probability
death_probability_distribution = Uniform_Disease_Property_Distribution({
    "lower_bound": 0.05,
    "upper_bound": 0.1
})

from disease_manipulator import Disease_Properties

# Creating the disease properties object
disease_properties = Disease_Properties(infection_rate_distribution=infection_rate_distribution, 
            immunity_distribution=immunity_distribution,
            disease_period_distribution=disease_period_distribution,
            death_probability_distribution=death_probability_distribution,
            incubation_period_distribution=incubation_period_distribution)
\end{lstlisting}

\hypertarget{run-the-simulation}{%
\subsection{Deploying the simulation}\label{run-the-simulation}}

Now we start working with the simulator class. The upcoming sections will illustrate the entire simulator's execution process.

\hypertarget{primary-settings}{%
\subsubsection{Primary settings}\label{primary-settings}}

The simulator starts with the population generator and disease properties
objects as base settings. Afterward, the generate\_model function steps in
to generate a simulation environment, such as people, families, and communities, as well as preparing the ground for
simulating in the following steps.

\begin{lstlisting}[language=Python]
from time_simulator import Simulator

# Instantiate the simulator object using previously prepared population generator and disease properties objects
simulator = Simulator(population_generator, disease_properties)

# Execute the model generation method
simulator.generate_model()
\end{lstlisting}

\hypertarget{end-time}{%
\paragraph{End time}\label{end-time}}

The simulation end time is crucial since it determines how long the simulation should keep going. Here we set a 60-day simulation, starting
from now.

\begin{lstlisting}[language=Python]
from datetime import datetime, timedelta
from time_handle import Time

end_time = Time(delta_time=timedelta(days=60), init_date_time=datetime.now())
\end{lstlisting}

\hypertarget{spread-period}{%
\paragraph{Spread period}\label{spread-period}}

Determining the spread period is crucial since it clarifies the simulation's granularity. To have a detailed
simulation, the user must set lower values where the spread sequence is investigated frequently. Otherwise, increasing the virus spread period causes a reduction in computational costs.

\begin{lstlisting}[language=Python]
spread_period = Time(delta_time=timedelta(hours=1))
\end{lstlisting}

\hypertarget{initial-infected-ids}{%
\paragraph{Initially infected people}\label{initial-infected-ids}}

Any pandemic must start from certain people, i.e., the initially
infected subjects. A list of id numbers represents the initially infected people.

\begin{lstlisting}[language=Python]
import random

initial_infected_ids = random.sample(range(500), 10)
\end{lstlisting}

\hypertarget{observers-list}{%
\paragraph{Observers}\label{observers-list}}

The observer is the module responsible for saving data into the
database. Using the observer, the data during the simulation can be stored
and later be used in plots, reasoning, etc. The simulator can handle a
list of observers with various trigger conditions.

\begin{lstlisting}[language=Python]
from observer import Observer
from conditions import Time_Period_Condition

# Build the observer object
observer = Observer(Time_Period_Condition(Time(delta_time=timedelta(hours=5))), True)

# Generate the observers list
observers_list = [observer]
\end{lstlisting}

\hypertarget{commands}{%
\paragraph{Commands}\label{commands}}

The command list is used to create a policy to contract the pandemic. A
simple strict command may be to quarantine all the communities. At this point, we leave the list empty to run a simple simulation.
\begin{lstlisting}[language=Python]
commands_list = []
\end{lstlisting}

\hypertarget{run-the-simulation-1}{%
\subsubsection{Simulate}\label{run-the-simulation-1}}

At this stage, the only remaining step is running the simulation. This
might take a while, depending on the population size and the total
end time. Other factors, such as the number of observers, are also
influential in determining the simulation time.

The report statistics can be varied from 0 (default) to 1 and 2 if there is a need
of reviewing more details of the simulation at the end.

\begin{lstlisting}[language=Python]
# Execute the simulation using the simulate method
simulator.simulate(end_time=end_time,
                    spread_period=spread_period,
                    initialized_infected_ids=initial_infected_ids,
                    commands=commands_list,
                    observers=observers_list,
                    report_statistics=2)
\end{lstlisting}

\hypertarget{plot-the-results}{%
\subsection{Plot the results}\label{plot-the-results}}

\begin{figure}
  \begin{subfigure}{\linewidth}
  \includegraphics[width=.5\linewidth]{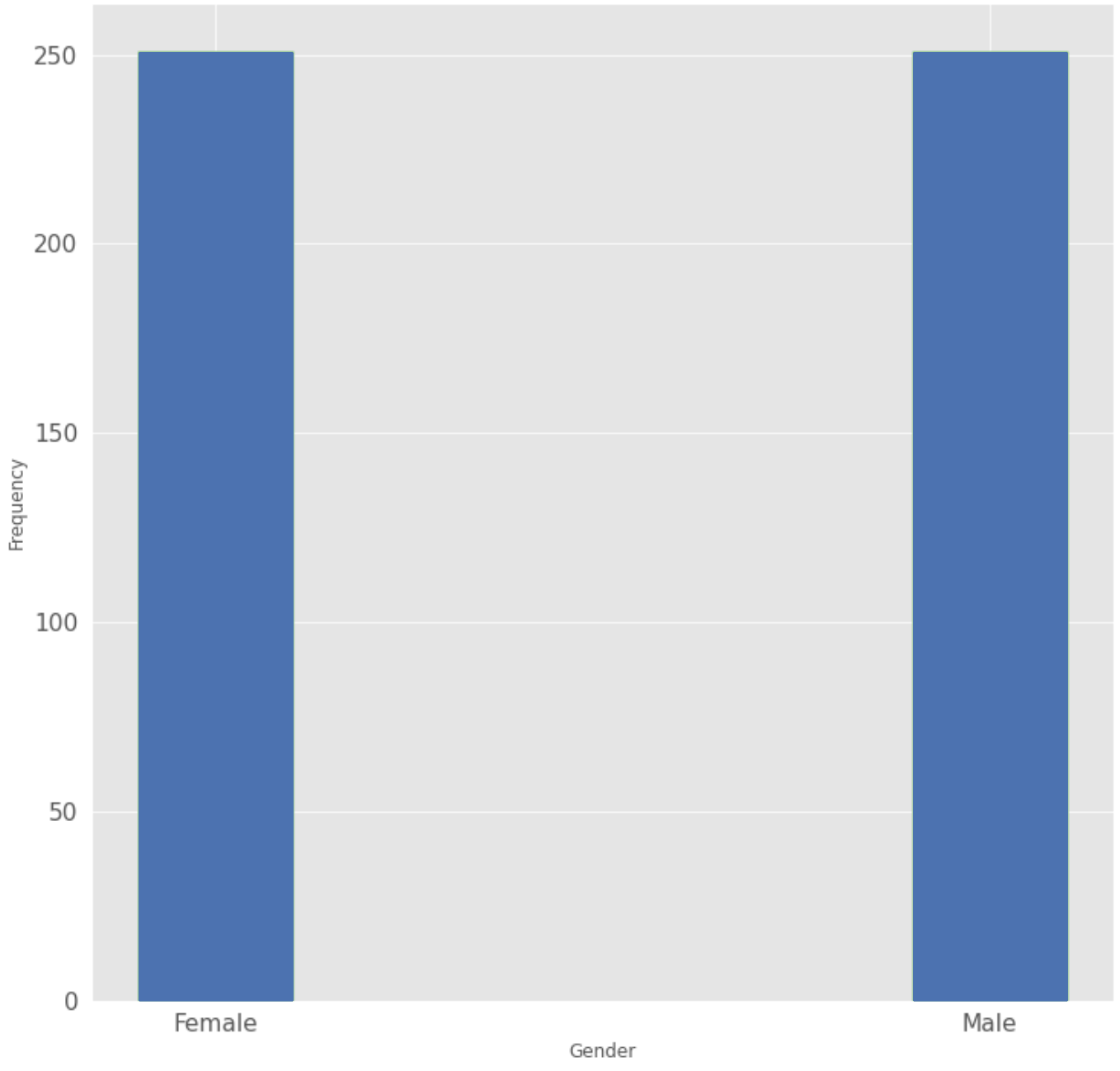}\hfill
  \includegraphics[width=.5\linewidth]{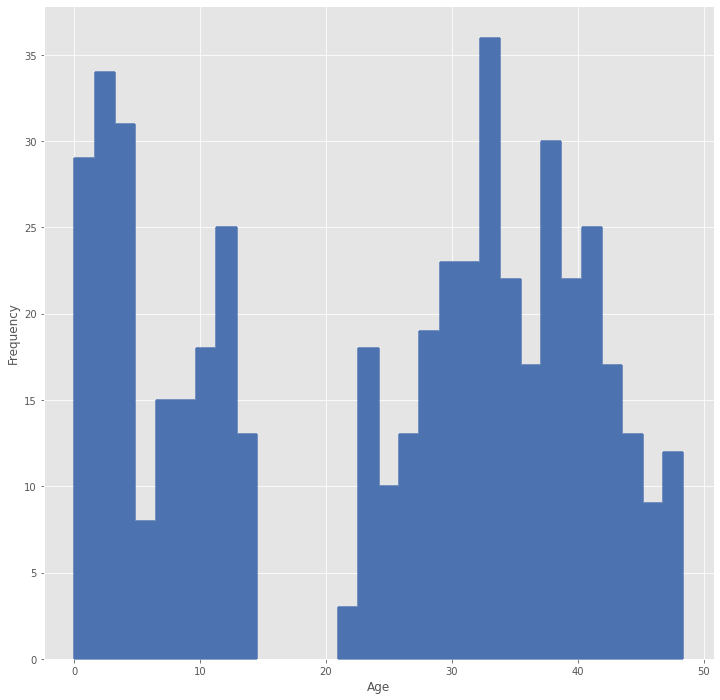}
  \caption{Gender and age distribution of the simulator's population}
  \end{subfigure}\par\medskip
  \begin{subfigure}{\linewidth}
  \includegraphics[width=.5\linewidth]{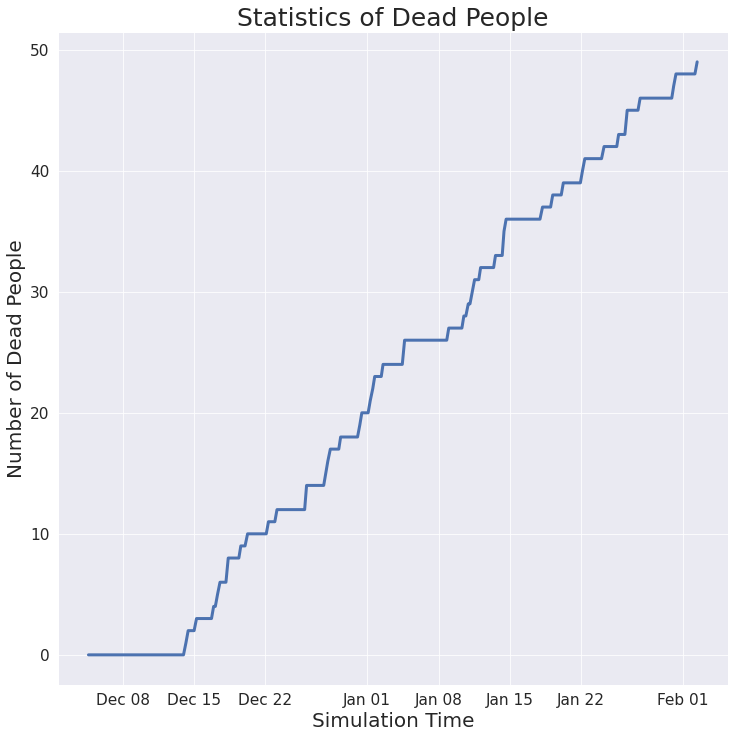}\hfill
  \includegraphics[width=.5\linewidth]{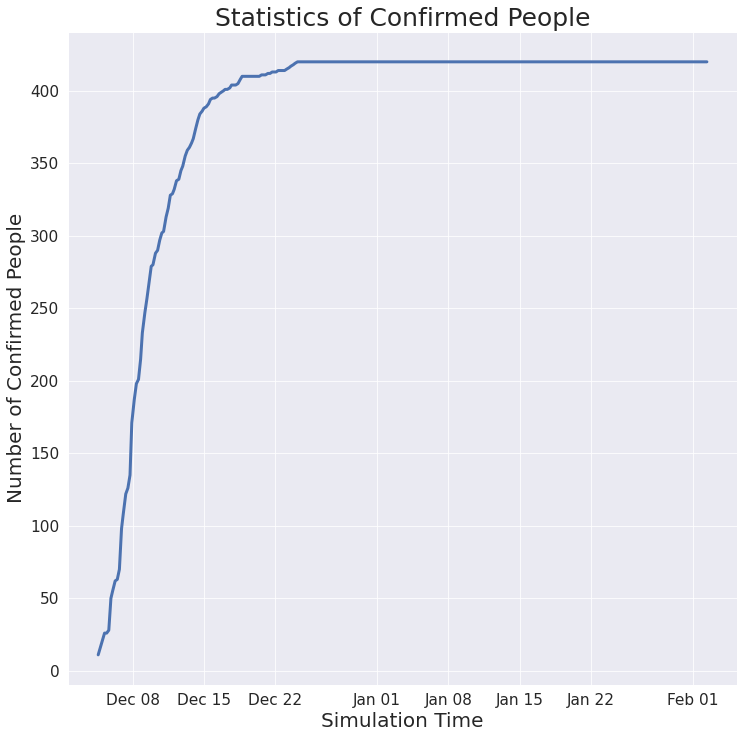}
  \caption{Statistics of dead and confirmed cases over time}
  \end{subfigure}\par\medskip
  \begin{subfigure}{\linewidth}
  \includegraphics[width=.5\linewidth]{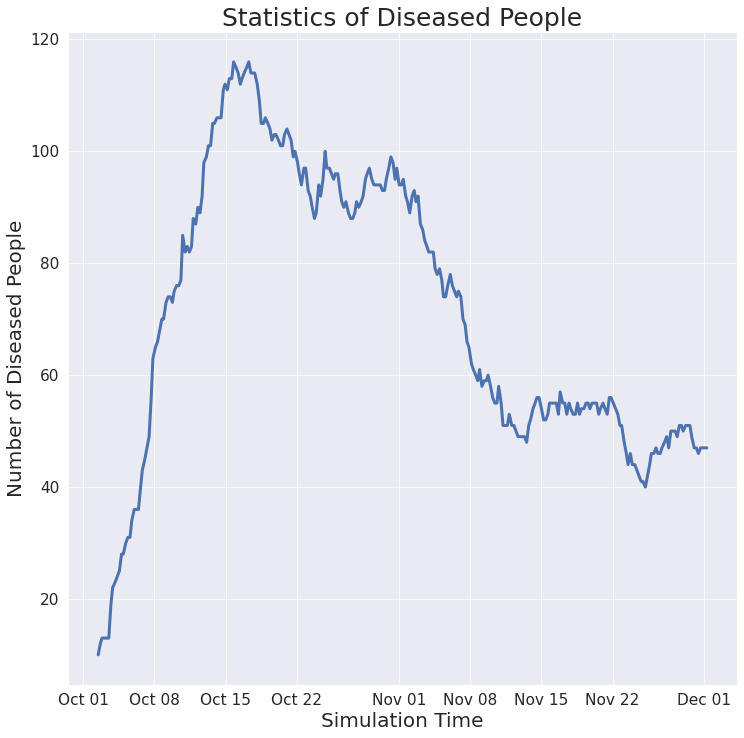}\hfill
  \includegraphics[width=.5\linewidth]{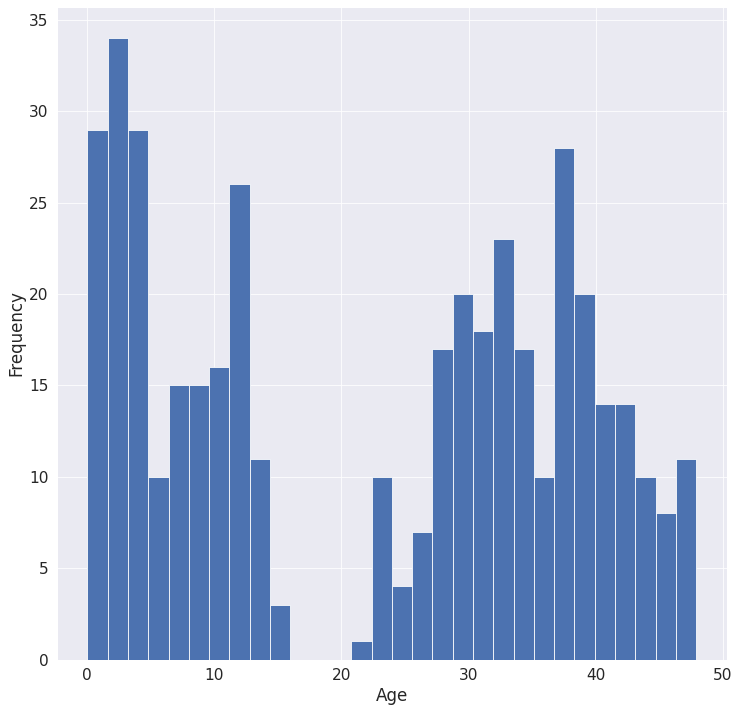}
  \caption{Statistics of active cases and frequency of confirmed cases at the end of the simulation time}
  \end{subfigure}
  \caption{The results of observer plots}
  \label{fig:plots_manual_simulation} 
\end{figure}

Here are some useful plots obtained by utilizing the observer's methods to evaluate and analyze the simulation's results. The data associated with a specific observer may be retrieved using
the simulator.database functions. However, the observer object provides
some useful plots and automatic data derivation without any need to be directly connected to the simulation database. The plots associated with the following code snippet are presented in \Cref{fig:plots_manual_simulation}.
\begin{lstlisting}[language=Python]
observer.plot_initial_bar_gender()
observer.plot_initial_hist_age()
observer.plot_initial_hist_health_condition()

from utils import Health_Condition, Infection_Status
observer.plot_disease_statistics_during_time(Health_Condition.DEAD)
observer.plot_disease_statistics_during_time(Health_Condition.HAS_BEEN_INFECTED)
observer.plot_final_hist_age(Health_Condition.HAS_BEEN_INFECTED)
observer.plot_disease_statistics_during_time(Health_Condition.IS_INFECTED)
\end{lstlisting}

\hypertarget{add-a-condition}{%
\subsection{Add a condition}\label{add-a-condition}}

A new condition can be designed and developed using the following structure and by
inheriting the Condition class. The is\_satisfied function determines
whether the condition is satisfied and is\_able\_to\_be\_removed
determines whether the condition is useless from now on or not.

\begin{lstlisting}[language=Python]
from conditions import Condition

# Build your own customized condition
class More_Than_X_Deaths_Condition(Condition):

    def __init__(self, x):
        super().__init__()
        self.x = x
        self.satisfied = False

    def is_satisfied(self, simulator: Simulator, end_time: Time):
        temp = self.x
        for person in simulator.people:
            if not person.is_alive:
                temp -= 1
        if temp <= 0:
            self.satisfied = True
            return [simulator.clock]
        return []

    def is_able_to_be_removed(self):
        return self.satisfied
\end{lstlisting}

The newly generated condition may be used in both observers and
commands. Here is an example of how to use the condition in an observer.

\begin{lstlisting}[language=Python]
from observer import Observer
observer = Observer(More_Than_X_Deaths_Condition(10), True)

# Generate the observers' list
observers_list = [observer]

simulator.simulate(end_time=end_time,
                    spread_period=spread_period,
                    initialized_infected_ids=initial_infected_ids,
                    commands=commands_list,
                    observers=observers_list)
\end{lstlisting}

\hypertarget{add-a-command}{%
\subsection{Add a command}\label{add-a-command}}

Similarly, as adding a condition, a developer can also add a command using
the Command class. New commands should follow the base class structure
and functions in order to work correctly. The action
that a particular command is supposed to take can be specified in the take\_action method.

\begin{lstlisting}[language=Python]
from commands import Command

# Build your own customized command
class Quarantine_Diseased_People(Command):
    def __init__(self, condition: Condition):
        super().__init__(condition)
        self.condition = condition

    def take_action(self, simulator: Simulator, end_time: Time):
        if self.condition.is_satisfied(simulator, end_time):
            for person in simulator.people:
                if person.infection_status is Infection_Status.CONTAGIOUS or \
                        person.infection_status is Infection_Status.INCUBATION:
                    person.quarantine()

    # This method is necessary due to automation purposes
    def to_json(self):
        return dict(name=self.__class__.__name__,
                    condition=self.condition)
\end{lstlisting}







\hypertarget{save-the-main-objects-as-json-configuration-files}{%
\subsection{Save the main objects as JSON configuration
files}\label{save-the-main-objects-as-json-configuration-files}}

The objects can be saved as JSON files. These files may also be employed
later to avoid preparations.

\begin{lstlisting}[language=Python]
from json_handle import Parser

# Build parser object
json_parser = Parser()

# Save population generator
json_parser.build_json(population_generator)
json_parser.save_json()

# Save disease properties
json_parser.build_json(disease_properties)
json_parser.save_json()

# Save population generator
json_parser.build_json(simulator)
json_parser.save_json()
\end{lstlisting}

\hypertarget{configured-simulation}{%
\section{Configured Simulation}\label{configured-simulation}}

This section is dedicated to explaining the automated setup procedure of the simulator. As opposed to the previous
section, the data required by the simulator is obtained by
prepared JSON configuration files.

\hypertarget{create-a-setting-folder}{%
\subsection{Create a setting folder}\label{create-a-setting-folder}}

A folder called 'data', located in the project's main directory, consists of
four major parts: json, figure, pickle, and sql. The sql folder, as appears of
its name, is related to the simulation database, and the database files are stored there. The json folder is used
to prepare the simulator's configuration files. By opening the json
folder, some samples are already designed and placed there under the respective folders.
These are some experimental configuration files that may be used to run a simulation.

For instance, open the folder named `test'. Under this folder, there are
three JSON files as explained below.
\begin{enumerate}
    \item Population\_Generator.json: This file consists of the information required to build an entire population generator object. 
    \item Disease\_Properties.json: This file consists of the information required to build an entire disease properties object. Parameters like the infection rate and immunity are sub-fields of this JSON file.
    \item Simulator.json: This file consists of the data required to call the simulate function in the Simulator class, including end\_time, spread\_period, commands, etc.
\end{enumerate}

\hypertarget{customized-json-files}{%
\subsubsection{Customized configuration files}\label{customized-json-files}}

To build a customized setting, the folder named `test' must be copied and
pasted as a new folder, and name it as you like. For instance, here,
we create a copy and call it `configured\_test'. Afterward, it is possible to change the JSON files' values to fit other criteria such as larger population, other diseases, and more complex population structures.

\begin{lstlisting}[language=Python]
import os
import sys
sys.path.insert(1, os.path.join(os.pardir, 'src'))

# Check if we are in 'test' folder right now
print(f'Current directory: {os.getcwd()}')

# Change directory to 'data' and then to 'json' folder
os.chdir(os.path.join(os.pardir, 'data', 'json'))
\end{lstlisting}

\begin{lstlisting}[language=Python]
# Build the configure_test folder if does not exists
try:
    os.mkdir('configured_test')
except FileExistsError:
    pass

# Determine source and destination for copy operation
destination = os.path.join(os.getcwd(), 'configured_test')
source = os.path.join(os.getcwd(), 'test')

# Copy the items in 'test' to 'configured_test'
import shutil
src_files = os.listdir(source)
for file_name in src_files:
    full_file_name = os.path.join(source, file_name)
    if os.path.isfile(full_file_name):
        shutil.copy(full_file_name, destination)
\end{lstlisting}

Now, we can try to make some changes to the JSON configuration files.
For the sake of simplicity, we change the population size to 800.
This can be done either manually, from an editor, or using a script like
the following.

\begin{lstlisting}[language=Python]
# Considering the last section, we are now in the 'json' folder, so we move to the 'data' folder again so that the configured test is accessible.
os.chdir(os.pardir)

# Import and initialize the parser
from json_handle import Parser
parser = Parser(folder_name='configured_test')

# Parse the Population_Generator.json in 'configured_test'
population_generator =  parser.parse_population_generator()
print(f'Population size before the change is: {population_generator.population_size}')

# Change the population size
population_generator.population_size = 800

# Save the new population generator as json
parser.build_json(population_generator)
parser.save_json()

# Parse the Population_Generator.json in 'configured_test'
population_generator =  parser.parse_population_generator()
print(f'Population size after the change is: {population_generator.population_size}')
\end{lstlisting}

\hypertarget{simulate-based-on-the-configured-data}{%
\subsection{Simulation}\label{simulate-based-on-the-configured-data}}

In this section, we simulate data based on the settings saved inside the
`configured\_test' folder.

\begin{lstlisting}[language=Python]
# Import and initialize the parser
from json_handle import Parser
parser = Parser(folder_name='configured_test')


# Load Simulator from JSON file
simulator = parser.parse_simulator()
simulator.generate_model()

# Check the population size after generation
print(f'Population size is: {len(simulator.people)}')
\end{lstlisting}

Additionally, the policy and simulation specifics must be obtained from
the Simulator.json as described below.

\begin{lstlisting}[language=Python]
# Load Simulator Data from JSON file
end_time, spread_period, initialized_infected_ids, commands, observers = parser.parse_simulator_data()

# Run the simulation
simulator.simulate(end_time=end_time,
                    spread_period=spread_period,
                    initialized_infected_ids=initialized_infected_ids,
                    commands=commands,
                    observers=observers)
\end{lstlisting}

We have completed the simulation, and the results may be obtained from
the database or statistics.

\begin{lstlisting}[language=Python]
from utils import Health_Condition
observers[0].plot_disease_statistics_during_time(Health_Condition.IS_INFECTED)
\end{lstlisting}

Finally, having access to a summary of the simulation is possibly either by
setting the report\_statistics option of the simulate function, or
separately calling in the statistics class.

\begin{lstlisting}[language=Python]
simulator.statistics.show_people_statistics(simulator=simulator)
\end{lstlisting}

\begin{lstlisting}
INFO - utils.py - 303 - show_people_statistics - 2020-12-04 12:58:56,722 - 
+----------------------------+-------+
|           People           | Count |
+============================+=======+
| Population Size            | 800   |
+----------------------------+-------+
| Confirmed (Active + Close) | 150   |
+----------------------------+-------+
| Total Death Cases          | 5     |
+----------------------------+-------+
| Total Recovered            | 795   |
+----------------------------+-------+
| Currently Active Cases     | 0     |
+----------------------------+-------+
\end{lstlisting}

\hypertarget{the-example-town}{%
\subsection{The example town}\label{the-example-town}}

In this section, our sample town, implemented under the 'json' folder with
50k population size, six family patterns, and major communities,
including schools and workplaces, gyms, and restaurants, is being
tested.

\hypertarget{parse-the-configuration-files}{%
\subsubsection{Parse the configuration
files}\label{parse-the-configuration-files}}

Prior to anything else, we have to parse the configuration files located
in the town folder, under data/json directory.

\begin{lstlisting}[language=Python]
# Import libs
import sys, time, os

sys.path.insert(1, os.path.join(os.pardir, 'src'))

# Import and initialize the parser
from json_handle import Parser
parser = Parser(folder_name='town')

# Load Simulator from JSON file
simulator = parser.parse_simulator()
\end{lstlisting}

\hypertarget{generate-and-save-the-model}{%
\subsubsection{Generate and save the
model}\label{generate-and-save-the-model}}

Since the model obtained by the generate\_model function in this
simulation is often large, we can utilize the simulator power to save the model
for later use by employing the simulator.save\_model method, and later
use it using the simulator.load\_model method.

\begin{lstlisting}[language=Python]
# Generate the simulation model
simulator.generate_model()

# Time the generation process
end_generate_model = time.time()

# Save the simulation model
simulator.save_model('town')

# Save the simulation model
simulator.load_model('town')
\end{lstlisting}

\hypertarget{simulate-the-town}{%
\subsubsection{Simulate the town}\label{simulate-the-town}}

After the model is generated, we simulate the town in this section.

\begin{lstlisting}[language=Python]
# Load Simulator Data from JSON file
end_time, spread_period, initialized_infected_ids, commands, observers = parser.parse_simulator_data()

# Run the simulation
simulator.simulate(end_time=end_time,
                    spread_period=spread_period,
                    initialized_infected_ids=initialized_infected_ids,
                    commands=commands,
                    observers=observers,
                    report_statistics=2)
\end{lstlisting}

\hypertarget{evaluate-the-results}{%
\subsubsection{Evaluate the results}\label{evaluate-the-results}}

\begin{figure}
  \begin{subfigure}{\linewidth}
  \includegraphics[width=.5\linewidth]{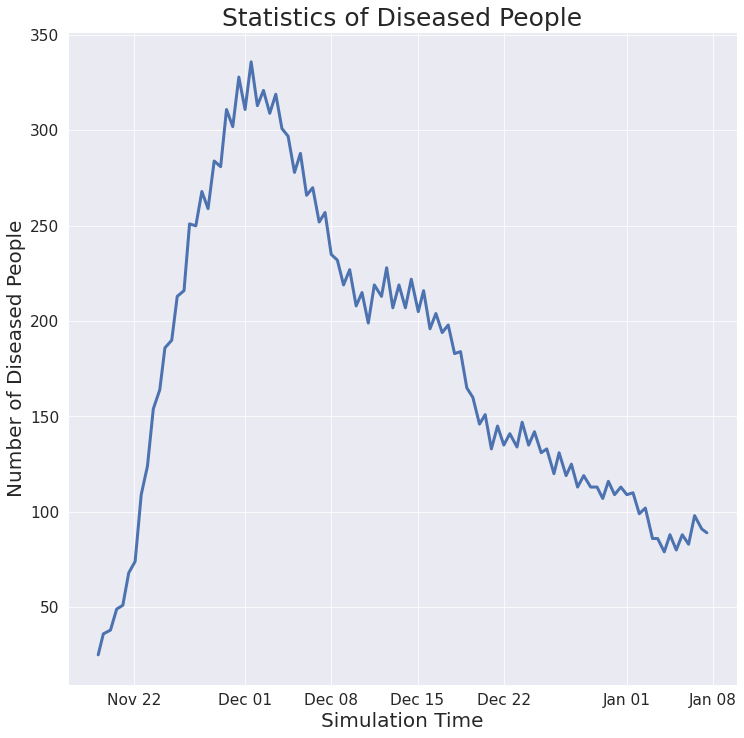}\hfill
  \includegraphics[width=.5\linewidth]{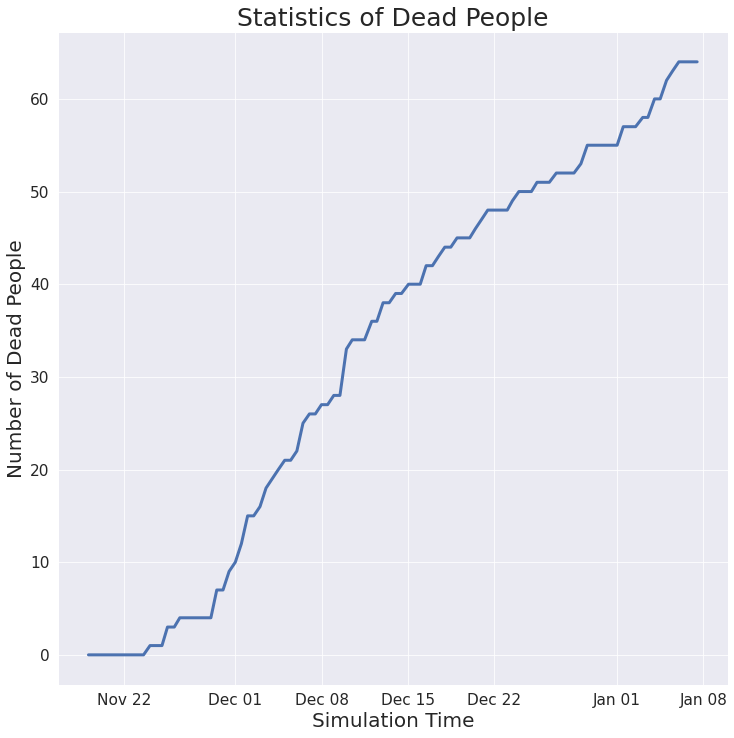}
  \caption{Statistics of dead and infected people over time}
  \end{subfigure}\par\medskip
  \begin{subfigure}{\linewidth}
  \includegraphics[width=.5\linewidth]{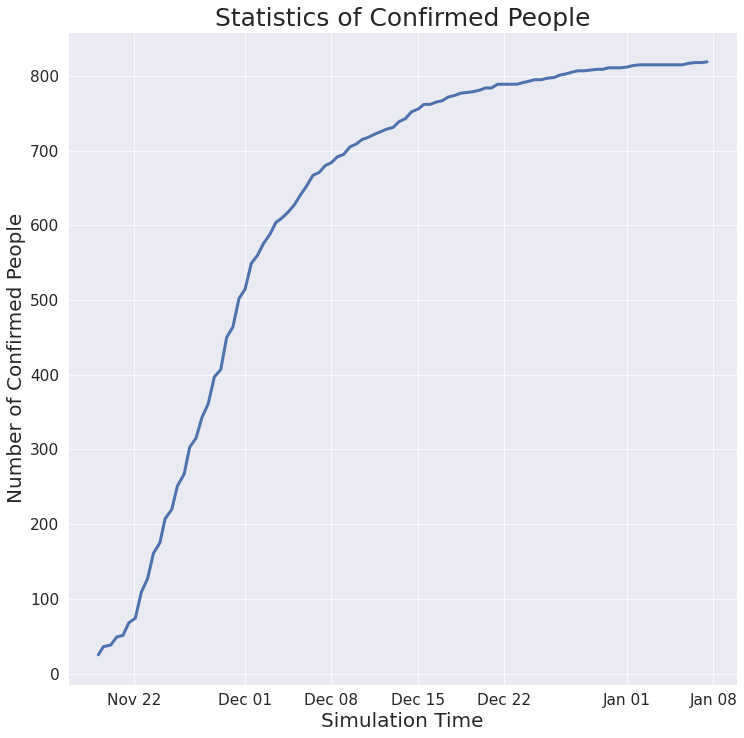}\hfill
  \includegraphics[width=.5\linewidth]{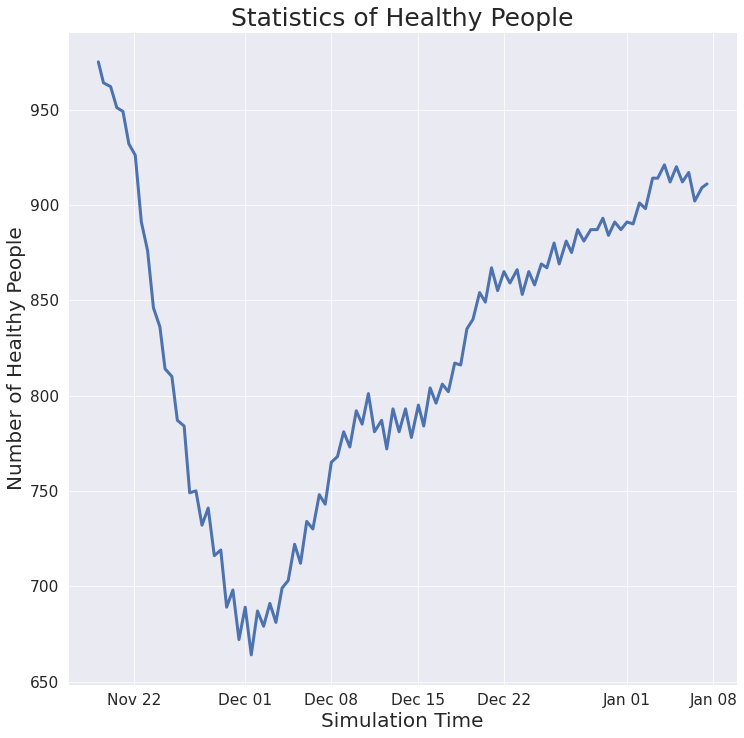}
  \caption{Statistics of healthy and confirmed cases over time}
  \end{subfigure}\par\medskip
  \caption{The results of observer plots}
  \label{fig:configure_simulation_plots}
\end{figure}

In the end, we present some plots in \Cref{fig:configure_simulation_plots} to illustrate the simulation's
results.

\begin{lstlisting}[language=Python]
observers[0].plot_disease_statistics_during_time(Health_Condition.IS_INFECTED)

observers[0].plot_disease_statistics_during_time(Health_Condition.DEAD)

observers[0].plot_disease_statistics_during_time(Health_Condition.HAS_BEEN_INFECTED)

observers[0].plot_disease_statistics_during_time(Health_Condition.IS_NOT_INFECTED)
\end{lstlisting}

\hypertarget{sanity-checks}{%
\section{Sanity checks}\label{sanity-checks}}

In this section, we demonstrate the procedure of conducting further
experiments in order to evaluate the basic functionality and sanity of
the simulator. Prior to engaging with this part, one should take a look
at Manual Simulation.

\hypertarget{import-the-necessary-libs}{%
\subsection{Import the necessary libraries}\label{import-the-necessary-libs}}

In the beginning, we import some necessary simulation libraries from the
code folder.

\begin{lstlisting}[language=Python]
import sys, os
sys.path.insert(1, os.path.join(os.pardir, 'src'))
\end{lstlisting}

\hypertarget{run-a-normal-simulation}{%
\subsection{Run a normal simulation}\label{run-a-normal-simulation}}

This is the most basic form of the simulation, with no commands or, in other words, no applied policies. In the first step, we initialize the
parser and load population generator and disease properties
configuration files from the individual JSON files.

\begin{lstlisting}[language=Python]
# Import Parser
from json_handle import Parser
parser = Parser('test')

# Load Population Generator from JSON file
population_generator = parser.parse_population_generator()

# Load Disease Properties from JSON file
disease_properties = parser.parse_disease_properties()
\end{lstlisting}

Then, simulator settings are parsed and loaded into the simulator. This
also includes the last two steps, so there is no need for the previous
code snippet, and it is generally used for explanation or debug purposes.

\begin{lstlisting}[language=Python]
# Load the simulator from JSON file
simulator = parser.parse_simulator()
simulator.generate_model()
\end{lstlisting}

Now we load simulator data as well.

\begin{lstlisting}[language=Python]
# Load simulator's data from JSON file
end_time, spread_period, initialized_infected_ids, _, observers = parser.parse_simulator_data()
\end{lstlisting}

We set the commands to an empty list and run the simulation.

\begin{lstlisting}[language=Python]
# Run the simulation
simulator.simulate(end_time=end_time,
                    spread_period=spread_period,
                    initialized_infected_ids=initialized_infected_ids,
                    commands=[], # No commands needed at this time

                    observers=observers)
\end{lstlisting}

\begin{figure}
  \begin{subfigure}{\linewidth}
  \includegraphics[width=.49\linewidth]{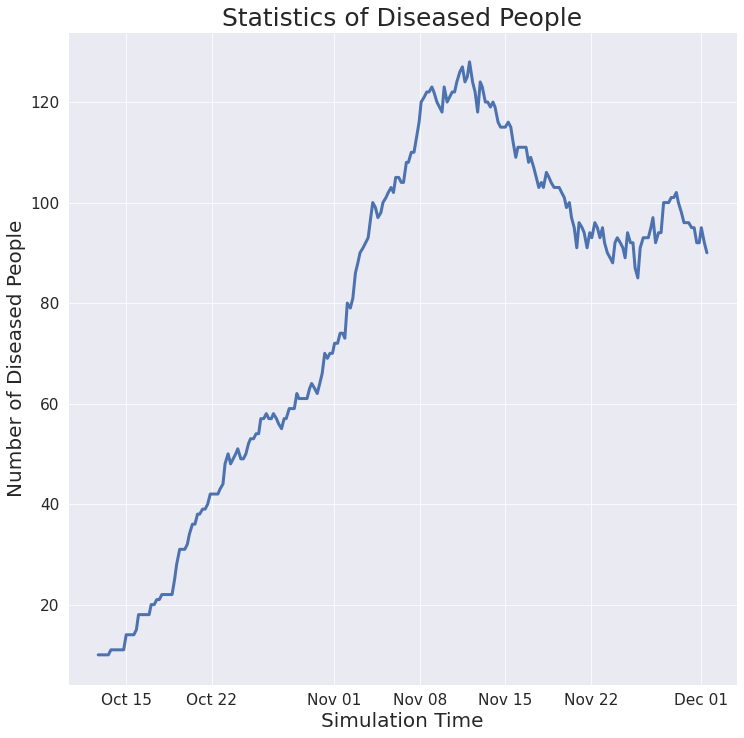}\hfill
  \includegraphics[width=.49\linewidth]{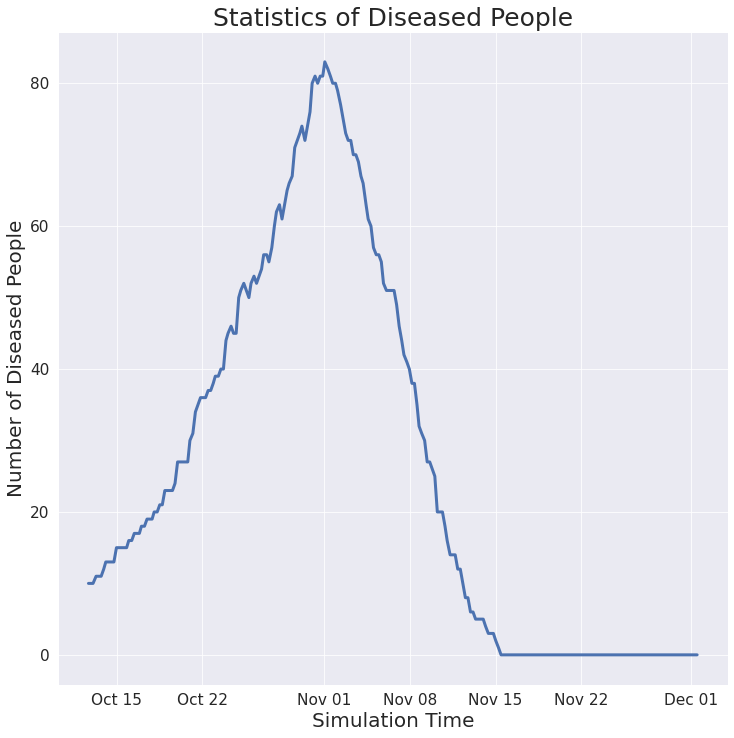}
  \caption{A plain simulation run- without interference- (left) and enforcing quarantine on everyone (right).}
  \label{fig:sanity_plain_quarantine_everyone}
  \end{subfigure}\par\medskip
  \begin{subfigure}{\linewidth}
  \includegraphics[width=.49\linewidth]{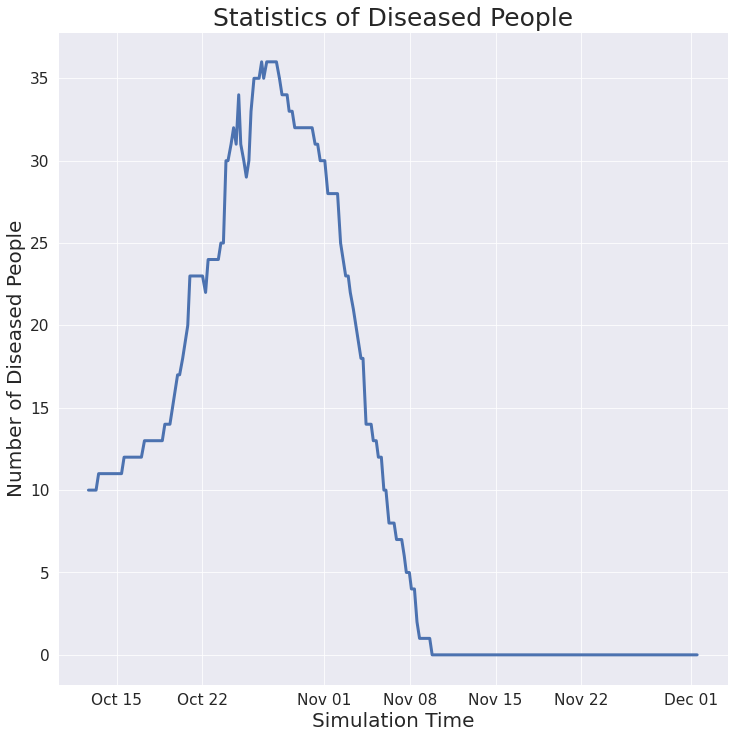}\hfill
  \includegraphics[width=.49\linewidth]{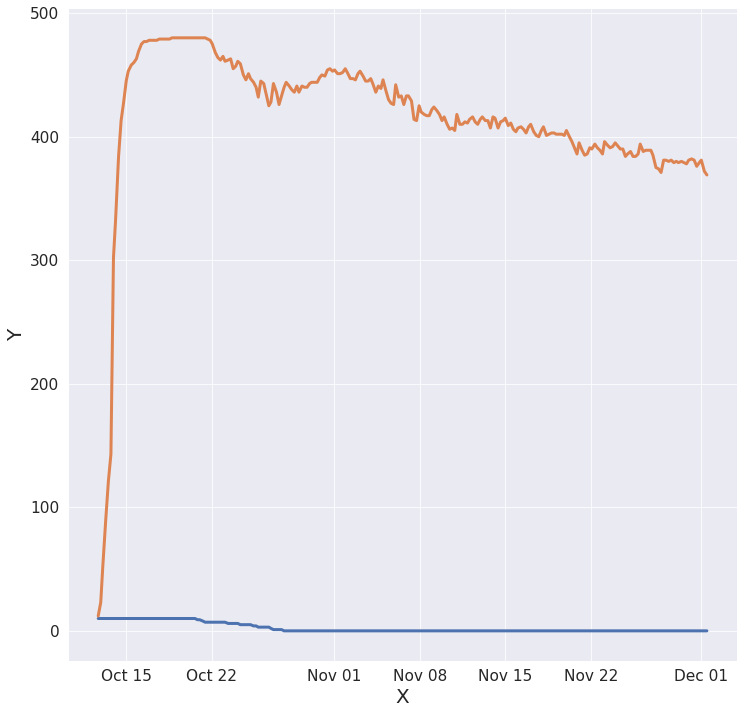}
  \caption{Enforcing quarantine on infected people (left) and effect of the infection rate (right).}
  \label{fig:sanity_quarantine_infecteds_infectious}   
  \end{subfigure}\par\medskip
  \begin{subfigure}{\linewidth}
  \includegraphics[width=.49\linewidth]{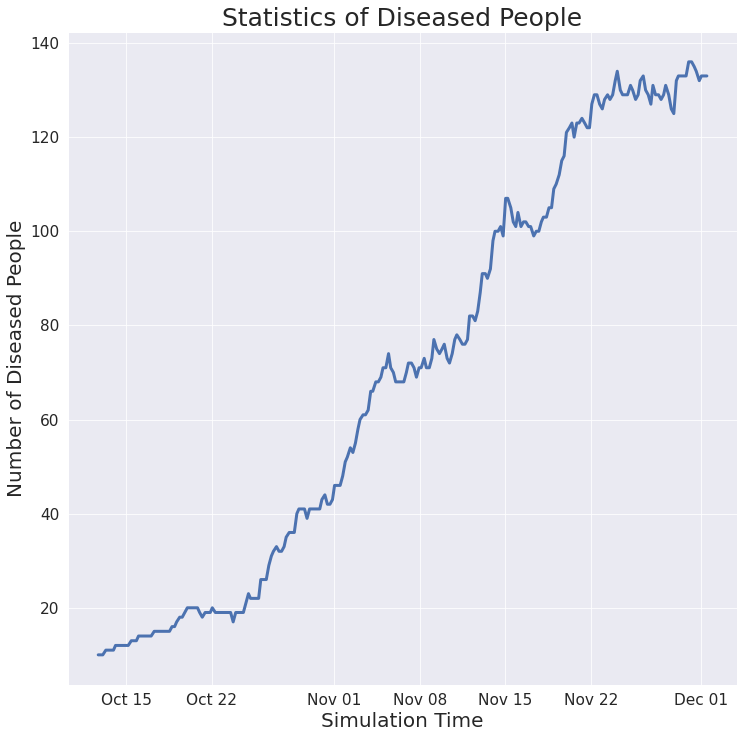}\hfill
  \includegraphics[width=.49\linewidth]{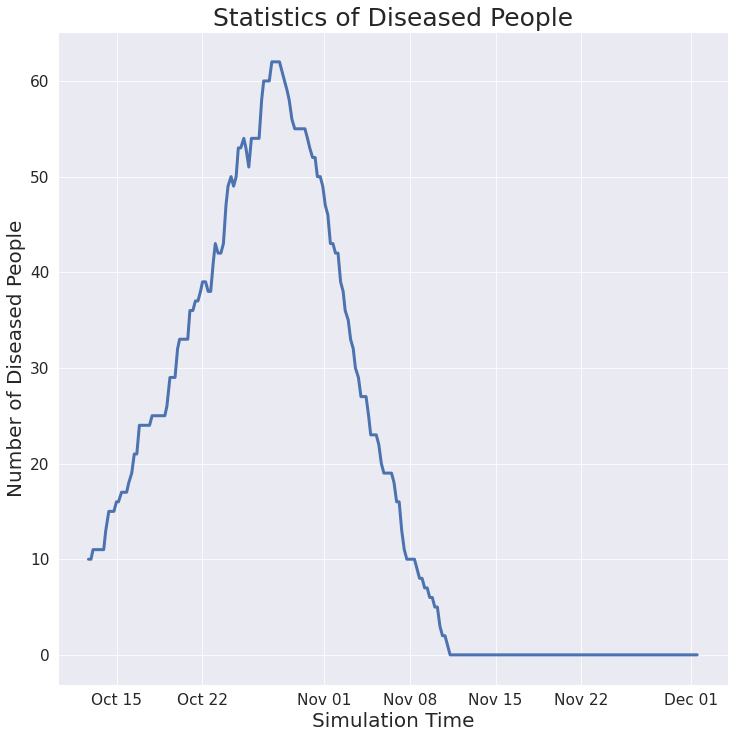}
  \caption{Effect of reducing immunity (left) and enforcing quarantine on all families (right).}
  \label{fig:sanity_immunity_quarantine_families}
  \end{subfigure}

  \caption{The results of observer plots}
  \label{fig:sanity_check_experiments}
\end{figure}

To observe the simulation results, we plot the curve of the active cases using
the observer object. The result appears in \Cref{fig:sanity_plain_quarantine_everyone}.

\begin{lstlisting}[language=Python]
from utils import Health_Condition
observers[0].plot_disease_statistics_during_time(Health_Condition.IS_INFECTED)
\end{lstlisting}

\hypertarget{simulate-with-a-quarantine-everyone-policy}{%
\subsection{Quarantine everyone}\label{simulate-with-a-quarantine-everyone-policy}}

In the next step, we add a policy to quarantine all the people after 20
days, a trivial form of quarantine, and see how the results change. You can compare the results from this section and \cref{simulate-with-a-quarantine-everyone-policy} in \Cref{fig:sanity_plain_quarantine_everyone}.

\begin{lstlisting}[language=Python]
# Import Parser
from json_handle import Parser
parser = Parser('test')

# Load the simulator from JSON file
simulator = parser.parse_simulator()
simulator.generate_model()

# Load simulator's data from JSON file
end_time, spread_period, initialized_infected_ids, _, observers = parser.parse_simulator_data()

# Build a general quarantine policy
from datetime import timedelta
from commands import Quarantine_Multiple_People
from conditions import Time_Point_Condition

commands = [Quarantine_Multiple_People(
                    condition=Time_Point_Condition(Time(timedelta(days=20))), 
                    ids=[i for i in range(500)])]

# Execute the simulation
simulator.simulate(end_time, 
                    spread_period, 
                    initialized_infected_ids, 
                    commands, 
                    observers, 
                    report_statistics=2)

# Plot the results
observers[0].plot_disease_statistics_during_time(Health_Condition.IS_INFECTED)
\end{lstlisting}

\hypertarget{simulate-with-quarantine-diseased-people}{%
\subsection{Quarantine infected
people}\label{simulate-with-quarantine-diseased-people}}

A more logical form of quarantine is to quarantine only the infected people at some point during the simulation, e.g., day 15, and naturally the
results, presented in \Cref{fig:sanity_quarantine_infecteds_infectious}, should be the same as \Cref{simulate-with-a-quarantine-everyone-policy} since the people who are not infected do not pose any threats in case they are not quarantined.

\begin{lstlisting}[language=Python]
# Import Parser
from json_handle import Parser
parser = Parser('test')

# Load the simulator from JSON file
simulator = parser.parse_simulator()
simulator.generate_model()

# Load simulator's data from JSON file
end_time, spread_period, initialized_infected_ids, _, observers = parser.parse_simulator_data()

# Build a policy
from datetime import timedelta
from commands import Quarantine_Diseased_People
from conditions import Time_Point_Condition
commands = [Quarantine_Diseased_People(
                    condition=Time_Point_Condition(Time(timedelta(days=15))))]

# Execute the simulation
simulator.simulate(end_time, 
                    spread_period, 
                    initialized_infected_ids, 
                    commands, 
                    observers, 
                    report_statistics=2)

# Plot the results
from utils import Health_Condition
observers[0].plot_disease_statistics_during_time(Health_Condition.IS_INFECTED)
\end{lstlisting}

\hypertarget{infectious-rate}{%
\subsection{Infection rate}\label{infectious-rate}}

Here, we set the infection rate to a meager amount and check the test
result. After setting the infection rate to another higher value, we compare the two curves in \Cref{fig:sanity_quarantine_infecteds_infectious}.

\begin{lstlisting}[language=Python]
# Import Parser (default constructor is the 'test' folder)
from json_handle import Parser
parser = Parser('test')

# Load the simulator from JSON file
simulator = parser.parse_simulator()

# Change the infection rate here (or just change in the respective JSON file and then parse the simulator)
from distributions import Uniform_Disease_Property_Distribution
simulator.disease_properties.infectious_rate_distribution = \
    Uniform_Disease_Property_Distribution(parameters_dict={"upper_bound":0.2, "lower_bound":0.1})

# Generate the simulation model
simulator.generate_model()

# Load simulator's data from JSON file
end_time, spread_period, initialized_infected_ids, _, observers = parser.parse_simulator_data()

# No policy is required
commands = []

# Execute the simulation
simulator.simulate(end_time, 
                    spread_period, 
                    initialized_infected_ids, 
                    commands, 
                    observers, 
                    report_statistics=2)
\end{lstlisting}

In the following code snippet, we increase the infection rate by assigning a
uniform distribution with much higher lower and upper bounds.

\begin{lstlisting}[language=Python]
# Load the simulator from JSON file
simulator = parser.parse_simulator()

# Change the infection rate here (or just change in the respective JSON file and then parse the simulator)
from distributions import Uniform_Disease_Property_Distribution
simulator.disease_properties.infectious_rate_distribution = \
    Uniform_Disease_Property_Distribution(parameters_dict={"upper_bound":0.95, "lower_bound":0.9})

# Generate the simulation model
simulator.generate_model()

# Load simulator's data from JSON file
end_time, spread_period, initialized_infected_ids, _, observers = parser.parse_simulator_data()

# No policy is required
commands = []

# Execute the simulation
simulator.simulate(end_time, 
                    spread_period, 
                    initialized_infected_ids, 
                    commands, 
                    observers, 
                    report_statistics=2)
\end{lstlisting}

The results of both lower and higher infection rates are shown in \Cref{fig:sanity_quarantine_infecteds_infectious}. A significant displacement in the peak of the curve is
observable that exactly matches our expectations from changing the
infection rate.

\begin{lstlisting}[language=Python]
from utils import Health_Condition
data_2 = observers[0].get_disease_statistics_during_time(Health_Condition.IS_INFECTED)

from plot_utils import Plot
Plot.plot_multiple_lines(data_1[1], [data_1[0], data_2[0]])
\end{lstlisting}

\hypertarget{decrease-immunity}{%
\subsection{Decrease immunity}\label{decrease-immunity}}

In this section, the immunity is decreased, and the results of the
simulation are shown in the following figure. With this amount of
immunity, almost every person should get infected. The pandemic
curve will also not become flat since there is a small generated immunity
after catching the infectious disease for the first time. The results of this experiment are observable in \cref{fig:sanity_immunity_quarantine_families}.

\begin{lstlisting}[language=Python]
# Import Parser
from json_handle import Parser
parser = Parser('test')

# Load the simulator from JSON file
simulator = parser.parse_simulator()

# Change the infection rate here (or just change in the respective JSON file and then parse the simulator)
from distributions import Uniform_Disease_Property_Distribution
simulator.disease_properties.immunity_distribution = \
    Uniform_Disease_Property_Distribution(parameters_dict={"upper_bound":0.03, "lower_bound":0.02})


# Generate the simulation model
simulator.generate_model()

# Load simulator's data from JSON file
end_time, spread_period, initialized_infected_ids, _, observers = parser.parse_simulator_data()

# No policy is required
commands = []

# Execute the simulation
simulator.simulate(end_time, 
                    spread_period, 
                    initialized_infected_ids, 
                    commands, 
                    observers, 
                    report_statistics=2)

# Plot the results
from utils import Health_Condition
observers[0].plot_disease_statistics_during_time(Health_Condition.IS_INFECTED)
\end{lstlisting}

\hypertarget{quarantine-all-the-families}{%
\subsection{Quarantine all
families}\label{quarantine-all-the-families}}

The simulator is capable of enforcing quarantines based on families, in
addition to persons and communities. In this part, we impose a full
quarantine over all the families and observe the results. This should
have the same effect as quarantining all the people. The result of this quarantine is depicted in \Cref{fig:sanity_immunity_quarantine_families}.

\begin{lstlisting}[language=Python]
# Import Parser
from json_handle import Parser
parser = Parser('test')

# Load the simulator from JSON file
simulator = parser.parse_simulator()
simulator.generate_model(is_parallel=False)

# Load simulator's data from JSON file
end_time, spread_period, initialized_infected_ids, _, observers = parser.parse_simulator_data()

# Build a policy
from datetime import timedelta
from commands import Quarantine_Multiple_Families
from conditions import Time_Point_Condition

commands = [Quarantine_Multiple_Families(
                    condition=Time_Point_Condition(Time(timedelta(days=15))), 
                    ids=[i for i in range(len(simulator.families))])]

# Execute the simulation
simulator.simulate(end_time, 
                    spread_period, 
                    initialized_infected_ids, 
                    commands, 
                    observers, 
                    report_statistics=2)
\end{lstlisting}
    
\section{Additional experiments}
\label{sec:additional_experiments}

In addition to the experiments presented in \Cref{sec:experiments}, another set of experiments has been developed here in order to demonstrate the simulator's ability to cope with any given criteria. The upcoming experiments are fundamentally the same as \Cref{sec:experiments}. However, the following differences in the configuration files are notable. 

\begin{enumerate}
    \item Incubation and disease period: the disease period and incubation rate are derived from uniform distributions of lower values (incubation period: Uniform [1, 2] and disease period: Uniform [5, 9]), as opposed to a normal distribution. 
    
    \item The population has a more simple structure but the same size, which means that people are divided into large communities instead of what is described in \Cref{tab:communities_information}.
    
    \item The simulation duration is shorter than before, four months instead of ten months in \Cref{sec:experiments}.
\end{enumerate}

\begin{figure}
  \begin{subfigure}{\linewidth}
  \includegraphics[width=.5\linewidth]{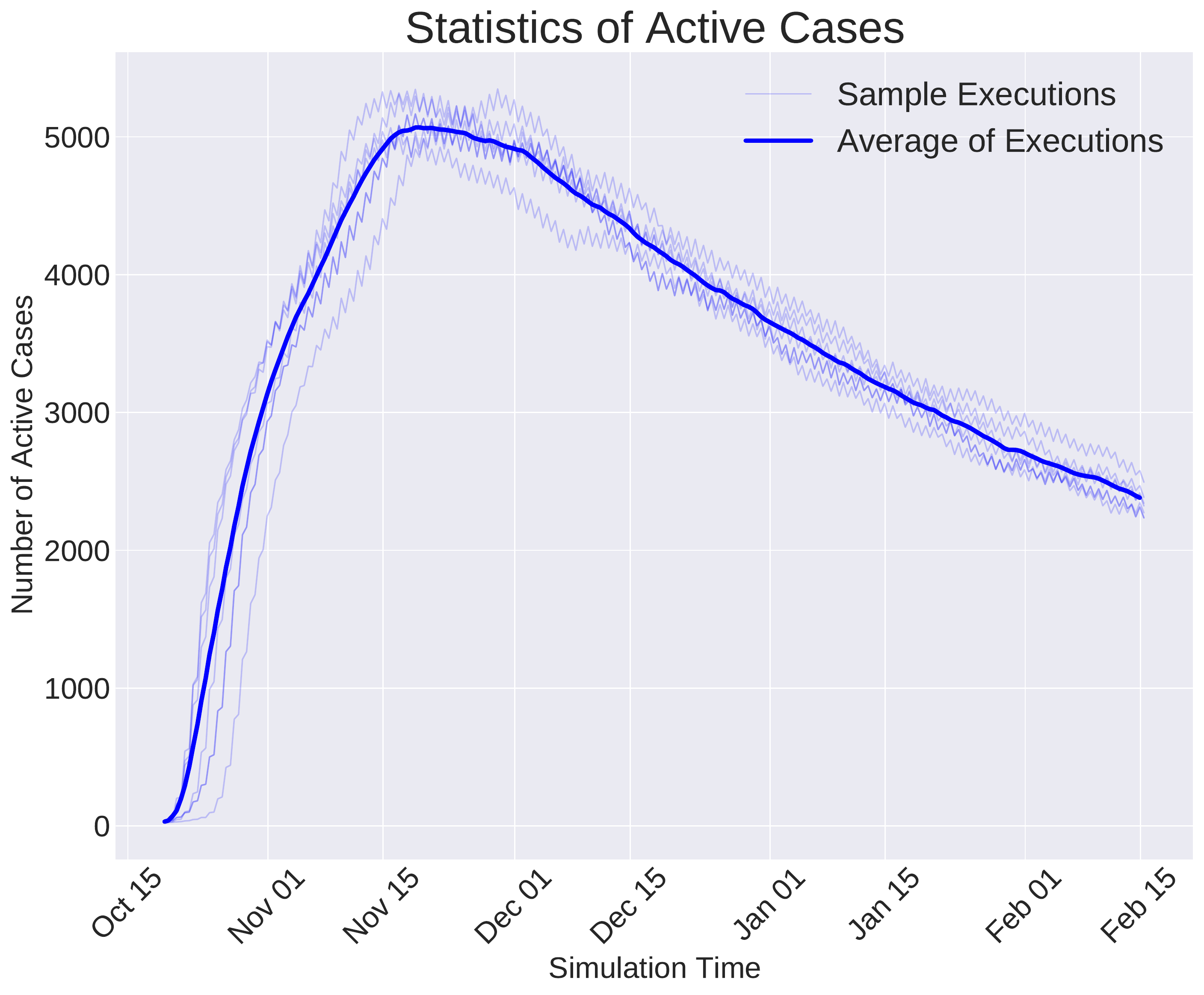}\hfill
  \includegraphics[width=.5\linewidth]{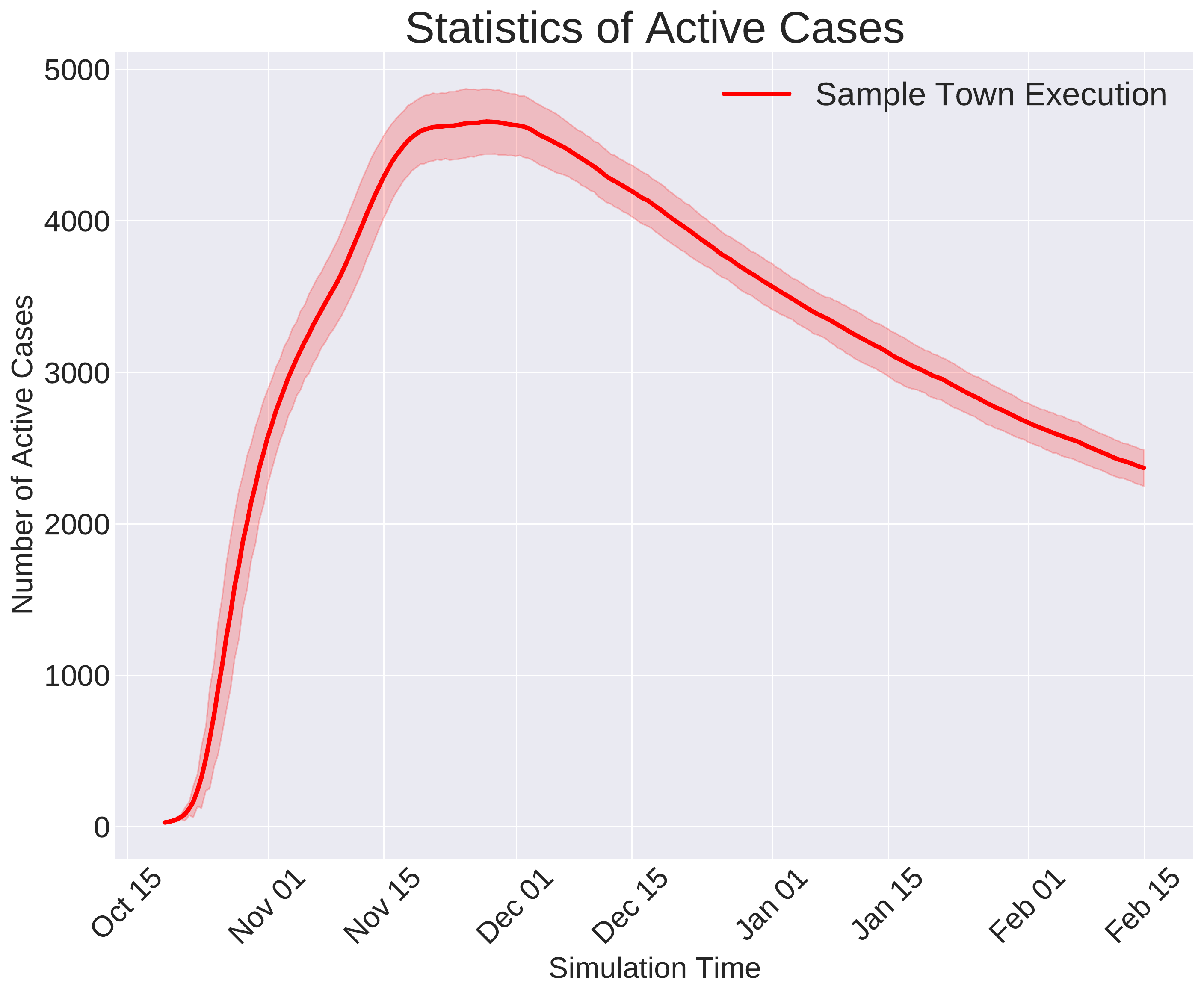}
  \caption{Left) Normal executions of the sample city. Due to the stochastic characteristics of the simulation, each run is slightly different compared to the others. Therefore, the blue line is introduced as the average of all executions, smoothed by a moving average of windows size 2. Right) A plain execution of the simulation without any interference, i.e., the virus spreads from the initially infected people to others without the presence of any prevention measures. Since the simulation has a stochastic nature, the error bands are displayed as a confidence measure when making deterministic conclusions.}
  \end{subfigure}\par\medskip
  \begin{subfigure}{\linewidth}
  \includegraphics[width=.5\linewidth]{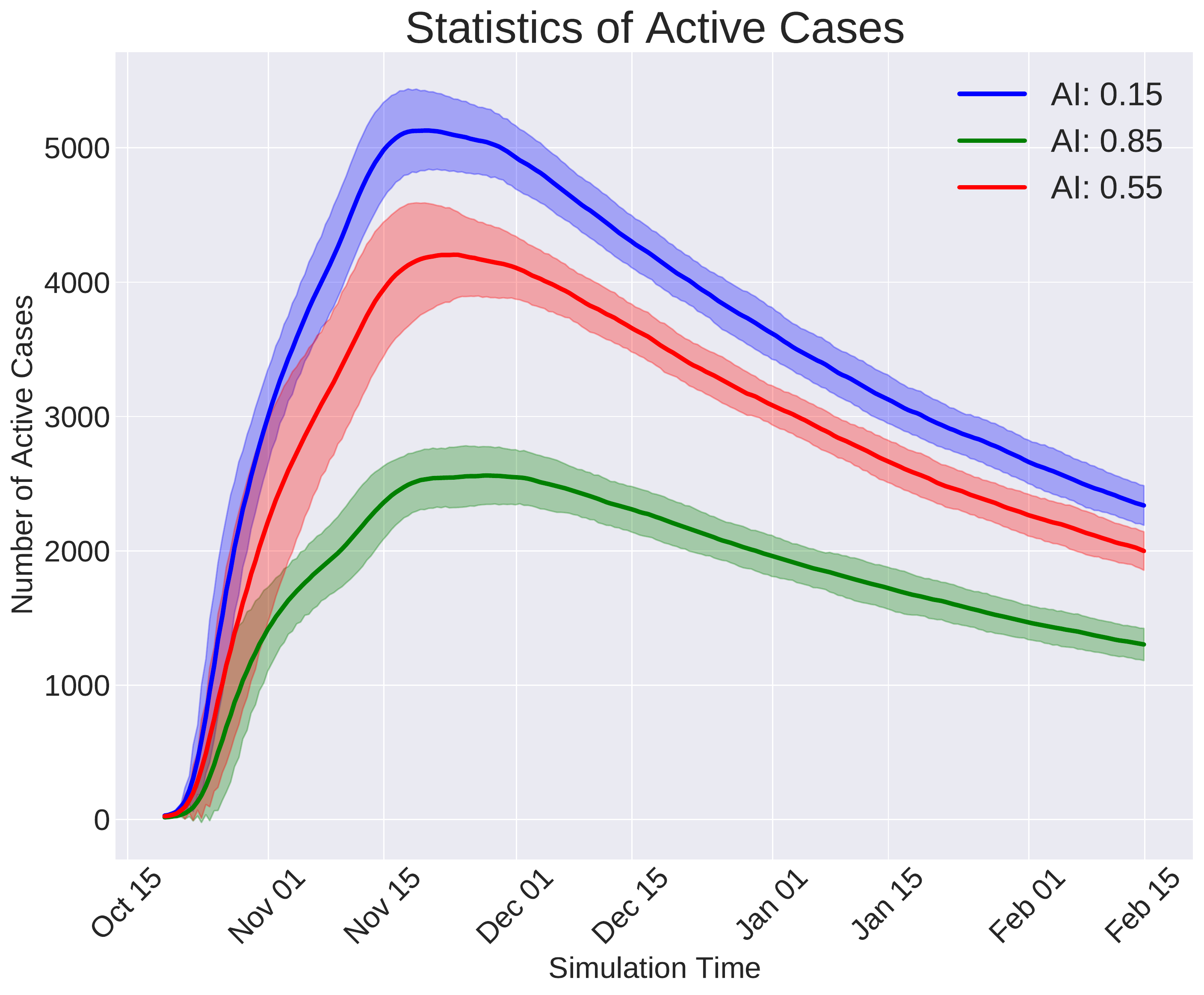}\hfill
  \includegraphics[width=.5\linewidth]{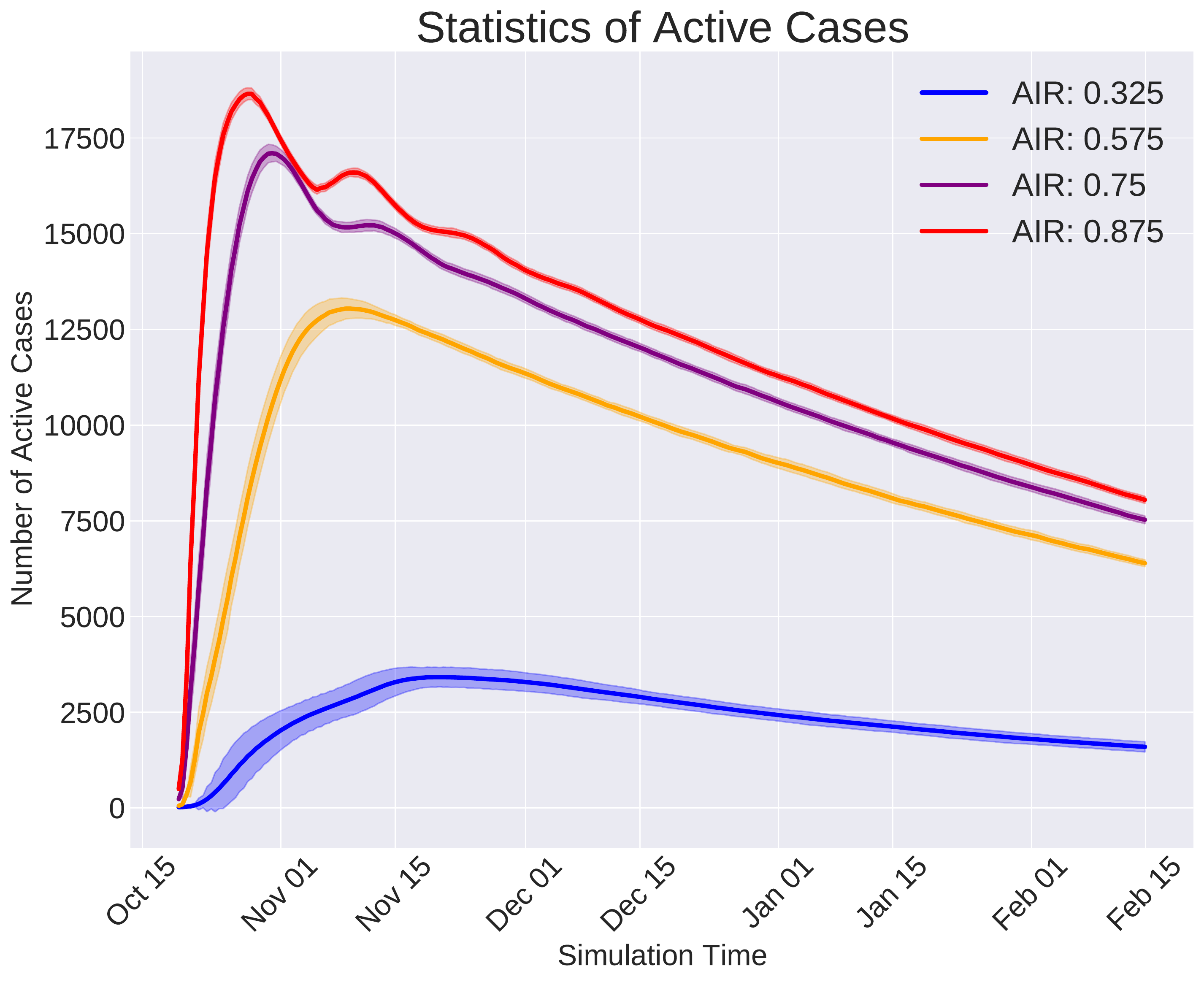}
  \caption{Left) The effect of the immunity rate on the simulation. In this set of experiments, the immunity rate is sampled from three different uniform distributions. The results indicate that larger immunity rates give rise to flatter curves. Note that AI (Average Immunity) is the mean of the uniform distribution from which the immunity rate of each curve is sampled. Right) Effect of infection rate on the statistic of active cases. Each curve corresponds to a specific infection rate distribution. The results indicate that the larger the infection rate gets, the higher the spread slope will be. As a result, it takes less time for the number of active cases to reach its peak. Notice that AIR stands for the Average Infection Rate, which is the mean distribution from which the infection rate is sampled.}
  \end{subfigure}\par\medskip

  \caption{First part of the additional experiment results, including disease properties and normal executions.}
    \label{fig:plots_additional_experiments_part1}
\end{figure}

\begin{figure}
  \begin{subfigure}{\linewidth}
  \includegraphics[width=.5\linewidth]{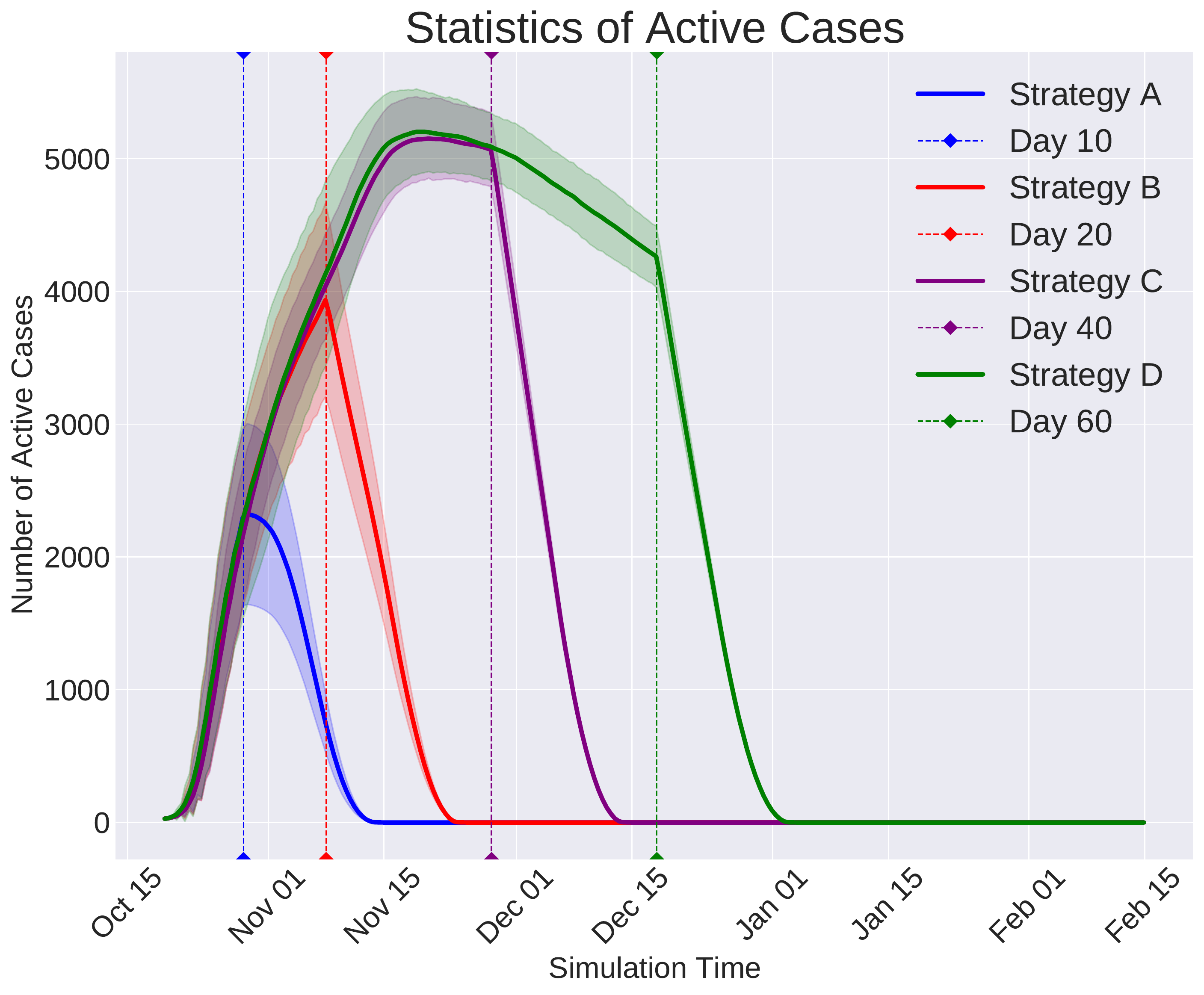}\hfill
  \includegraphics[width=.5\linewidth]{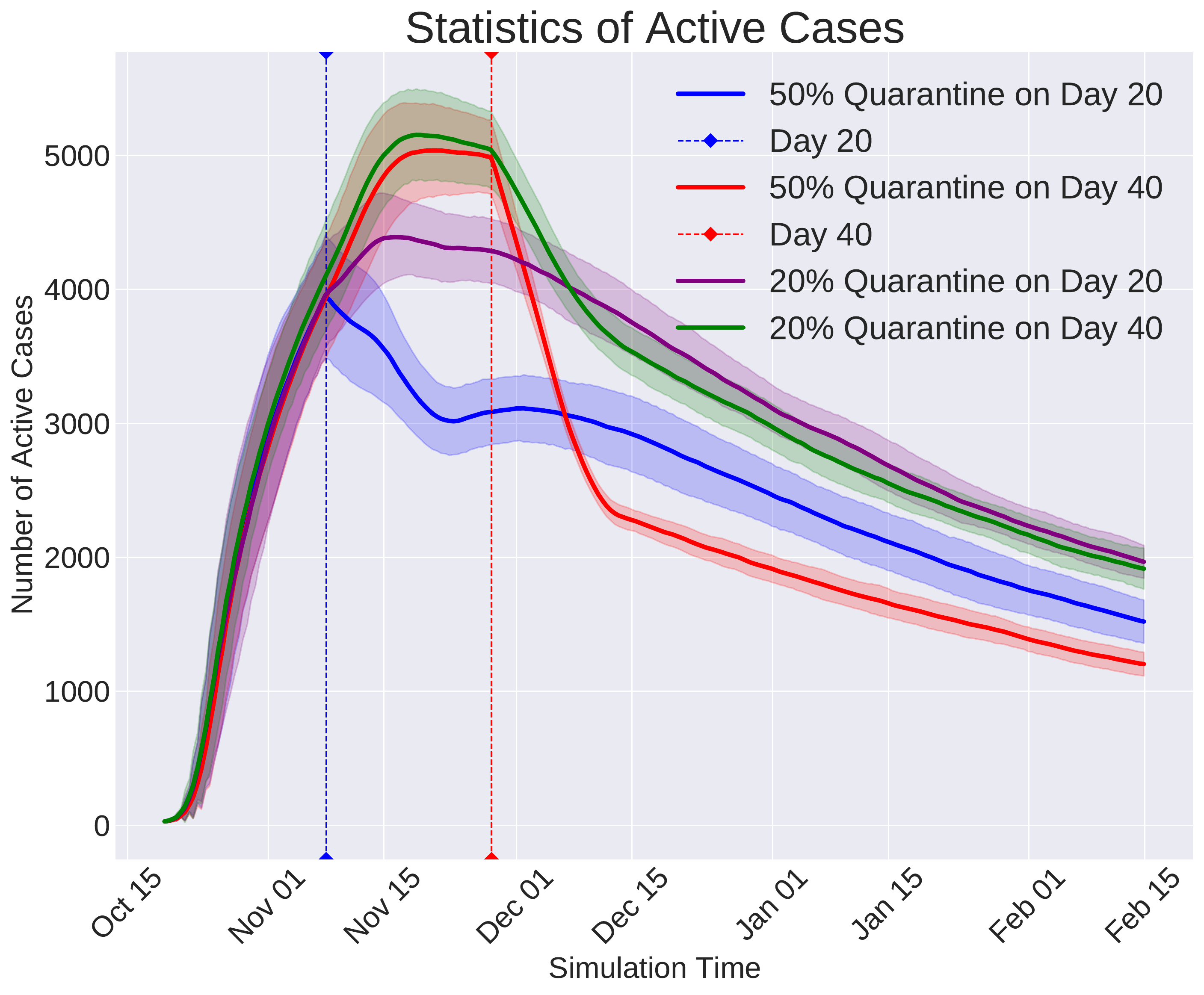}
  \caption{Left) Quarantine the currently infected people activated by time point conditions. This experiment focuses on enforcing universal quarantines on a specific day of the simulation. Quarantines may be applied at any time point during a simulation. In this figure, the quarantines are mandated both before and after the active cases' curve reaches its maximum value. A, B, C, and D strategies enforce a quarantine after 40, 60, 100, and 200 days after the beginning of the outbreak. Right) Quarantine the currently infected people activated by time point conditions and detection error. This experiment focuses on enforcing a probabilistic quarantine after a specific number of days. In this experiment, the process of quarantining the infected people suffers from an error in the detection of active cases; therefore, changing the last part's deterministic nature into a probabilistic one.}
  \end{subfigure}\par\medskip
  \begin{subfigure}{\linewidth}
  \includegraphics[width=.5\linewidth]{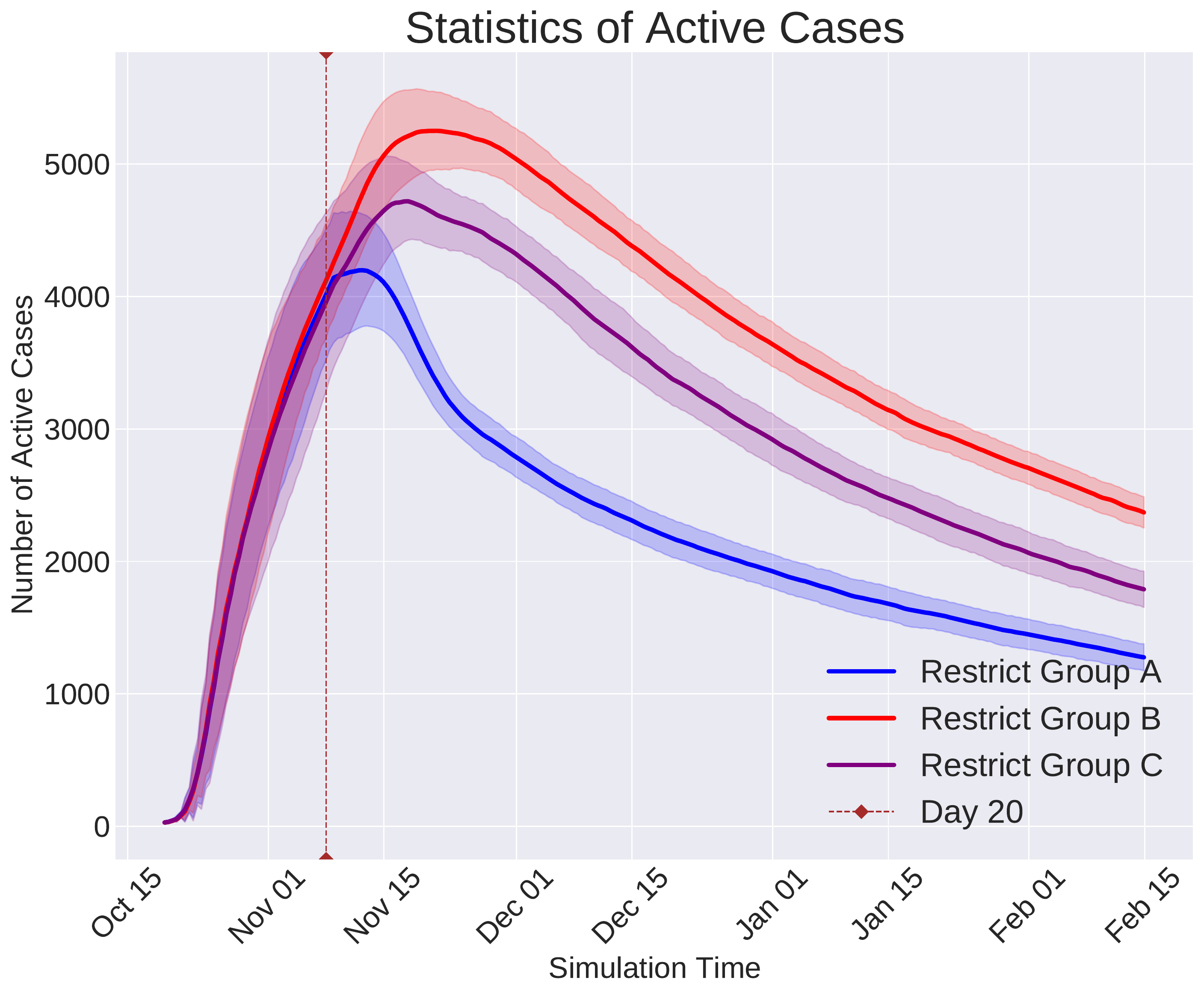}\hfill
  \includegraphics[width=.5\linewidth]{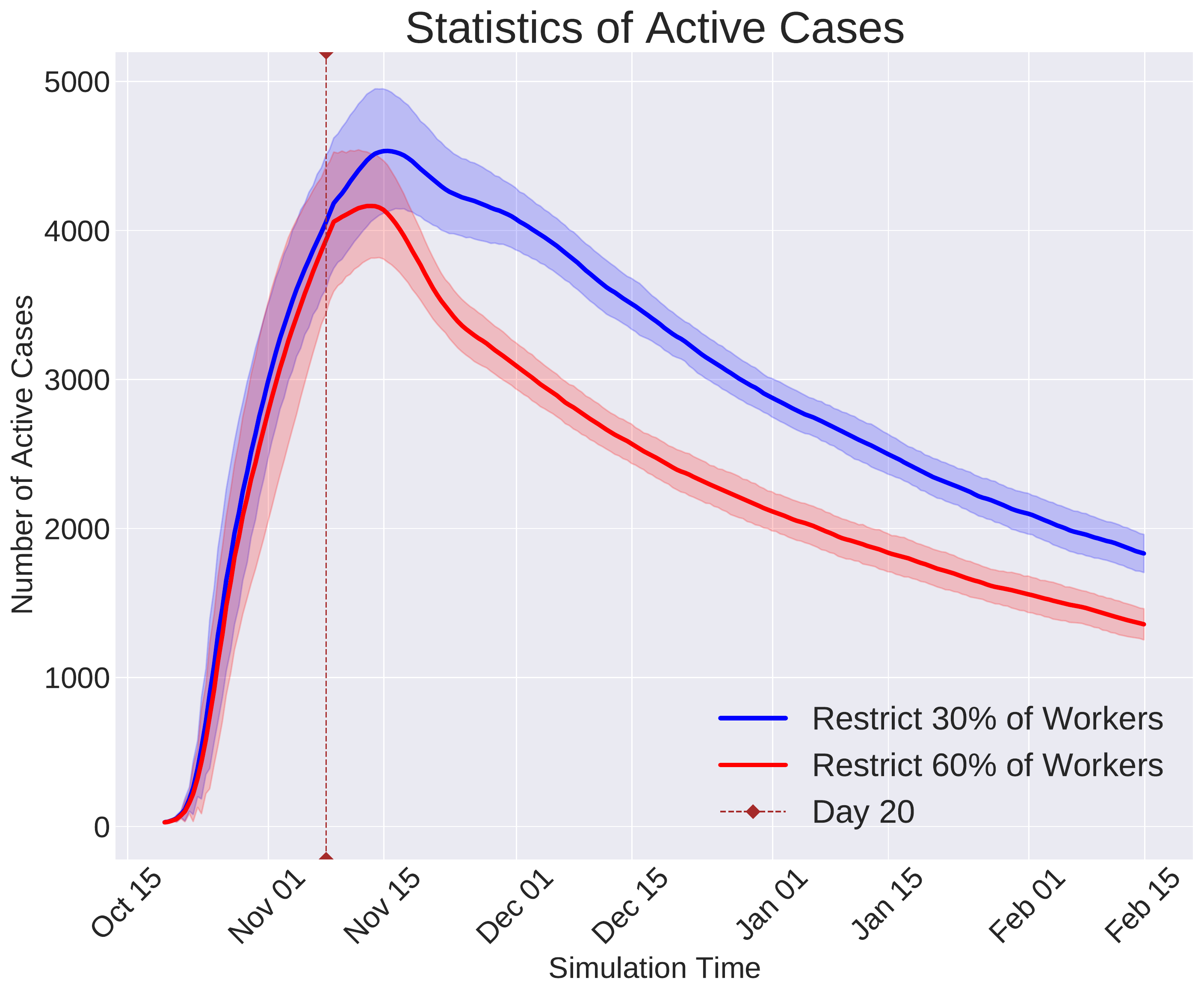}
  \caption{Left) Quarantine specific sectors of society. Aiming at the importance of quarantining various economic sectors of society, this experiment is dedicated to displaying the outcomes of this type of quarantine on the population and the graph of currently infected cases. Group A includes all the workspaces of any size. Group B consists of gyms, restaurants, and cinemas, while group C is focused on more public groups like malls and public transportation. Right) Quarantine specific roles of the society. The simulator is also capable of contracting the virus spread by enforcing restrictions on certain roles in society. Specifically, in this experiment, we quarantine only workers of any kind, e.g., industry and offices. The effect is quite tremendous since the workplaces, in general, have a considerable transmission potential due to highly possible and repetitive close contacts.}
  \end{subfigure}\par\medskip

  \caption{Second part of the additional experiment results, including exemplary restriction measures.}
    \label{fig:plots_additional_experiments_part2}
\end{figure}

\begin{figure}
  \begin{subfigure}{\linewidth}
  \includegraphics[width=.5\linewidth]{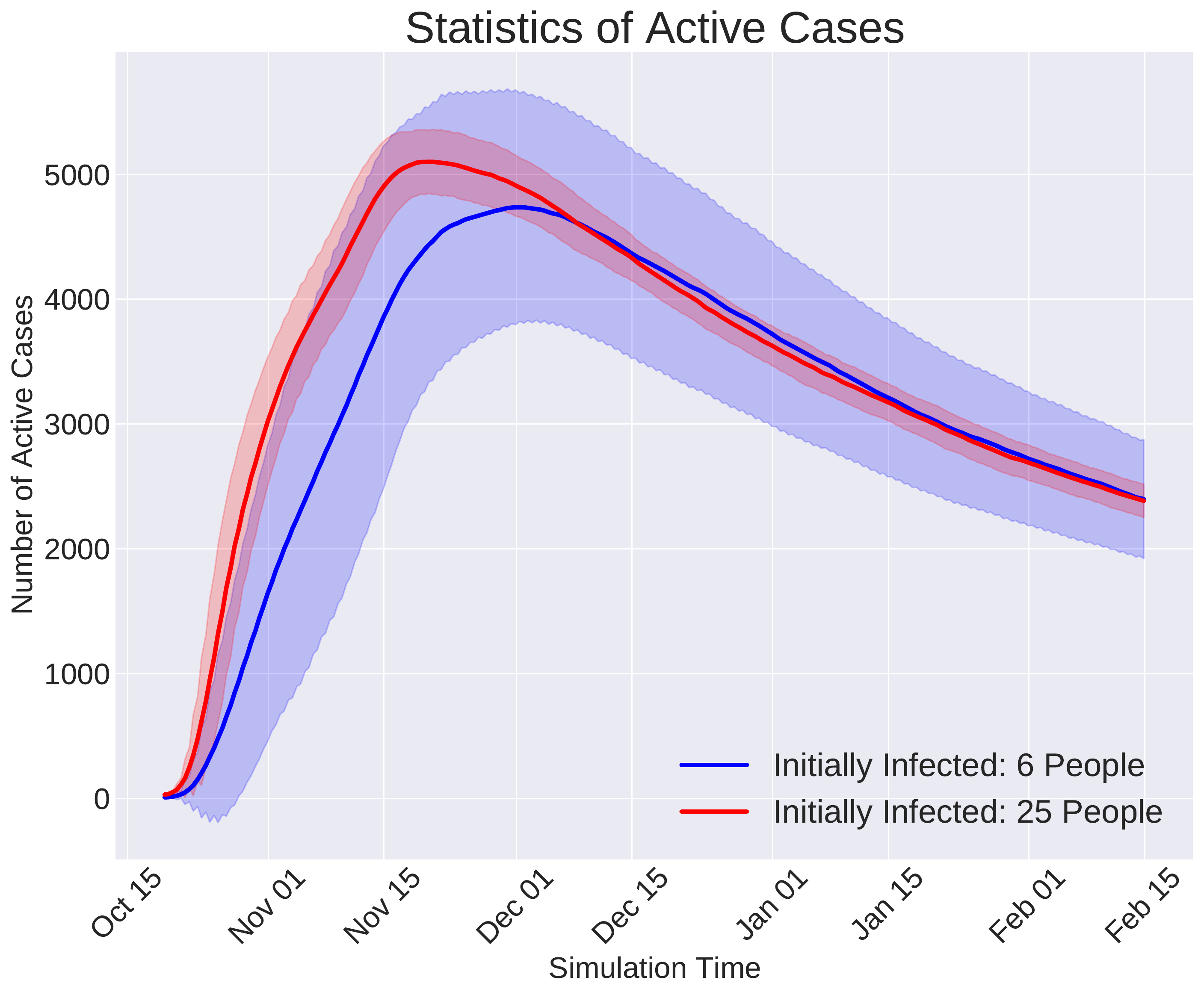}\hfill
  \includegraphics[width=.5\linewidth]{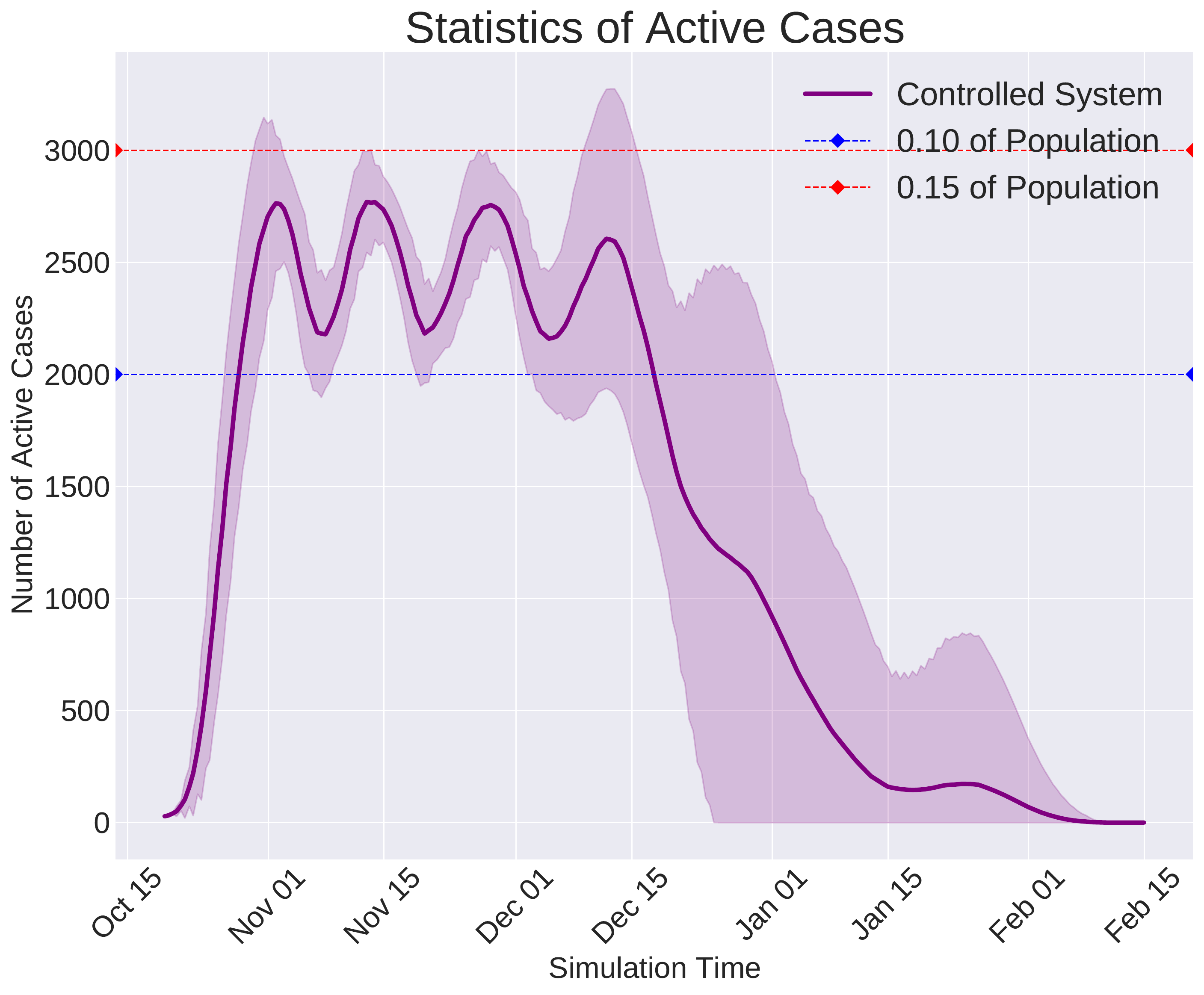}
  \caption{Left) The effect of the number of initially infected people. The confidence intervals are expectedly wider for smaller initially infected set because it results in some communities without an initial spreader and consequently a less homogeneous spread of the disease. Right) The bang-bang controller. In this experiment, two different conditions are in place. One triggers quarantining the infected people; the other one triggers lifting the quarantine. Whenever the ratio of currently infected cases to the population reaches $15\%$, quarantines are enforced, and when the ratio declines below $10\%$, the quarantine is lifted.}
  \end{subfigure}\par\medskip
  \caption{Last part of the additional experiment results, including bang-bang controller and effect of initially infected individuals.}
  \label{fig:plots_additional_experiments_part3}

\end{figure}

The results presented in \Cref{fig:plots_additional_experiments_part1}, \Cref{fig:plots_additional_experiments_part2}, and \Cref{fig:plots_additional_experiments_part3}, imply that the simulator is flexible with regard to various population and disease structures. Moreover, as appears in the mentioned experiments, our inference in a larger population is quite the same as what we observe in a simpler structure, indicating that the simulator is expandable to any population size if enough information is available about the overall structure. 

\clearpage

\end{document}